\documentclass[11pt]{article}
 \usepackage{amsmath, amsthm, amssymb}	% for the aligned equations
 \usepackage{graphicx}    % needed for including graphics e.g. EPS, PS
\usepackage[long, nodayofweek, 24hr]{datetime}

\usepackage[
	pdfpagemode={UseOutlines},
	pdfstartview={FitH},
	colorlinks,
	urlcolor={blue},
	pdftitle={On the Interpretation of Spectral Red-Shift in Astrophysics: A Survey of Red-Shift Mechanisms - II},
	pdfauthor={Louis Marmet},
	pdfsubject={Red-shift mechanisms},
	pdfkeywords={red-shift, redshift, red-shift mechanisms, redshift mechanisms, spectral,  cosmological, Big Bang},
	pdfdisplaydoctitle
]{hyperref}

\begin{document}

\title{On the Interpretation of Spectral Red-Shift in Astrophysics:\\A Survey of Red-Shift Mechanisms - II}

\author{Louis Marmet\footnote{Adjunct professor, Physics and Astronomy, York University, Toronto ON, Canada  M3J 1P3}}

%\date{Date: \today\ / Time: \currenttime}
\date{\today}

\newcounter{mechanism_number}

\maketitle

\begin{abstract}
This paper gives a compilation of physical mechanisms which have been proposed to explain the spectral red-shift of distant astronomical objects.  Over sixty selected mechanisms are listed for the purpose of quantitative comparisons.  Some of these mechanisms may not completely account for all astrophysical observations but are relevant to particular situations such as the solar limb red-shift, the red-shift of quasars or the cosmological red-shift.  However, this paper focuses mainly on the \ref{count:mechanism_number} mechanisms giving a quantitative description of the cosmological redshift-distance relationship.  A description is given for each mechanism including its properties, limits of applicability, functional relationships and comments.
\end{abstract}

\section{Introduction}

After the first observations of the spectral red-shift of nebulae in the early 20$^{\mbox{th}}$ century, it took Hubble a little over ten years to formulate the empirical \emph{Redshift-Distance Law} of galaxies.  The spectra of galaxies were redshifted in agreement with the Doppler effect and the observed linear relation between distance and red-shift, now called the Hubble law, seemed to imply large receding velocities of distant galaxies and an enormous amount of kinetic energy.  These velocities also indicated that all matter originated from a dense and hot state, the ``Big~Bang'', several billion years ago.  This was recognized in 1929 by Fritz Zwicky who, in order to avoid what was considered ``extraordinary implications'', suggested another mechanism to explain the observed red-shift\cite{Zwicky773.1929} where photons lose energy as they propagate through space.  Working with Einstein's equations of general relativity, Friedman and Lema\^itre found solutions in which space was expanding.  Their solutions became accepted as the mechanism producing the observed red-shifts and are today part of the current favoured hypothesis of the formation of the universe.

However the cosmology implied by general relativity was never completely accepted by Hubble.  In a series of lectures he gave in 1936, he explains that ``the familiar [Doppler] interpretation of red-shifts seems to imply a strange and dubious universe, very young and very small''\cite{Hubble1937.68}.  %p.26
Today, the universe is estimated to be ten times larger and older than Hubble's universe and the cosmological red-shift is explained by the expansion of space.  Even then, the mechanism behind many red-shift observations remains unclear --- the expansion of space does not explain the solar limb red-shift, the ``K-effect'', compact groups of galaxies with one discordant red-shift member, intrinsic red-shifts of quasars, quantization of red-shifts, alignments of galaxies showing the ``Finger of God'', and the high degree of uniformity of the CMB radiation.  As a result of these anomalies, many mechanisms and models have been proposed in an attempt to provide an explanation\cite{Gray.DunningDavies4085.2008}.  Some of these mechanisms are based on an idea similar to Zwicky's while others are based on the Doppler red-shift, space expansion, a Big~Bang scenario and even `new physics'.

This paper lists many of these hypotheses and attempts to provide the basic quantitative predictions made by each, so that comparisons between the many possibilities are possible.  The main focus of this paper is to extend an earlier review\cite{MarmetL315.2009} of red-shift mechanisms related to the cosmological distance to the observed objects.  Other red-shift mechanisms, not necessarily related to distance, are given a short description at the end of the paper.  They may explain, for example, intrinsic red-shifts observed in quasars.

The red-shift mechanisms are classified under seven categories based on how space, time, matter and light combine to produce the red-shift:
\begin{enumerate}
	\item A time-dependent distance or metric of space,
	\item a time-dependent property of gravity,
	\item a time-dependent property of matter,
	\item a time-dependent property of light or an interaction of light with itself,
	\item a time-independent geometry of space and time,
	\item a time-independent property of a field, gravitational or other,
	\item an interaction between light and matter.
\end{enumerate}
In some cases, a subset of mechanisms are selected within a category and some functional relationships are plotted as a function of red-shift.  This is presented for a quick comparison with $\Lambda$-Cold-Dark-Matter Cosmology and other mechanisms within the same category.

Because of the wide range of conditions under which red-shifts are observed in astronomy, it is likely that more than one mechanism is at play.  Every red-shift mechanisms listed here may have a contribution to the cosmological red-shift.  This attempt to give objective reviews of the mechanisms is not without some risks.  I cannot understand each of these models in all their details.  Some descriptions are limited to the available space and my understanding of the physics involved in the mechanism.  This point is especially important for the functional relationships given here: in the original articles, the functions are often expressed differently or not provided at all.  I had the chance to be helped by several proponents of these mechanisms, but the reader is encouraged to study the original work.

An important question is how can one determine if a theory is ``better'' than another.  The answer to this is not easy to obtain and is highly dependent on personal preferences!  However, one necessary first step is to have quantitative predictions from all the theories to be compared.  A closer agreement with experimental data is certainly a step in the right direction, but not a sufficient condition to accept a theory.  This paper is an attempt to providing quantitative predictions from several theories.  Some other desirable properties of a scientific theory are that it has few adjustable parameters, no ad hoc hypothesis, does not contradict existing observations and can be confirmed or refuted experimentally with repeatable measurements\cite{Dilworth128.2009}.

Even today, ``we seem to face, as once before in the days of Copernicus, a choice between a small, finite universe, and a universe indefinitely large plus a new principle of nature''~\cite{Hubble1937.68}.  In this paper, I use the spelling of 'red-shift' used by Hubble in honour of his long term insight into the future.

\section{Definitions}

In order to simplify the list of the many different mechanisms presented here, this section gives definitions used in the majority of cases.  In the case of a disagreement with the definitions given in this section, the specific details are given in the description of the mechanism.

\subsection{Units and constants}

The SI System of Units\cite{BIPM180.2006} is used in this paper.  Although this is the system of units currently used in physics, the explanation of some red-shift mechanisms is based on different definitions of the meter or the second.  As a result, the descriptions given below might represent these mechanisms as I see them through the tinted glasses of the SI System.  These units are defined locally to an observer in the following way: ``The metre is the length of the path travelled by light in vacuum during a time interval of $1/299\,792\,458$ of a second''.  The speed of light is thus $c = 299\,792\,458$~m/s.  ``The second is the duration of $9\,192\,631\,770$ periods of the radiation corresponding to the transition between the two hyperfine levels of the ground state of the caesium~133 atom''~\cite{BIPM22.2006}.

Other constants used in this paper have the following values.  Planck's constant is $h = 6.626\times10^{-34}$~Js.  $H_0$ is Hubble's parameter at the present epoch.  It's exact value is not important in this paper, but current estimates give $H_0 = 67.8$~(km/s)/Mpc or $2.20\times10^{-18}$/s.  Throughout the paper, I use the dimensionless Hubble parameter $h_{100}$ defined so that $H_0 = 100 h_{100}$~(km/s)/Mpc.  The Hubble distance is $D_H = c/H_0$.  The subscript $0$ is used for values, observed or defined, at the present epoch on Earth.

The comoving distance $d_C$ is a useful quantity in the description of the mechanisms.  Although not measured directly experimentally, it describes our intuitive notion of distance: how many times a ruler fits between two points in space.  Its precise definition depends on the cosmology.  When necessary, the comoving distance will be defined in the description of the mechanisms.

\subsection{Experimental considerations: Observables and inferred quantities}

In this paper, the mechanisms are described in terms of an experimental procedure which can be followed to obtain the predicted results.  The sky is observed with a detector behind a collecting and imaging device such as a telescope.  Experimental observations in astronomy are limited to measurements of three observables:
\begin{itemize}
	\item \textbf{The frequency}:  The frequency $\nu_0$ of light is measured with a detector that responds to light having a frequency within a range $\nu_0$ to $\nu_0+\mbox{d}\nu_0$.  The frequency $\nu_0$ is usually measured with a heterodyne technique which selects light at the desired frequency, or it is measured with a dispersive device, such as a prism or a grating, which selects light at a desired wavelength $\lambda_0$ from which the frequency $\nu_0=c/\lambda_0$ is computed.
	\item \textbf{The angular size}:  The angular size $\theta_0$ is the angle of an object detected by the imaging device.  It is relevant to objects that are large enough or separated enough to be resolved.  At large distances, this applies to clusters, galaxies and isolated objects such as quasars.
	\item \textbf{The spectral irradiance}:  When light is detected, the spectral irradiance $I_0(\nu_0)$~[W~m$^{-2}$~Hz$^{-1}$]~\cite{note1} is the power received by the detector per unit area (perpendicular to the light beam) in the frequency range $\nu_0$ to $\nu_0+\mbox{d}\nu_0$~\cite{Kawasaki1159.2004}.  The index of refraction is assumed to be near one.  The total irradiance is the spectral irradiance integrated over all frequencies $I_0 = \int_0^\infty{I_0(\nu_0) \mbox{d}\nu_0}$~\cite{Condon.Ransom534.2008}.  The irradiance is not to be confused with the luminosity $L$.
\end{itemize}
The following quantities are defined independently of a cosmological model.  However, their interpretation requires that the object at a cosmological distance has the same properties as those of a well known object.  In evaluating these properties, it may be necessary to use a specific cosmology.
\begin{itemize}
	\item \textbf{The red-shift} $z$:  The red-shift is $z \equiv \Delta\lambda/\lambda = (\lambda_0-\lambda) / \lambda = (\nu / \nu_0)-1$, where $\lambda$ and $\lambda_0$ are the emitted and observed wavelengths respectively, and $\nu$ and $\nu_0$ are the emitted and observed frequencies respectively, of an identified spectral line.  Here, the red-shift is determined with respect to an observer on Earth.  (It is assumed that $z$ is independent of wavelength.  However this is not absolutely necessary:  A wavelength dependent $z(\lambda)$ could be useful if measurements of different red-shifts for the same object observed at different wavelengths were substantiated.)
	\item \textbf{Angular diameter distance} $d_A$:  The angular diameter distance $d_A$ is defined as the ratio of an object's physical transverse size to its observed angular size $\theta_0$.  The object's physical transverse size is inferred by comparisons with similar objects.
	\item \textbf{The luminosity}:  The luminosity $L$ is the total power radiated by the source.  A measurement of the luminosity of a distant object requires a knowledge of the distance and spatial geometry between the source and the observer.
	\item \textbf{Luminosity distance} $d_L$:  The luminosity distance is the distance at which an object would lie based on its observed luminosity in the absence of any absorption.  The luminosity distance $d_L$ is defined by the relation $d_L = \sqrt{L / (4\pi S_0)}$ where $L$ is the luminosity (total power emitted) of the object and $S_0 = \int_{source}I_0\mbox{d}\Omega_0$ is the measured flux from the object, that is, the irradiance (power per unit area at the observer) integrated over the solid angle subtended by the source.  The luminosity of the object $L$ is inferred by comparisons with similar objects.
	\item \textbf{Magnitude} $m - M$:  The distance modulus is related to the luminosity distance through $m - M = 5\log_{10}[d_L/D_H] + C$~\cite{Mannheim266.2005}.  % p. 45.
  The constant $C$ depends of the type of object which is observed.  Although it is derived from the luminosity distance, this expression might be more familiar to some readers.
	\item \textbf{Surface Brightness} $\left\langle SB\right\rangle$:  The surface brightness is related to the angular distance and the luminosity distance through $\left\langle SB\right\rangle \propto (d_A/d_L)^2$~\cite{Hartnett3500.2007}.
  Although it is derived from the angular and the luminosity distances, this expression shown explicitly what to expect for the Tolman surface brightness test.  When some authors do not agree with this definition, their equation is listed.  Note that this definition is different from the usage where the magnitude is used in the definition of the surface brightness.
	\item \textbf{Time dilation factor} $F_\tau$:  The time dilation factor $F_\tau$ is obtained by observing the temporal variation of the electromagnetic field (e.g. irradiance decrease after the explosion of a supernova).  The dilated time $F_\tau$ is defined as the ratio of the observed duration of an event to the duration of that same event as would be measured at the object.   The time duration at the object is inferred by comparisons with similar objects.
	\item \textbf{Spectral line blurring} $\eta_z$:  Some mechanisms involve an interaction of light with the intergalactic medium.  Quantization may produce an increase in the linewidth of spectral lines due to the statistical variations in the number of interactions.  Usually interpreted as a result of the temperature of the emitting or absorbing medium, an increase in the observed linewidth could also result from the red-shift mechanism.  For a null linewidth at the emitter, the spectral line bluring is defined as the ratio of the observed Full-Width-at-Half-Maximum linewidth $\delta \lambda_0$ to the observed wavelength of the spectral line $\lambda_0$: $\eta_z = \delta \lambda_0/\lambda_0$.  When real spectra are considered, this apparent linewidth increase is to be added (convolution) to the linewidth of the spectral lines in the measured spectrum.
	\item \textbf{Angular spread} $\Delta_\theta$:  Some mechanisms involve a statistical interaction of light with the intergalactic medium.  This may produce an angular spread of the light with each collision.  The angular spread $\Delta_\theta$ is the observed angular size of a point source.  This value can be different from zero if there is a certain diffusion of light.  The measurement of this angle assumes that the light originates from a point source, otherwise the angular size of the object must be deconvoluted.  The angular size may not always be available experimentally.
\end{itemize}
Only measurable quantities are used here.  Other distances such as the comoving distance or the comoving volume are convenient for theoretical derivations but cannot be measured in practice.

%                                                 <<<<<<<<<<----------
\subsection{\label{sec:format}Format Used to Describe the Red-Shift Mechanisms}

The format used for each red-shift mechanism is as follows:  A description of the red-shift mechanism at the undergraduate physics level is given in the first paragraph, completed by a description at the graduate physics level.  When possible, I have tried to contact the authors of the models to have their agreement on this description.  In order to keep this paper within a reasonable length, only short explanations are given to explain each mechanism.  References to published work in journals or web pages are provided.

\noindent \textbf{Conditions, limits of applicability and restrictions:}

The domain of applicability of the red-shift mechanism.  This could be for example: the solar limb red-shift, the red-shift of quasars, and the cosmological red-shift.  The assumptions required to make the mechanism work, adjustable parameters, and conflicts with currently accepted theories.

\noindent \textbf{Functional relationships:}

Functional relations between various quantities for large red-shifts, measured or inferred: $z$, $d_A$, $d_L$, $m - M$, $\left\langle SB\right\rangle$, $F_\tau$, and $\eta_z$.  Unless otherwise specified, $\Delta_\theta = 0$.  Preferably, the quantities will be given as functions of the red-shift.  When possible, the equation were taken directly from a published reference or confirmed by the author.  However, when I had to derive some of these equations and some doubt remained in my mind (because e.g. the equations have not yet been verified by the original author or against any reference) the symbol $=^?$ is used in the equation.

\noindent \textbf{Discussion and comments:}

Any experiment that can distinguish the model from other models can be described here.  If the model has the same properties as another model, an explanation will be given as to why it is not just another description of the same mechanism.  My personal comments are given here and do not necessarily reflect the opinion of the author of the mechanism.
%
%                                                 ---------->>>>>>>>>>

% ####################################################################################### %
% Computer formulas for calculations of distances, standard Big~Bang theory see http://kobesearch.cpan.org/htdocs/Astro-Cosmology/Astro/Cosmology.html
% More equations for distances, \Omega, etc. http://icecube.wisc.edu/~halzen/notes/week1-3.pdf 
% Use ``Reference-frame red-shifts''?? http://www.bautforum.com/questions-answers/34778-redshift-category-terminology.html#post601167

\section{Red-Shifts produced by a time-dependent distance or metric of space}
The red-shift mechanisms listed in this section depend on the changing metric of space between the emitter and the detector.  They can involve a simple change in distance or a more complex space curvature.  The properties of space are selected to match the observed Hubble law.  For all the mechanisms in this section
\begin{align*}
	d_L &= (1 + z)^2 d_A,\\ % Ref: from red-shift of a blackbody spectrum, McKinley, Relativistic Transformation Light Power, AmJPhys 1979. \cite{McKinley602.1979}
	F_\tau &= 1 + z,\\ % Ref: from relativity: time dilation independent of frequency.
	\eta_z &= 0,\\ % Ref: no wavelength spread in relativity.
	\Delta_\theta &= 0. % Ref: no angular spread in relativity.
\end{align*}
The relationship between $d_A$ and $d_L$ is $d_L = (1 + z)^2 d_A$ resulting from the loss of luminosity due to the red-shift (a factor $\sqrt{1 + z}$), the lower rate at which the photons reach the observer (a factor $\sqrt{1 + z}$) and the reduction of the solid angle (a factor $1 + z$)~\cite{McKinley602.1979}.

%  Other definitions: $$\Omega_M \equiv \frac{8\pi G\rho_0} {3H_0^2}\ \ \ \ \ \ \ \ \ \Omega_\Lambda \equiv \frac{\Lambda c^2} {3H_0^2}$$ $$1 + z=\frac{a_0} {a}$$ where $a$ is the time dependent cosmic scale factor and $a_0$ is its value at the present epoch.

%                                                 <<<<<<<<<<----------1842
\stepcounter{mechanism_number}
\subsection{\label{sec:mec.doppler}Doppler Effect}
Galaxies move away from us with velocities that produce a red-shift $z$ proportional to their distance in a Euclidian universe.  The red-shift is a result of the Doppler effect.  With the theory of special relativity included in this mechanism, the red-shift as a function of velocity is $1 + z = \sqrt{(c+v)/(c-v)}$.

In the cosmology of Alfv\'en and Klein\cite{Alfven1990.5}, the red-shift is produced by a recessional velocity of the galaxies.

\textbf{Conditions, limits of applicability and restrictions:}

One adjustable parameter is needed, the proportionality constant between the red-shift and the distance.  The mechanism neglects the masses in the universe which would curve space.
% Conditions and limits of applicability: Solar limb red-shift? Quasars? Discordant red-shifts? Scale: intergalactic, interstellar, interplanetary red-shift? The assumptions required to make the mechanism work, adjustable parameters, and conflicts with currently accepted theories. Adjustable parameters with density last.

\noindent \textbf{Functional relationships:}
\begin{align*}
  d_A &= D_H z,\\ % made up by me
%	d_L &= (1 + z)^2 d_A,\\ % Ref: from red-shift of a blackbody spectrum, McKinley, Relativistic Transformation Light Power, AmJPhys 1979. \cite{McKinley602.1979}
	m - M &= 10\log_{10}[1 + z] + 5\log_{10}[z] + C,\\ % Ref: Mannheim340.2006, arXiv:astro-ph/0505266v2 p. 45.
	\left\langle SB\right\rangle &\propto (1 + z)^{-4}. % Ref: astro-ph/0603500
%	F_\tau &= 1 + z,\\ % red-shift applies to all frequencies, even as frequency -> 0 
%	\eta_z &= 0,\\ % red-shift applies to all frequencies equally
%	\Delta_\theta &= 0. % nothing for light to diffuse on
\end{align*}

%\textbf{Discussion and comments:}

%
%                                                 ---------->>>>>>>>>>

%                                                 <<<<<<<<<<----------1922
\stepcounter{mechanism_number}
\subsection{\label{sec:mec.flrw}Standard Model of modern Cosmology}
In this model by Alexander Friedman and Georges Lema\^itre, the metric expansion of space increases the wavelength of the light over time.  This effect is present everywhere in the universe.

The Friedmann-Lema\^itre-Robertson-Walker (FLRW) metric is an exact solution of Einstein's field equations of general relativity.  Included in this model is de Sitter's original model which is a solution of FLRW cosmology with a slightly different metric.

\textbf{Conditions, limits of applicability and restrictions:}

These relations are valid for the general case $(\Omega_M, \Omega_k)$~\cite{Hogg116.2000}.
% Conditions, limits of applicability and restrictions: Solar limb red-shift? Quasars? Discordant red-shifts? Scale: intergalactic, interstellar, interplanetary red-shift? The assumptions required to make the mechanism work, adjustable parameters, and conflicts with currently accepted theories. Adjustable parameters with density last.

\noindent \textbf{Functional relationships:}
\begin{align*}
% z &\neq f(\lambda),\\
	d_A &= \frac{D_H}{(1 + z)^2} \frac{2[2-\Omega_M(1-z)-(2-\Omega_M)(1+\Omega_M z)^{1/2}]}{\Omega_M^2},\\ % Ref: Hogg116.2000
%	d_L &= (1 + z)^2 d_A,\\ % Ref: Hogg116.2000
	m - M &= 10\log_{10}[1 + z] + 5\log_{10}[d_A/D_H] + C,\\ % Ref: Mannheim340.2006, arXiv:astro-ph/0505266v2 p. 45.
	\left\langle SB\right\rangle &\propto (1 + z)^{-4}. % Ref: astro-ph/0603500
%	F_\tau &= 1 + z,\\ % red-shift applies to all frequencies, even as frequency -> 0 
%	\eta_z &= 0,\\ % red-shift applies to all frequencies equally
%	\Delta_\theta &= 0. % nothing for light to diffuse on
\end{align*}

\noindent \textbf{Discussion and comments:}

As Einstein commented to Lema\^itre: ``Vos calculs sont corrects, mais votre physique est abominable''~\cite{Deprit370.1984}.  The de Sitter universe is based on imaginary fabrication of a repulsive force varying directly with distance\cite{Reber18.1977}.
%
%                                                 ---------->>>>>>>>>>

%                                                 <<<<<<<<<<----------
\stepcounter{mechanism_number}
\subsection{\label{sec:mec.lcdm}\texorpdfstring{$\Lambda$}{Lambda}-Cold-Dark-Matter Cosmology}

$\Lambda$ is the cosmological constant introduced by Einstein.  Dark matter is described as being cold, non-baryonic, dissipationless and collisionless.  The model assumes a nearly scale-invariant spectrum of primordial perturbations and a universe without spatial curvature.

The model uses the FLRW (Friedmann-Lema\^itre-Robertson-Walker) metric, the Friedmann equations and the cosmological equations of state to describe the universe from right after the inflationary epoch to the present.\\
Formulas kindly provided by Philip Mannheim.  See also Hogg\cite{Hogg116.2000}.

\textbf{Conditions, limits of applicability and restrictions:}

In the case where $\Lambda \neq 0$, distances are calculated from the comoving distance
$$d_C = D_H \int_0^z \frac{dz'}{\sqrt{\Omega_M(1 + z')^3+\Omega_k(1 + z')^2+\Omega_\Lambda}}.$$
  For the general case, $d_C$ must be integrated numerically.

  A possible fit to experimental data gives: $\Omega_\Lambda = 0.73$, $\Omega_M = 0.23$, and $\Omega_k = 0.04$.
% Conditions, limits of applicability and restrictions: Solar limb red-shift? Quasars? Discordant red-shifts? Scale: intergalactic, interstellar, interplanetary red-shift? The assumptions required to make the mechanism work, adjustable parameters, and conflicts with currently accepted theories. Adjustable parameters with density last.

\noindent \textbf{Functional relationships:}
\begin{center}
% z &\neq f(\lambda),\\
	\begin{tabular}{r | l l}
 & \parbox{1.8in}{$$\frac{D_H \sinh\left[\sqrt{\Omega_k}  d_C/D_H \right]} {(1 + z)\sqrt{\Omega_k}}$$}, & for $\Omega_k>0$ \\
$d_A =$ & \parbox{1.1in}{$$d_C/(1 + z)$$}, & for $\Omega_k=0$ \\
 & \parbox{1.8in}{$$\frac{D_H \sin \left[\sqrt{-\Omega_k} d_C/D_H \right]} {(1 + z)\sqrt{-\Omega_k}}$$}, & for $\Omega_k<0$ \\
	\end{tabular} % Ref: Mannheim and Hogg116.2000
\end{center}
\begin{align*}
%	d_L &= (1 + z)^2 d_A,\\ % Ref: Mannheim and Hogg116.2000
	m - M &= 10\log_{10}[1 + z] + 5\log_{10}[d_A/D_H] + C,\\ % Ref: Mannheim340.2006, arXiv:astro-ph/0505266v2 p. 45.
	\left\langle SB\right\rangle &\propto (1 + z)^{-4}. % Ref: astro-ph/0603500
%	F_\tau &= 1 + z,\\ % red-shift applies to all frequencies, even as frequency -> 0 
%	\eta_z &= 0,\\ % red-shift applies to all frequencies equally
%	\Delta_\theta &= 0. % nothing for light to diffuse on
\end{align*}

\noindent \textbf{Discussion and comments:}

This theory suffers from several problems:
\begin{itemize}
  \item It contains three adjustable parameters which are not based on known constants of nature.
  \item The assumption of spatial isotropy and spatial homogeneity is postulated, which makes the theoretical universe unstable.
  \item The interpretation of red-shifts as the expansion of space requires that the universe has a curvature to fit the measured galaxy distributions.  However the curvature can only be the result of a lot more mass than is currently observed.  An ad hoc hypothesis is added: dark matter.
  \item The rate of expansion of the universe seems to be accelerating.  An ad hoc hypothesis is added: dark energy.
  \item The time required for galaxies to evolve conflicts with the smoothness of the CMB.  An ad hoc hypothesis is added: inflation.
  \item This model does not explain any of the anomalous or intrinsic red-shifts.
\end{itemize}
%
%                                                 ---------->>>>>>>>>>

%                                                 <<<<<<<<<<----------1992
\stepcounter{mechanism_number}
\subsection{\label{sec:mec.yilmaz}General time-varying Yilmaz theory of Gravity}
The Yilmaz gravitational theory is a refinement of the Einstein's General theory of Relativity.  It incorporates all of the principles of the Einstein theory except for the gravitational field equation.
 % Voir http://www.olduniverse.com/ pour plus de details
%Yilmaz, H. (1992). ``Toward a field theory of gravitation''. Nuovo Cimento B107 (8): 941–960. Bibcode 1992NCimB.107..941Y. doi:10.1007/BF02899296.

%\textbf{Conditions, limits of applicability and restrictions:}

\noindent \textbf{Functional relationships:}
\begin{align*}
% z &\neq f(\lambda),\\
%	d_A &= \mbox{ not available},\\ % Ref: ?
	d_L &=^? (1 + z)^2 d_A,\\ % Ref: ? There is expansion in the Yilmaz theory, so this equation should be correct
	m - M &= 5\log_{10}[d_L/D_H] + C,\\ % Ref: Mannheim340.2006, arXiv:astro-ph/0505266v2 p. 45.
	\left\langle SB\right\rangle &\propto \left(d_A/d_L\right)^2. % Ref: astro-ph/0603500
%	F_\tau &= 1 + z,\\ % red-shift applies to all frequencies, even as frequency -> 0 
%	 &= 0,\\ % red-shift applies to all frequencies equally
%	\Delta_\theta &= 0. % nothing for light to diffuse on
\end{align*}
A functional relationship for $d_A$ is not available.

\textbf{Discussion and comments:}

The Yilmaz theory does not yield any physically impossible concepts like Black Holes or other singularities, which Einstein absolutely rejected.
%Recent data indicate a Hubble expansion rate of about 20 km/sec per million light years of galaxy distance (65 km/sec per mega-parsec). For the gravitational curvature of our universe to produce this Hubble expansion rate, the Yilmaz theory predicts that the universe should have an average mass density equivalent to 9.6 hydrogen atoms per cubic meter. This is consistent with recent astronomical measurements, which give average mass densities equivalent to 3.0 to 7.2 hydrogen atoms per cubic meter. 
%
%                                                 ---------->>>>>>>>>>

%                                                 <<<<<<<<<<----------2005
\stepcounter{mechanism_number}
\subsection{\label{sec:mec.mannheim}Conformal Gravity}
In Mannheim's model, a four-dimensional theory of quantum gravity is proposed to fit the accelerating universe data\cite{Mannheim266.2005, Mannheim340.2006}.  The cosmological constant is taken as being very large, but not as part of the standard of the Newton-Einstein theory.

%  So, the red-shift is a result of the expanding metric of space?  However, it is done consistently with particle physics?

\textbf{Conditions, limits of applicability and restrictions:}

A possible fit to experimental data gives a deceleration parameter $q_0 = -0.37$.
% Conditions, limits of applicability and restrictions: Solar limb red-shift? Quasars? Discordant red-shifts? Scale: intergalactic, interstellar, interplanetary red-shift? The assumptions required to make the mechanism work, adjustable parameters, and conflicts with currently accepted theories. Adjustable parameters with density last.

\noindent \textbf{Functional relationships:}
\begin{align*}
% z &\neq f(\lambda),\\
	d_A &= - \frac{D_H}{q_0} \left(1-\sqrt{ 1+q_0-\frac{q_0}{(1 + z)^2} }\right),\\ % Ref: ?
%	d_L &= (1 + z)^2 d_A,\\ % Ref: ?
	m - M &= 10\log_{10}[1 + z] + 5\log_{10}[d_A/D_H] + C,\\ % Ref: Mannheim340.2006, arXiv:astro-ph/0505266v2 p. 45.
	\left\langle SB\right\rangle &\propto (1 + z)^{-4}. % Ref: astro-ph/0603500
%	F_\tau &= 1 + z,\\ % red-shift applies to all frequencies, even as frequency -> 0 
%	\eta_z &= 0,\\ % red-shift applies to all frequencies equally
%	\Delta_\theta &= 0. % nothing for light to diffuse on
\end{align*}

%\textbf{Discussion and comments:}

%
%                                                 ---------->>>>>>>>>>

\vspace{3cm}

% Open Office Calc: graph 33cm x 16cm
% PSP: resize 6.9in 300pixel/in, save as .eps colour, preview
\begin{figure}[h]
\centerline{\includegraphics[width=6.9in]{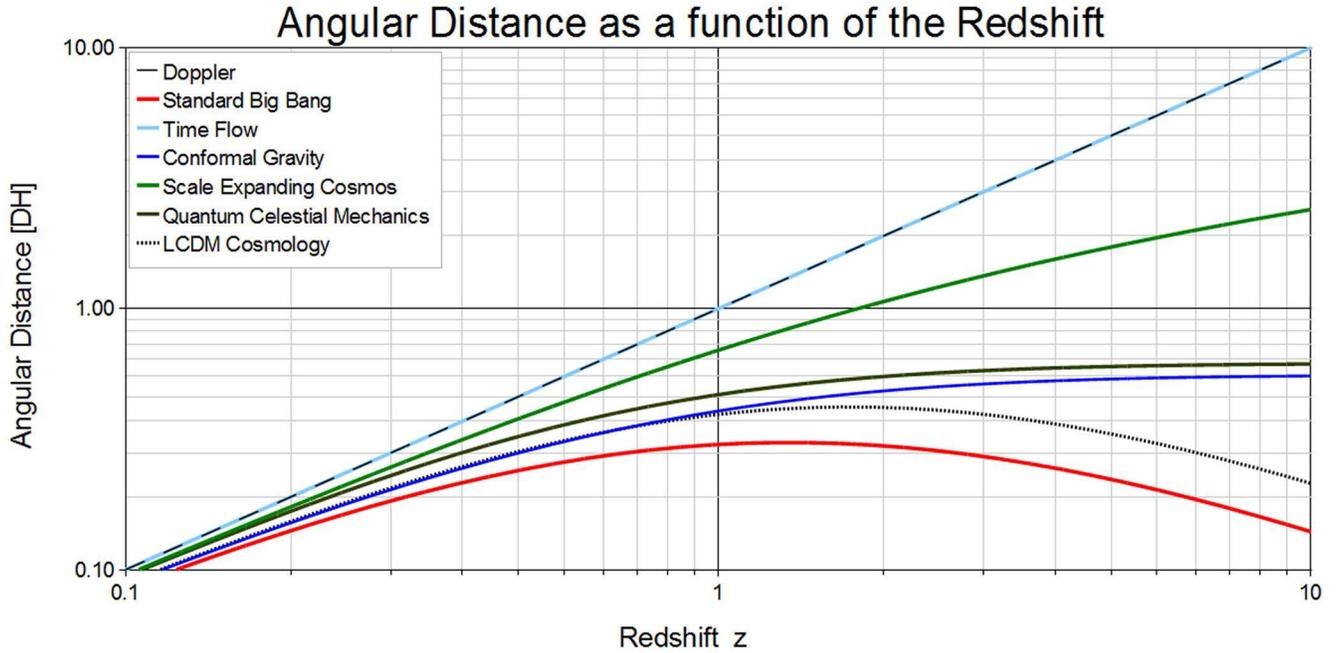}}
\caption{Angular distance in units of $D_H$ for a red-shift produced by a time-dependent metric of space.}
\label{fig:dAmetric}
\end{figure}

\clearpage

\begin{figure}
\centerline{\includegraphics[width=6.9in]{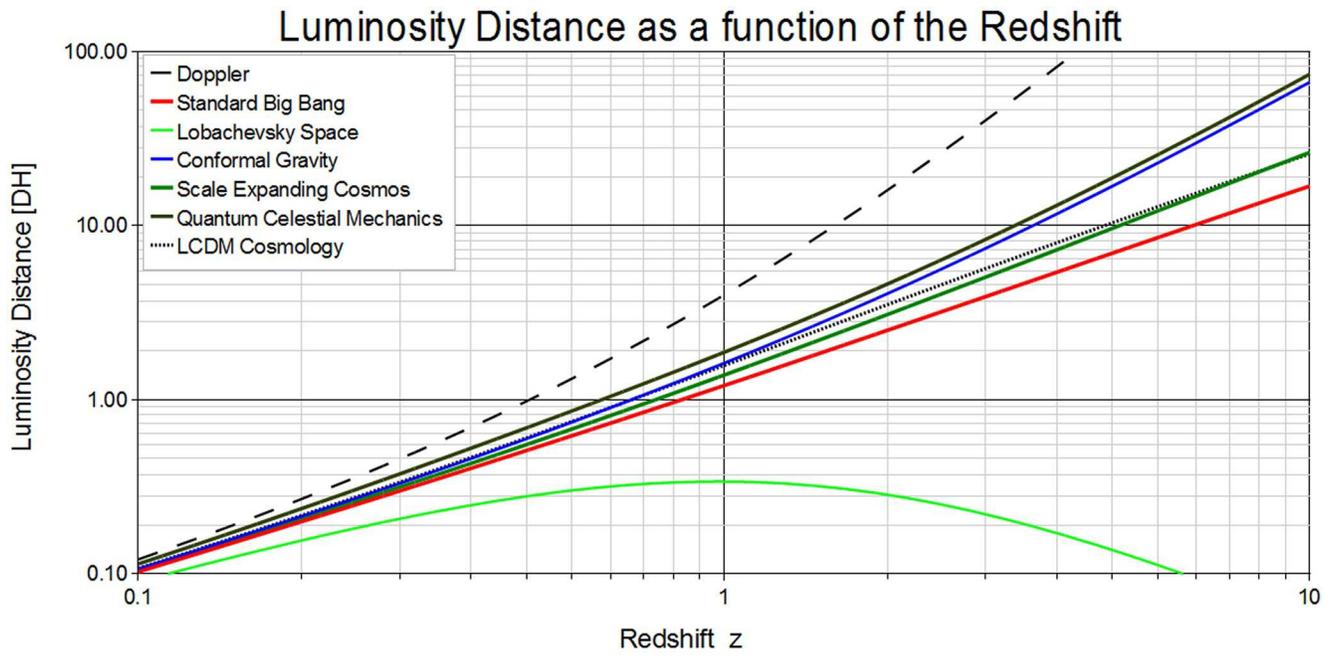}}
\caption{Luminosity distance in units of $D_H$ for a red-shift produced by a time-dependent metric of space.}
\label{fig:dLmetric}
\end{figure}

\begin{figure}
\centerline{\includegraphics[width=6.9in]{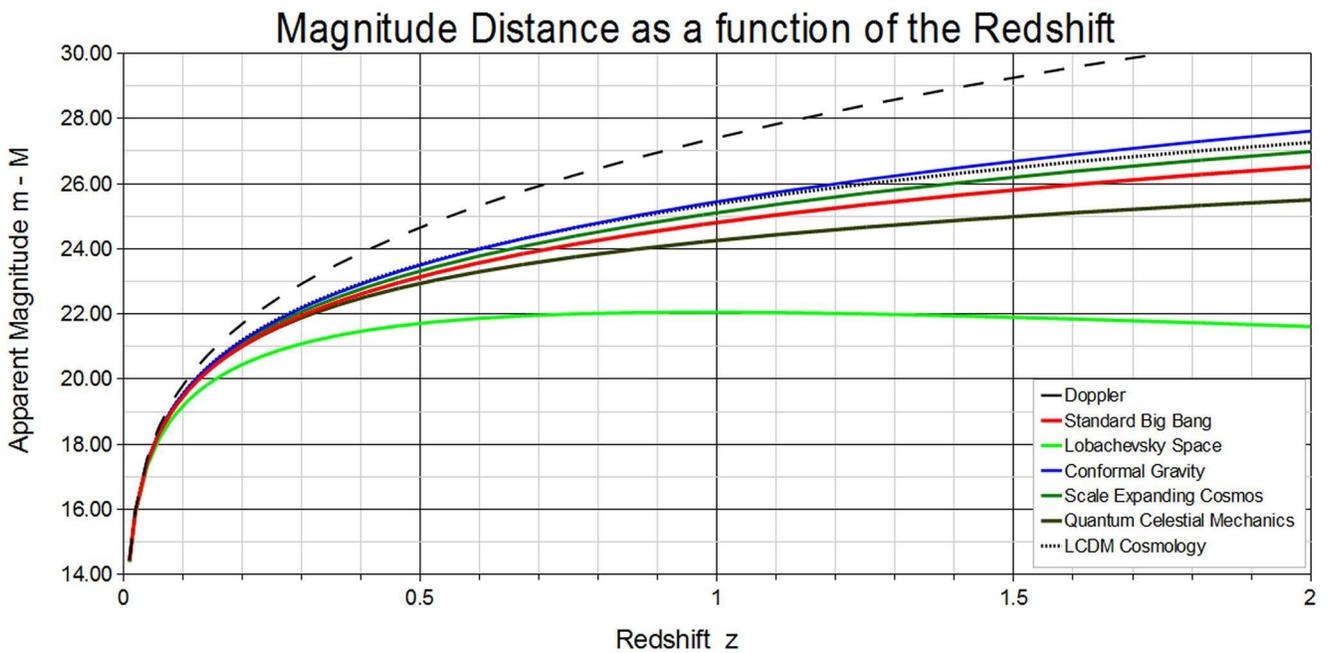}}
\caption{Magnitude distance for a red-shift produced by a time-dependent metric of space.}
\label{fig:dMmetric}
\end{figure}

\clearpage

% ####################################################################################### %
\section{Red-Shifts produced by a time-dependent property of gravity}
The red-shift mechanisms listed in this section depend on the changing properties of gravity.  For all the mechanisms in this section
\begin{align*}
	F_\tau &= 1 + z,\\ % red-shift applies to all frequencies, even as frequency -> 0 
	\eta_z &= 0,\\ % red-shift applies to all frequencies equally
	\Delta_\theta &= 0. % nothing for light to diffuse on
\end{align*}

%                                                 <<<<<<<<<<----------1938
\stepcounter{mechanism_number}
\subsection{\label{sec:mec.dirac}Dirac's Large Number Hypothesis}
Dirac's cosmology was based on the principle ``that any two of the very large dimensionless numbers (of the order $10^{39}$ and $10^{78}$), occurring in Nature, are connected by a simple mathematical relation in which the coefficients are of the order of magnitude unity''~\cite{Dirac199.1938}.  From this principle Dirac deduced that the gravitational 'constant' $G \propto (epoch)^{-1}$.  In order to describe a cosmological model according to the principles of the general theory of relativity he then changed the units of distance and time so as to make $G$ constant\cite{Gilbert192.1961, Ray.Gosh21.2007}.

%  Dirac proposed that instead of the space between the galaxies expanding, as general relativity predicts, all space is expanding because the basic scale of all objects from electrons to galaxy clusters grows with time, due to an unknown physical law.   From Eric Lerner ``the Big~Bang never happened'', p. 429
% There is space expansion in this theory: http://adsabs.harvard.edu/full/1986AuJPh..39..339L

\textbf{Conditions, limits of applicability and restrictions:}

  In one version of his cosmology, the observed expansion imitates the Doppler effect.  Thus one parameter is required: the proportionality constant between the red-shift and the distance.
% Conditions, limits of applicability and restrictions: Solar limb red-shift? Quasars? Discordant red-shifts? Scale: intergalactic, interstellar, interplanetary red-shift? The assumptions required to make the mechanism work, adjustable parameters, and conflicts with currently accepted theories. Adjustable parameters with density last.

\noindent \textbf{Functional relationships:}
\begin{align*}
% z &\neq f(\lambda),\\
	d_A &= D_H z,\\ % Ref: if Dirac's model is like the Doppler effect
	d_L &=^? (1 + z)^2 d_A,\\ % Ref: from red-shift of a blackbody spectrum, McKinley, Relativistic Transformation Light Power, AmJPhys 1979.
	m - M &= 10\log_{10}[1 + z] + 5\log_{10}[z] + C,\\ % Ref: Mannheim340.2006, arXiv:astro-ph/0505266v2 p. 45.
	\left\langle SB\right\rangle &\propto (1 + z)^{-4}. % Ref: astro-ph/0603500
%	F_\tau &= 1 + z,\\ % red-shift applies to all frequencies, even as frequency -> 0 
%	\eta_z &= 0,\\ % red-shift applies to all frequencies equally
%	\Delta_\theta &= 0. % nothing for light to diffuse on
\end{align*}

\noindent \textbf{Discussion and comments:}

  One issue of Dirac's cosmology is that the model introduces an ad hoc hypothesis: ``the continuous creation of matter is needed''.  A more serious objection is that this is entirely based on numerology.
%
%                                                 ---------->>>>>>>>>>

%                                                 <<<<<<<<<<----------2001
\stepcounter{mechanism_number}
\subsection{\label{sec:mec.booth}Machian General Relativity}
  In Booth's model, a scale-invariant form of the field equations of General Relativity is postulated, based on the replacement of the Newtonian gravitational constant by an formulation based explicitly on Mach's principle\cite{Booth7.2001}.

%\textbf{Conditions, limits of applicability and restrictions:}

% Conditions, limits of applicability and restrictions: Solar limb red-shift? Quasars? Discordant red-shifts? Scale: intergalactic, interstellar, interplanetary red-shift? The assumptions required to make the mechanism work, adjustable parameters, and conflicts with currently accepted theories. Adjustable parameters with density last.

\noindent \textbf{Functional relationships:}
\begin{align*}
% z &\neq f(\lambda),\\
%	d_A &= \mbox{ not available},\\ % Ref: ?
	d_L &=^? D_H (1 + z) \log{(1 + z)},\\ % Ref: ?
	m - M &= 5\log_{10}[d_L/D_H] + C,\\ % Ref: ?
	\left\langle SB\right\rangle &\propto \left(d_A/d_L\right)^2. % Ref: astro-ph/0603500
%	F_\tau &= 1 + z,\\ % red-shift applies to all frequencies, even as frequency -> 0 
%	\eta_z &= 0,\\ % red-shift applies to all frequencies equally
%	\Delta_\theta &= 0. % nothing for light to diffuse on
\end{align*}
A functional relationship for $d_A$ is not available.

%\textbf{Discussion and comments:}

%
%                                                 ---------->>>>>>>>>>

% ####################################################################################### %
\section{Red-Shifts produced by a time-dependent property of matter}
  The red-shift mechanisms listed in this section depend on the changing properties of matter.  The properties of matter are selected to match the observed Hubble law.  For all the mechanisms in this section
\begin{align*}
	\eta_z &= 0,\\ % red-shift applies to all frequencies equally
	\Delta_\theta &= 0. % nothing for light to diffuse on
\end{align*}

%                                                 <<<<<<<<<<----------
\stepcounter{mechanism_number}
\subsection{\label{sec:mec.qssc}Quasi Steady State Cosmology}
%Jayant Narlikar/ Fred Hoyle/ Halton Arp
% or is it Bondi, Gold and Hoyle?? Burbidge
  In this model supported by Narlikar, Hoyle, Arp \emph{et al.}, the mass of matter increases with the square of its age.  Since the spectroscopic red-shift varies inversely as the mass, the observed red-shift depends on the age difference between the observer and source.  J,~Narlikar found the solution $m = at^2$ ($a$ is a constant) to the generalized equations of general relativity.  The effect explains particularly well the red-shifts of quasars.  Also, the speed of light makes distant galaxies appear younger.  Frequency shift $\Delta\lambda/\lambda$ independent of $\lambda$.
% $c \times 39$~(km/s)/Mpc or $12$~(km/s)/Myear
% See R.G. Vishwakarma,...

\textbf{Conditions, limits of applicability and restrictions:}

  The following definition is used:
$$H(z) = H_0[\Omega_{\Lambda 0} - \Omega_{k0}(1 + z)^2+\Omega_{m0}(1 + z)^3+\Omega_{c0}(1 + z)^4]^{1/2},$$
where $\Omega_{\Lambda 0}-\Omega_{k0}+\Omega_{m0}+\Omega_{c0}=1$.  Extinction due to the whisker dust is not included in the equations below.
% Conditions, limits of applicability and restrictions: Solar limb red-shift? Quasars? Discordant red-shifts? Scale: intergalactic, interstellar, interplanetary red-shift? The assumptions required to make the mechanism work, adjustable parameters, and conflicts with currently accepted theories. Adjustable parameters with density last.

\noindent \textbf{Functional relationships:}
\begin{align*}
% z &\neq f(\lambda),\\
	d_A &= (1 + z)^{-1}  \int_0^z \frac{\mbox{d}z'}{H(z')},\\ % Ref: ?
	d_L &=^? \sqrt{1 + z} d_A,\\ % Ref: ?
	m - M &= 10\log_{10}[1 + z] + 5\log_{10}[d_A/D_H] + C,\\ % Ref: Mannheim340.2006, arXiv:astro-ph/0505266v2 p. 45.
	\left\langle SB\right\rangle &\propto (1 + z)^{-4},\\ % Ref: astro-ph/0603500
	F_\tau &= 1 + z. % Ref: ?
%	\eta_z &= 0,\\ % red-shift applies to all frequencies equally
%	\Delta_\theta &= 0. % nothing for light to diffuse on
\end{align*}

\noindent \textbf{Discussion and comments:}

  This model introduces an ad hoc hypothesis (which satisfies Einsteinian physics): ``the mass of matter increasing as the square of its age''.

  An objection to this model is based on experimental results.  In Machian physics, the red-shift is related to the age of matter through
$$\frac{1 + z_1}{1 + z_0}=\frac{t_0^2}{t_1^2},$$
where $z_0$ is the red-shift of matter created $t_0$ years ago and where $z_1$ is the red-shift of matter created $t_1$ years ago.  For two particles created $t$ years ago within a small time interval $\Delta t$, the red-shifts satisfy $(1 + z_1)/(1 + z_0)=1+2\Delta t/t$.  Recent measurements of the resonant frequency of a single aluminium ion have been reported to a relative error smaller than $5.2\times 10^{-17}$~\cite{Rosenband.Bergquist319.2008}.  Different ions have been measured, all giving the same result within measurement errors.  Assuming these ions were created $10$ billion years ago ($t=3.2\times 10^{17} s$), the difference in age for these ions is obtained from solving $1+5.2\times 10^{-17}=(1 + z_1)/(1 + z_0)=1+2\Delta t/(3.2\times 10^{17})$, which gives $\Delta t = 8.2 s$.  Unless these different ions were all created within $10$~seconds of each other, $10$ billion years ago, the solution $m = at^2$ is not compatible with experimental results.

  Some other arguments (biased toward the Big Bang cosmology) are given by Wright\cite{Wright1997.1}.
%
%                                                 ---------->>>>>>>>>>

%                                                 <<<<<<<<<<----------1986
\stepcounter{mechanism_number}
\subsection{\label{sec:mec.laviolette}Hadron mass variation}
  In Laviolette's model, the red-shift is explained by hadrons increasing in mass as they get older\cite{Laviolette310.1986}.

%\textbf{Conditions, limits of applicability and restrictions:}

% Conditions, limits of applicability and restrictions: Solar limb red-shift? Quasars? Discordant red-shifts? Scale: intergalactic, interstellar, interplanetary red-shift? The assumptions required to make the mechanism work, adjustable parameters, and conflicts with currently accepted theories. Adjustable parameters with density last.

\noindent \textbf{Functional relationships:}
\begin{align*}
% z &\neq f(\lambda),\\
	d_A &=^? D_H \ln(1 + z)/(1 + z),\\ % Ref: ?
	d_L &=^? \sqrt{1 + z} d_A,\\ % Ref: ?
	m - M &= 5\log_{10}[d_L/D_H] + C,\\ % Ref: Mannheim340.2006, arXiv:astro-ph/0505266v2 p. 45.
	\left\langle SB\right\rangle &\propto \left(d_A/d_L\right)^2. % Ref: astro-ph/0603500
%	F_\tau &= \mbox{ not available}. % Ref: ?
%	\eta_z &= 0,\\ % red-shift applies to all frequencies equally
%	\Delta_\theta &= 0. % nothing for light to diffuse on
\end{align*}
A functional relationship for $F_\tau$ is not available.

\textbf{Discussion and comments:}

  This model introduces an ad hoc hypothesis: ``the mass of the hadrons increases with time''.
%
%                                                 ---------->>>>>>>>>>

%                                                 <<<<<<<<<<----------
\stepcounter{mechanism_number}
\subsection{\label{sec:mec.schmitz}Exponential decrease in the radius of nucleons}
  In Schmitz's model, the comoving distance is $D_C = R_0 \ln(1 + z)$, where $R_0$ is a constant\cite{Schmitz1.20yy}.

%\textbf{Conditions, limits of applicability and restrictions:}

% Conditions, limits of applicability and restrictions: Solar limb red-shift? Quasars? Discordant red-shifts? Scale: intergalactic, interstellar, interplanetary red-shift? The assumptions required to make the mechanism work, adjustable parameters, and conflicts with currently accepted theories. Adjustable parameters with density last.

\noindent \textbf{Functional relationships:}
\begin{align*}
% z &\neq f(\lambda),\\
	d_A &= D_H \ln(1 + z)/(1 + z),\\ % Ref: ?
	d_L &=^? \sqrt{1 + z} d_A,\\ % Ref: ?
	m - M &= 5\log_{10}[d_L/D_H] + C,\\ % Ref: Mannheim340.2006, arXiv:astro-ph/0505266v2 p. 45.
	\left\langle SB\right\rangle &\propto \left(d_A/d_L\right)^2,\\ % Ref: astro-ph/0603500
	F_\tau &= 1 + z. % Ref: ?
%	\eta_z &= 0,\\ % red-shift applies to all frequencies equally
%	\Delta_\theta &= 0. % nothing for light to diffuse on
\end{align*}

\textbf{Discussion and comments:}

  This model introduces an ad hoc hypothesis: ``the radius of the nucleon decreases exponentially with time''.
%
%                                                 ---------->>>>>>>>>>

\clearpage

% Open Office Calc: graph 33cm x 16cm
% PSP: resize 6.9in 300pixel/in, save as .eps colour, preview
\begin{figure}
\centerline{\includegraphics[width=6.9in]{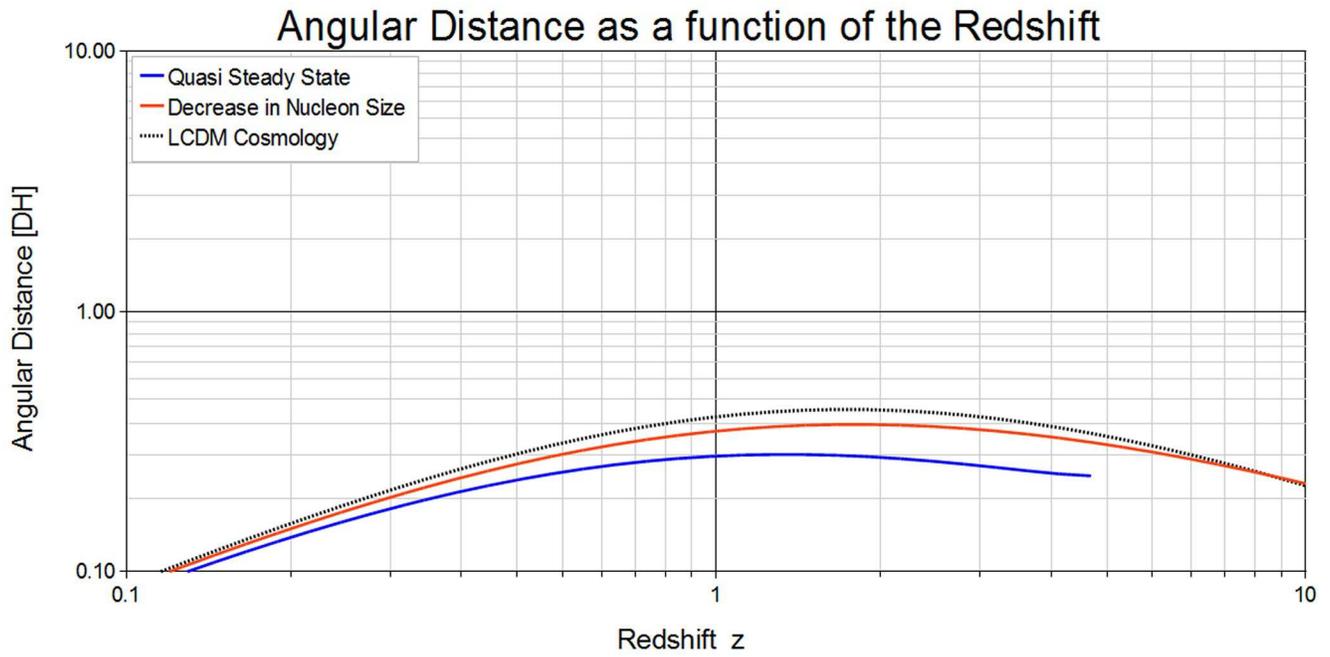}}
\caption{Angular distance in units of $D_H$ for a red-shift produced by a time-dependent property of matter.  ``LCDM Cosmology'' is included for comparison.}
\label{fig:dAmatter}
\end{figure}

\begin{figure}
\centerline{\includegraphics[width=6.9in]{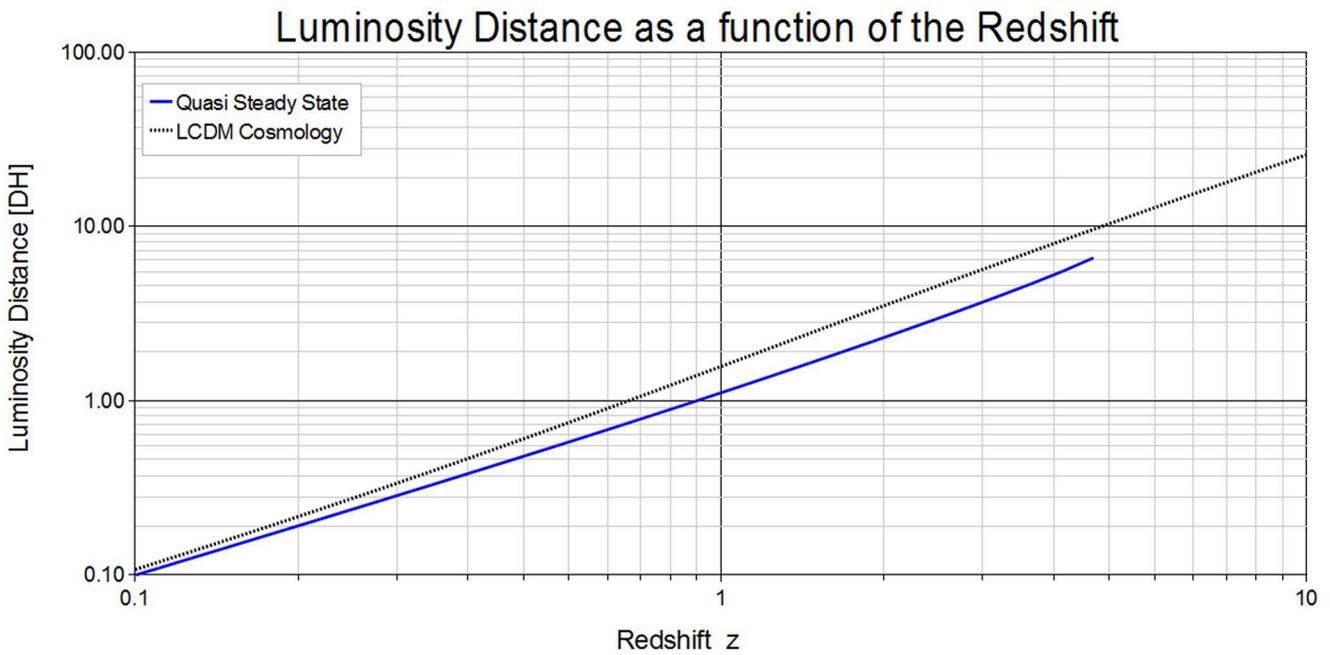}}
\caption{Luminosity distance in units of $D_H$ for a red-shift produced by a time-dependent property of matter.  ``LCDM Cosmology'' is included for comparison.}
\label{fig:dLmatter}
\end{figure}

\clearpage

\begin{figure}
\centerline{\includegraphics[width=6.9in]{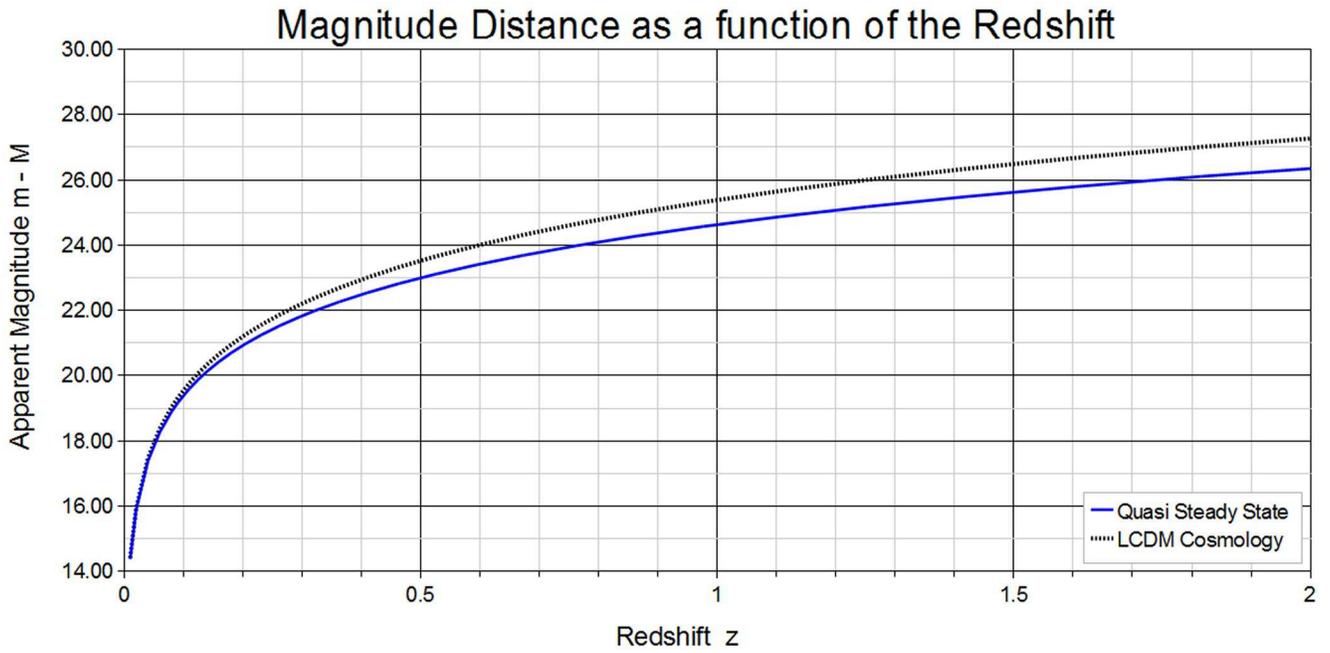}}
\caption{Magnitude distance for a red-shift produced by a time-dependent property of matter.  ``LCDM Cosmology'' is included for comparison.}
\label{fig:dMmatter}
\end{figure}

\clearpage

% ####################################################################################### %
\section{Red-Shifts produced by a time-dependent property of light or an interaction of light with itself}
The red-shift mechanisms listed in this section depend on a changing property of light or an interaction between photons and other virtual particles, ether or empty space.  They do not depend on the presence of matter.  These are sometimes referred to as ``tired light'' mechanisms.

  For many of these mechanisms, $F_\tau = 1 + z$ because the mechanism is completely independent of the frequency of light.  Therefore, even temporal variations of the order of hours and days are 'redshifted'.  This makes these temporal variations appear to evolve more slowly.

%                                                 <<<<<<<<<<----------1954
\stepcounter{mechanism_number}
\subsection{\label{sec:mec.freundlich}Finlay-Freundlich Hypothesis}
  Loss of energy by observed photons traversing a radiation field via a photon-photon interaction\cite{Alpher1962.367}.  An empirical formula is proposed.

\textbf{Conditions, limits of applicability and restrictions:}

	$T$ is defined as the temperature of the radiation field.

  No mechanism given except for a proposal that the energy lost reappears as lower frequency radiation, or as neutrino pairs from the exchange of a graviton between two photons.
% Conditions, limits of applicability and restrictions: Solar limb red-shift? Quasars? Discordant red-shifts? Scale: intergalactic, interstellar, interplanetary red-shift? The assumptions required to make the mechanism work, adjustable parameters, and conflicts with currently accepted theories. Adjustable parameters with density last.

\noindent \textbf{Functional relationships:}
\begin{align*}
% z &\neq f(\lambda),\\
	z &= A T^4 d_A,\\ % Ref: ?
	d_L &=^? \sqrt{1 + z} d_A,\\ % Ref: ?
	m - M &= 5\log_{10}[d_L/D_H] + C,\\ % Ref: Mannheim340.2006, arXiv:astro-ph/0505266v2 p. 45.
	\left\langle SB\right\rangle &\propto \left(d_A/d_L\right)^2. % Ref: astro-ph/0603500
%	F_\tau &= \mbox{ not available},\\ % Ref: ?
%	\eta_z &= \mbox{ not available},\\ % Ref: ?
%	\Delta_\theta &= \mbox{ not available}. % Ref: ?
\end{align*}
Functional relationships for $F_\tau, \eta_z$ and $\Delta_\theta$ are not available.

\noindent \textbf{Discussion and comments:}

  A proposal was made to attempt to detect the photon interaction in an experiment combining the high sensitivity for resonant absorption associated with the M\"ossbauer effect with the high-radiation fields from a thermonuclear fusion device\cite{Ward858.1961}.

  The desired answer is placed into the hypothesis, so the correct result is inevitable\cite{Reber18.1977}.
%
%                                                 ---------->>>>>>>>>>

%                                                 <<<<<<<<<<----------
\stepcounter{mechanism_number}
\subsection{\label{sec:mec.lewis}Photon Decay}
  In Lewis' model\cite{Lewis1.19yy}, light follows the photon decay equation:
$$\nabla^2\phi = \frac{1}{c^2}\left(\frac{k}{\hbar}\frac{\partial\phi}{\partial t} + \frac{\partial^2\phi}{\partial t^2}\right).$$

%\textbf{Conditions, limits of applicability and restrictions:}

% Conditions, limits of applicability and restrictions: Solar limb red-shift? Quasars? Discordant red-shifts? Scale: intergalactic, interstellar, interplanetary red-shift? The assumptions required to make the mechanism work, adjustable parameters, and conflicts with currently accepted theories. Adjustable parameters with density last.

\textbf{Functional relationships:}
\begin{align*}
% z &\neq f(\lambda),\\
	d_A &=^? c\ln(1 + z)/(68.7\mbox{~(km/s)/Mpc}),\\ % Ref: ?
	d_L &=^? \sqrt{1 + z} d_A,\\ % Ref: ?
	m - M &= 5\log_{10}[d_L/D_H] + C,\\ % Ref: Mannheim340.2006, arXiv:astro-ph/0505266v2 p. 45.
	\left\langle SB\right\rangle &\propto \left(d_A/d_L\right)^2,\\ % Ref: astro-ph/0603500
	F_\tau &= 1 + z,\\ % Ref: ? Mechanism completely independent of light frequency
	\eta_z &\approx 0,\\ % Ref: ?
	\Delta_\theta &\approx 0. % Ref: ?
\end{align*}

\textbf{Discussion and comments:}

  This model introduces an ad hoc hypothesis: ``the decay equation of the photon''.
%
%                                                 ---------->>>>>>>>>>

%                                                 <<<<<<<<<<----------
\stepcounter{mechanism_number}
\subsection{\label{sec:mec.stolmar}Effect of Half-Life}
  In Stolmar's model, the photon simply has a decay lifetime similar to radioactive or unstable particles.  In this case, it is the photon's energy that decreases with distance.
$$R = H_d \ln(1 + z)/\ln(2)$$
where $H_d=D_H \ln(2)$ is the Hubble wavelength-doubling distance\cite{Stolmar1.19yy}.

%\textbf{Conditions, limits of applicability and restrictions:}

% Conditions, limits of applicability and restrictions: Solar limb red-shift? Quasars? Discordant red-shifts? Scale: intergalactic, interstellar, interplanetary red-shift? The assumptions required to make the mechanism work, adjustable parameters, and conflicts with currently accepted theories. Adjustable parameters with density last.

\noindent \textbf{Functional relationships:}
\begin{align*}
% z &\neq f(\lambda),\\
	d_A &= D_H \ln(1 + z),\\ % ad hoc hypothesis
	d_L &=^? \sqrt{1 + z} d_A,\\ % Ref: ?
	m - M &= 5\log_{10}[1 + z] + 5\log_{10}[\ln(1 + z)] + C,\\ % Ref: Mannheim340.2006, arXiv:astro-ph/0505266v2 p. 45.
	\left\langle SB\right\rangle &\propto (1 + z)^{-2},\\ % Ref: astro-ph/0603500
	F_\tau &= 1 + z,\\ % Ref: ? Mechanism completely independent of light frequency
	\eta_z &=^? 0,\\ % Ref: ?
	\Delta_\theta &= 0. % Ref: ?
\end{align*}

\textbf{Discussion and comments:}

  This model introduces an ad hoc hypothesis: ``the photon loses energy with distance''.  No specific mechanism seems to be given to explain why or how the photon decays.
%
%                                                 ---------->>>>>>>>>>

%                                                 <<<<<<<<<<----------
\stepcounter{mechanism_number}
\subsection{\label{sec:mec.caswell}Heisenberg Effect}
  In Caswell's model, the red-shift is due to the Heisenberg Uncertainty Principle when applied to photons and extrapolated across the universe.  The photons lose energy to maintain the vacuum of space at $2.7$~K due to a quantum mechanical phenomenon in the vacuum of space.  The standard equation $E=hc/\lambda$ is used and $E$ when differentiated with respect to $\lambda$ shows the relationship between $\mbox{d}E$ and $\mbox{d}\lambda$.  This can be rearranged to give $\mbox{d}\lambda/\lambda^2 = -\mbox{d}E/hc$.  When the uncertainty principle is adapted for em radiation and modified to take account of the dual polarization of photons (c.f. Planck Radiation) the red-shift factor $7.9\times 10^{-27}$m$^{-1}$ is deduced.  This provides the fractional apparent increase in wavelength per metre of travel\cite{Caswell1.19yy}.

\textbf{Conditions, limits of applicability and restrictions:}

  No adjustable parameters: calculated from $h^2c^2/(8\pi)$, the red-shift is equivalent to an apparent expansion rate of $73$~(km/s)/Mpc. % to review...
% Conditions, limits of applicability and restrictions: Solar limb red-shift? Quasars? Discordant red-shifts? Scale: intergalactic, interstellar, interplanetary red-shift? The assumptions required to make the mechanism work, adjustable parameters, and conflicts with currently accepted theories. Adjustable parameters with density last.

\noindent \textbf{Functional relationships:}
\begin{align*}
% z &\neq f(\lambda),\\
	d_A &= c\ln(1 + z)/(73\mbox{~(km/s)/Mpc}),\\ % Ref: ?
	d_L &=^? \sqrt{1 + z} d_A,\\ % Ref: ?
	m - M &= 5\log_{10}[d_L/D_H] + C,\\ % Ref: Mannheim340.2006, arXiv:astro-ph/0505266v2 p. 45.
	\left\langle SB\right\rangle &\propto \left(d_A/d_L\right)^2,\\ % Ref: astro-ph/0603500
	F_\tau &=^? 1. % Ref: ?
%	\eta_z &= \mbox{ not available},\\ % Ref: ?
%	\Delta_\theta &= \mbox{ not available}. % Ref: ?
\end{align*}
Functional relationships for $\eta_z$ and $\Delta_\theta$ are not available.

% \textbf{Discussion and comments:}

%
%                                                 ---------->>>>>>>>>>

%                                                 <<<<<<<<<<----------1981
\stepcounter{mechanism_number}
\subsection{\label{sec:mec.broberg}Interaction Between the Photon Energy and Vacuum Space}
  Broberg's mechanism describes the actual group velocity of an electromagnetic wave in vacuum slightly smaller than $c$.  As the photon propagates, a small amount of energy, equal to $H_0 h$, is lost each time the photon travels the distance equal to its wavelength\cite{Broberg223.1981, Broberg333.1993}.

\textbf{Conditions, limits of applicability and restrictions:}

  The following postulate is made: ``The energy of the universe is quantized in integer numbers of a system-invariant elementary energy quantum''.  This hypothesis is added: ``The elementary energy quantum in the universe is equal to the product of Planck's and Hubble's constants''.
% Conditions, limits of applicability and restrictions: Solar limb red-shift? Quasars? Discordant red-shifts? Scale: intergalactic, interstellar, interplanetary red-shift? The assumptions required to make the mechanism work, adjustable parameters, and conflicts with currently accepted theories. Adjustable parameters with density last.

\noindent \textbf{Functional relationships:}
\begin{align*}
% z &\neq f(\lambda),\\
	d_A &= D_H z,\\ % Ref: Equation (3) on p. 335 in Broberg333.1993
	d_L &=^? \sqrt{1 + z} d_A,\\ % Ref: ?
	m - M &= 5\log_{10}[d_L/D_H] + C,\\ % Ref: Mannheim340.2006, arXiv:astro-ph/0505266v2 p. 45.
	\left\langle SB\right\rangle &\propto \left(d_A/d_L\right)^2,\\ % Ref: astro-ph/0603500
	F_\tau &= 1 + z. % Ref: ? Mechanism completely independent of light frequency
%	\eta_z &= \mbox{ not available},\\ % Ref: ?
%	\Delta_\theta &= \mbox{ not available}. % Ref: ?
\end{align*}
Functional relationships for $\eta_z$ and $\Delta_\theta$ are not available.

\noindent \textbf{Discussion and comments:}

  This model introduces an ad hoc hypothesis: ``The elementary energy quantum in the universe is equal to the product of Planck's and Hubble's constants''.
%
%                                                 ---------->>>>>>>>>>

%                                                 <<<<<<<<<<----------1988
\stepcounter{mechanism_number}
\subsection{\label{sec:mec.vigier}Interaction of a massive Photon with Vacuum Particles}
% http://redshift.vif.com
  Pecker and Vigier introduce a new tired-light mechanism involving an interaction between a massive photon and Dirac's vacuum particles.  A new term in the equations of quantum mechanics is hypothesized which cause the vacuum itself to absorb energy\cite{Vigier19.1988}.

%\textbf{Conditions, limits of applicability and restrictions:}

% Conditions, limits of applicability and restrictions: Solar limb red-shift? Quasars? Discordant red-shifts? Scale: intergalactic, interstellar, interplanetary red-shift? The assumptions required to make the mechanism work, adjustable parameters, and conflicts with currently accepted theories. Adjustable parameters with density last.

\noindent \textbf{Functional relationships:}
\begin{align*}
% z &\neq f(\lambda),\\
	d_A &=^? c \ln(1 + z)/(77.5\mbox{~(km/s)/Mpc}),\\ % Ref: ?
	d_L &=^? \sqrt{1 + z} d_A,\\ % Ref: ?
	m - M &= 5\log_{10}[d_L/D_H] + C,\\ % Ref: Mannheim340.2006, arXiv:astro-ph/0505266v2 p. 45.
	\left\langle SB\right\rangle &\propto \left(d_A/d_L\right)^2,\\ % Ref: astro-ph/0603500
	F_\tau &= 1 + z. % Ref: ? Mechanism completely independent of light frequency
%	\eta_z &= \mbox{ not available},\\ % Ref: ?
%	\Delta_\theta &= \mbox{ not available}. % Ref: ?
\end{align*}
Functional relationships for $\eta_z$ and $\Delta_\theta$ are not available.

\textbf{Discussion and comments:}

  This model introduces an ad hoc hypothesis: ``the massive photon interacts with the vacuum''.
%
%                                                 ---------->>>>>>>>>>

%                                                 <<<<<<<<<<----------1998
\stepcounter{mechanism_number}
\subsection{\label{sec:mec.hannon}Hannon Red-Shift} % doit etre reclassifier si dependent de la gravite
  In Hannon's paper, the nature and conventional explanation of the cosmological red-shift are briefly reviewed.  A simple alternative explanation is offered\cite{Hannon576.1998}.
% seems to be related to light interacting with itself, read http://forums.randi.org/archive/index.php/t-136128.html

%\textbf{Conditions, limits of applicability and restrictions:}

% Conditions, limits of applicability and restrictions: Solar limb red-shift? Quasars? Discordant red-shifts? Scale: intergalactic, interstellar, interplanetary red-shift? The assumptions required to make the mechanism work, adjustable parameters, and conflicts with currently accepted theories. Adjustable parameters with density last.

\noindent \textbf{Functional relationships:}
\begin{align*}
% z &\neq f(\lambda),\\
%	d_A &= \mbox{ not available},\\ % Ref: ?
	d_L &=^? \sqrt{1 + z} d_A,\\ % Ref: ?
	m - M &= 5\log_{10}[d_L/D_H] + C,\\ % Ref: Mannheim340.2006, arXiv:astro-ph/0505266v2 p. 45.
	\left\langle SB\right\rangle &\propto \left(d_A/d_L\right)^2. % Ref: astro-ph/0603500
%	F_\tau &= \mbox{ not available},\\ % Ref: ?
%	\eta_z &= \mbox{ not available},\\ % Ref: ?
%	\Delta_\theta &= \mbox{ not available}. % Ref: ?
\end{align*}
Functional relationships for $d_A, F_\tau, \eta_z$ and $\Delta_\theta$ are not available.

%\textbf{Discussion and comments:}

%
%                                                 ---------->>>>>>>>>>

%                                                 <<<<<<<<<<----------1990
\stepcounter{mechanism_number}
\subsection{\label{sec:mec.gray}Finite Conductivity of Space}
  Gray's mechanism is based on a recent proposal by Monti that background space may have a very small but finite conductivity.  The theory is based upon a relativistically invariant, semi-superconductive ether of the Einstein-Dirac genre.  It is shown that the Hubble red-shift may be interpreted as the resonant relaxation of all four time constants of the vacuum state, given by all time-like ratios and products of the vacuum constitutive parameters\cite{Gray436.1990}.

%\textbf{Conditions, limits of applicability and restrictions:}

% Conditions, limits of applicability and restrictions: Solar limb red-shift? Quasars? Discordant red-shifts? Scale: intergalactic, interstellar, interplanetary red-shift? The assumptions required to make the mechanism work, adjustable parameters, and conflicts with currently accepted theories. Adjustable parameters with density last.

\noindent \textbf{Functional relationships:}
\begin{align*}
% z &\neq f(\lambda),\\
%	d_A &= \mbox{ not available},\\ % Ref: ?
	d_L &=^? \sqrt{1 + z} d_A,\\ % Ref: ?
	m - M &= 5\log_{10}[d_L/D_H] + C,\\ % Ref: Mannheim340.2006, arXiv:astro-ph/0505266v2 p. 45.
	\left\langle SB\right\rangle &\propto \left(d_A/d_L\right)^2,\\ % Ref: astro-ph/0603500
	F_\tau &= 1 + z. % Ref: ? Mechanism completely independent of light frequency
%	\eta_z &= \mbox{ not available},\\ % Ref: ?
%	\Delta_\theta &= \mbox{ not available}. % Ref: ?
\end{align*}
Functional relationships for $d_A, \eta_z$ and $\Delta_\theta$ are not available.

%\textbf{Discussion and comments:}

%
%                                                 ---------->>>>>>>>>>

%                                                 <<<<<<<<<<----------1992
\stepcounter{mechanism_number}
\subsection{\label{sec:mec.shtyrkov}Variable Light Velocity}
  In this model, Shtyrkov assumes that the vacuum has some small internal friction due to a real viscosity.  Such a resistance is introduced in the wave equation for a photon with a damping term with some constant vacuum damping constant $\gamma$.  As a result, the wavelength increases and the speed of light decrease as a function of propagation distance, and the frequency remains unchanged\cite{Shtyrkov4.1992, Shtyrkov327.1993}.

\textbf{Conditions, limits of applicability and restrictions:}

  To match the cosmological red-shift, $\gamma = 1/D_H$.
% Conditions, limits of applicability and restrictions: Solar limb red-shift? Quasars? Discordant red-shifts? Scale: intergalactic, interstellar, interplanetary red-shift? The assumptions required to make the mechanism work, adjustable parameters, and conflicts with currently accepted theories. Adjustable parameters with density last.

\noindent \textbf{Functional relationships:}
\begin{align*}
% z &\neq f(\lambda),\\
%	d_A &= \mbox{ not available},\\ % Ref: ?
	d_L &=^? \sqrt{1 + z} d_A,\\ % Ref: ?
	m - M &= 5\log_{10}[d_L/D_H] + C,\\ % Ref: Mannheim340.2006, arXiv:astro-ph/0505266v2 p. 45.
	\left\langle SB\right\rangle &\propto \left(d_A/d_L\right)^2,\\ % Ref: astro-ph/0603500
	F_\tau &= 1 + z. % Ref: ? Mechanism completely independent of light frequency
%	\eta_z &= \mbox{ not available},\\ % Ref: ?
%	\Delta_\theta &= \mbox{ not available}. % Ref: ?
\end{align*}
Functional relationships for $d_A, \eta_z$ and $\Delta_\theta$ are not available.

\noindent \textbf{Discussion and comments:}

  This model introduces an ad hoc hypothesis: ``the vacuum has some small internal friction due to a real viscosity''.
%
%                                                 ---------->>>>>>>>>>

%                                                 <<<<<<<<<<----------1993
\stepcounter{mechanism_number}
\subsection{\label{sec:mec.dart}Quantized Electromagnetic Radiation}
  In the model proposed by Dart, electromagnetic radiation is quantized in increments of $H_0 h$ and possesses a zero-point energy of $H_0 h/2$, in the same manner as a quantum mechanical oscillator\cite{Dart5.1993}.
% How does this produce a red-shift?

%\textbf{Conditions, limits of applicability and restrictions:}

% Conditions, limits of applicability and restrictions: Solar limb red-shift? Quasars? Discordant red-shifts? Scale: intergalactic, interstellar, interplanetary red-shift? The assumptions required to make the mechanism work, adjustable parameters, and conflicts with currently accepted theories. Adjustable parameters with density last.

\noindent \textbf{Functional relationships:}
\begin{align*}
% z &\neq f(\lambda),\\
%	d_A &= \mbox{ not available},\\ % Ref: ?
	d_L &=^? \sqrt{1 + z} d_A,\\ % Ref: ?
	m - M &= 5\log_{10}[d_L/D_H] + C,\\ % Ref: Mannheim340.2006, arXiv:astro-ph/0505266v2 p. 45.
	\left\langle SB\right\rangle &\propto \left(d_A/d_L\right)^2,\\ % Ref: astro-ph/0603500
	F_\tau &= 1 + z. % Ref: ? Mechanism completely independent of light frequency
%	\eta_z &= \mbox{ not available},\\ % Ref: ?
%	\Delta_\theta &= \mbox{ not available}. % Ref: ?
\end{align*}
Functional relationships for $d_A, \eta_z$ and $\Delta_\theta$ are not available.

\textbf{Discussion and comments:}

  This model introduces an ad hoc hypothesis: ``electromagnetic radiation is quantized in increments of $H_0 h$''.
%
%                                                 ---------->>>>>>>>>>

%                                                 <<<<<<<<<<----------1995
\stepcounter{mechanism_number}
\subsection{\label{sec:mec.stensch}Pulsating Photons}
  The theory presented by Stensch defines and analyses photons as elementary pulse beats of the universe.  They travel at the speed of light and pulsate while propagating through space.  This pulsating sets up electromagnetic fields whose harmonics are used to derive an alternative interpretation of Planck's classical blackbody spectrum\cite{Stensch553.2006, Stensch194.1995, Stensch207.1996, Stensch154.2001}.
% How does that produce a red-shift?

%\textbf{Conditions, limits of applicability and restrictions:}

% Conditions, limits of applicability and restrictions: Solar limb red-shift? Quasars? Discordant red-shifts? Scale: intergalactic, interstellar, interplanetary red-shift? The assumptions required to make the mechanism work, adjustable parameters, and conflicts with currently accepted theories. Adjustable parameters with density last.

\noindent \textbf{Functional relationships:}
\begin{align*}
% z &\neq f(\lambda),\\
%	d_A &= \mbox{ not available},\\ % Ref: ?
	d_L &=^? \sqrt{1 + z} d_A,\\ % Ref: ?
	m - M &= 5\log_{10}[d_L/D_H] + C,\\ % Ref: Mannheim340.2006, arXiv:astro-ph/0505266v2 p. 45.
	\left\langle SB\right\rangle &\propto \left(d_A/d_L\right)^2,\\ % Ref: astro-ph/0603500
	F_\tau &= 1 + z. % Ref: ? Mechanism completely independent of light frequency
%	\eta_z &= \mbox{ not available},\\ % Ref: ?
%	\Delta_\theta &= \mbox{ not available}. % Ref: ?
\end{align*}
Functional relationships for $d_A, \eta_z$ and $\Delta_\theta$ are not available.

\textbf{Discussion and comments:}

  This model introduces an ad hoc hypothesis: ``photons pulsate while propagating through space and set up electromagnetic fields which cause some energy loss''.
%
%                                                 ---------->>>>>>>>>>

%                                                 <<<<<<<<<<----------1998
\stepcounter{mechanism_number}
\subsection{\label{sec:mec.rothwarf}Degenerate Fermion Fluid}
  An aether model based upon a degenerate Fermion fluid, composed primarily of electrons and positrons in a negative energy state relative to the null state or true vacuum, is proposed and its consequences are explored for physics and cosmology. The model provides both insight and quantitative results for the derivation of Hubble's law; red-shifts; $\gamma$-ray bursters.  In the model, the speed of light decreases with time on the scale of the age of the universe\cite{Rothwarf444.1998}.

%\textbf{Conditions, limits of applicability and restrictions:}

% Conditions, limits of applicability and restrictions: Solar limb red-shift? Quasars? Discordant red-shifts? Scale: intergalactic, interstellar, interplanetary red-shift? The assumptions required to make the mechanism work, adjustable parameters, and conflicts with currently accepted theories. Adjustable parameters with density last.

\noindent \textbf{Functional relationships:}
\begin{align*}
% z &\neq f(\lambda),\\
%	d_A &= \mbox{ not available},\\ % Ref: ?
	d_L &=^? \sqrt{1 + z} d_A,\\ % Ref: ?
	m - M &= 5\log_{10}[d_L/D_H] + C,\\ % Ref: Mannheim340.2006, arXiv:astro-ph/0505266v2 p. 45.
	\left\langle SB\right\rangle &\propto \left(d_A/d_L\right)^2,\\ % Ref: astro-ph/0603500
	F_\tau &= 1 + z. % Ref: ? Mechanism completely independent of light frequency
%	\eta_z &= \mbox{ not available},\\ % Ref: ?
%	\Delta_\theta &= \mbox{ not available}. % Ref: ?
\end{align*}
Functional relationships for $d_A, \eta_z$ and $\Delta_\theta$ are not available.

\textbf{Discussion and comments:}

  A key assumption is that the speed of light is the Fermi velocity of the degenerate electron-positron plasma that dominates the aether.
%
%                                                 ---------->>>>>>>>>>

%                                                 <<<<<<<<<<----------2001
\stepcounter{mechanism_number}
\subsection{\label{sec:mec.marcinkowski}Quantized Emission Red-Shifts in Quasar Radiation}
  In Marcinkowski's mechanism, thermodynamic theory is applied to a needle-shaped aging optical quasar photon to deduce aging infrared radiation.  The conservation laws require a number of infrared photons emitted during a burst of aging radiation.  These bursts create a sequence of quantized energy levels (and wavelengths) for the aging optical quasar photon as it propagates through space.  A distribution of quasar sources at different distances then produces a group of observed quantized red-shifts arriving here on Earth\cite{Marcinkowski183.2001}.

%\textbf{Conditions, limits of applicability and restrictions:}

% Conditions, limits of applicability and restrictions: Solar limb red-shift? Quasars? Discordant red-shifts? Scale: intergalactic, interstellar, interplanetary red-shift? The assumptions required to make the mechanism work, adjustable parameters, and conflicts with currently accepted theories. Adjustable parameters with density last.

\noindent \textbf{Functional relationships:}
\begin{align*}
% z &\neq f(\lambda),\\
%	d_A &= \mbox{ not available},\\ % Ref: ?
	d_L &=^? \sqrt{1 + z} d_A,\\ % Ref: ?
	m - M &= 5\log_{10}[d_L/D_H] + C,\\ % Ref: Mannheim340.2006, arXiv:astro-ph/0505266v2 p. 45.
	\left\langle SB\right\rangle &\propto \left(d_A/d_L\right)^2. % Ref: astro-ph/0603500
%	F_\tau &= \mbox{ not available},\\ % Ref: ?
%	\eta_z &= \mbox{ not available},\\ % Ref: ?
%	\Delta_\theta &= \mbox{ not available}. % Ref: ?
\end{align*}
Functional relationships for $d_A, F_\tau, \eta_z$ and $\Delta_\theta$ are not available.

\textbf{Discussion and comments:}

  The theory is adjusted to agree with the first observed quantized red-shift.  This model introduces an ad hoc hypothesis: ``a needle-shaped aging optical quasar photon aging as it propagates through space''.
%
%                                                 ---------->>>>>>>>>>

%                                                 <<<<<<<<<<----------2004
\stepcounter{mechanism_number}
\subsection{\label{sec:mec.driscoll}Photon Structure}
   In Driscoll's model, it is assumed that the photon has electric dipole moment P normal to its spin, rotating at the photon frequency f and radiating classically\cite{Driscoll46.2004, Driscoll341.2007}.
%  If the photon has an electric dipole moment (EDM) rotating about its spin axis at photon frequency ?, it must radiate electromagnetic energy at a rate accounting simply for the observed 'accelerated' Hubble red-shift in a nonexpanding, noncontracting universe unbounded in 3 ? 1 dimensions, subject to accelerations only by local aggregations of observable mass. The EDM along the spin axis is zero.

%\textbf{Conditions, limits of applicability and restrictions:}

% Conditions, limits of applicability and restrictions: Solar limb red-shift? Quasars? Discordant red-shifts? Scale: intergalactic, interstellar, interplanetary red-shift? The assumptions required to make the mechanism work, adjustable parameters, and conflicts with currently accepted theories. Adjustable parameters with density last.

\noindent \textbf{Functional relationships:}
\begin{align*}
% z &\neq f(\lambda),\\
	d_A &=^? D_H \ln(1 + z),\\ % Ref: ?
	d_L &=^? \sqrt{1 + z} d_A,\\ % Ref: ?
	m - M &= 5\log_{10}[d_L/D_H] + C,\\ % Ref: Mannheim340.2006, arXiv:astro-ph/0505266v2 p. 45.
	\left\langle SB\right\rangle &\propto \left(d_A/d_L\right)^2,\\ % Ref: astro-ph/0603500
	F_\tau &= 1 + z. % Ref: ? Mechanism completely independent of light frequency
%	\eta_z &= \mbox{ not available},\\ % Ref: ?
%	\Delta_\theta &= \mbox{ not available}. % Ref: ?
\end{align*}
Functional relationships for $\eta_z$ and $\Delta_\theta$ are not available.

\textbf{Discussion and comments:}

  If the photon has an electric dipole moment, a laser beam should easily be bent by a strong electric field gradient.  To my knowledge, this has not been demonstrated experimentally.
%
%                                                 ---------->>>>>>>>>>

%                                                 <<<<<<<<<<----------2004
\stepcounter{mechanism_number}
\subsection{\label{sec:mec.alfonsofaus}Mass Boom}
  In the mechanism proposed by Alfonso-Faus, the gravitational masses and the speed of light vary with cosmological time (the masses increase and the speed of light decreases) in a manner consistent with Einstein's theory of general relativity.  This effect is a linear increase of gravitational masses with cosmological time, with a corresponding decrease of the speed of light and the gravitational constant.  By integrating Einstein's cosmological equations and finding the solution for the cosmological scale factor to be $a(t) = \mbox{constant} \times t^2$, an accelerated expansion for the universe is implied.  This is the interpretation given to recent observations obtained from the Supernova Type Ia\cite{Alfonso-Faus2004.133}.

\textbf{Conditions, limits of applicability and restrictions:}

% Conditions, limits of applicability and restrictions: Solar limb red-shift? Quasars? Discordant red-shifts? Scale: intergalactic, interstellar, interplanetary red-shift? The assumptions required to make the mechanism work, adjustable parameters, and conflicts with currently accepted theories. Adjustable parameters with density last.

\noindent \textbf{Functional relationships:}
\begin{align*}
% z &\neq f(\lambda),\\
%	d_A &= \mbox{ not available},\\ % Ref: ?
	d_L &=^? \sqrt{1 + z} d_A,\\ % Ref: ?
	m - M &= 5\log_{10}[d_L/D_H] + C,\\ % Ref: Mannheim340.2006, arXiv:astro-ph/0505266v2 p. 45.
	\left\langle SB\right\rangle &\propto \left(d_A/d_L\right)^2,\\ % Ref: astro-ph/0603500
	F_\tau &= 1 + z,\\ % red-shift applies to all frequencies, even as frequency -> 0 
	\eta_z &= 0,\\ % red-shift applies to all frequencies equally
	\Delta_\theta &= 0. % nothing for light to diffuse on
\end{align*}
A functional relationshis for $d_A$ is not available.

\noindent \textbf{Discussion and comments:}

  This model introduces an ad hoc hypothesis: ``the gravitational masses and the speed of light vary with cosmological time''.

  In the lab system no such time variations can be detected unless we are dealing with cosmological observations (e.g., the red-shift).
%
%                                                 ---------->>>>>>>>>>

%                                                 <<<<<<<<<<----------2005
\stepcounter{mechanism_number}
\subsection{\label{sec:mec.sanejouand}Varying-speed-of-light hypothesis}
  The varying-speed-of-light hypothesis, based on a paper by Wold\cite{Wold217.1935}, is that the speed of light decreases by nearly 2~cm/s per year.  It provides an alternative explanation for the red-shift--distance relationship of type Ia supernovae which is nowadays given in terms of a new form of (dark) energy of unknown origin.

  Sanejouand\cite{Sanejuand8.2005, Sanejuand9.2009} discusses the empirical evidences in favor of the hypothesis that the speed of light decreases with time according to $c(t) = c_0 + a_c t$, where $a_c \approx -H_0c_0$ produces a good fit for the red-shift--magnitude relationship of type Ia supernovae.  Lunar laser ranging data, the so-called Pioneer anomaly, as well as an alternative explanation for both the apparent time-dilation of remote events and the apparent acceleration of the Universe are thought to be consistent with this hypothesis.

%\textbf{Conditions, limits of applicability and restrictions:}

% Conditions, limits of applicability and restrictions: Solar limb red-shift? Quasars? Discordant red-shifts? Scale: intergalactic, interstellar, interplanetary red-shift? The assumptions required to make the mechanism work, adjustable parameters, and conflicts with currently accepted theories. Adjustable parameters with density last.

\noindent \textbf{Functional relationships:}
\begin{align*}
% z &\neq f(\lambda),\\
	d_A &= D_Hz(1 + z/2),\\ % derived from first principles: speed of light is c(t)=c_0-a_ct where t is time in the past (now is t=0, yesterday is t=1, etc.)  Then d_A=c_0t -a_ct^2/2.  Time flows faster today, wavelength is constant, so light appears to have a longer wavelength because it oscillates more slowly.  1 + z=c(t)/c_0
	d_L &= D_Hz(1 + z/2),\\ % Ref: inferred from d_g = -c_0^2/a_c (z+0.5z^2), talk ay Lyon by Sanejouand
	m - M &= 5\log_{10}[d_L/D_H] + C,\\ % Ref: Mannheim340.2006 arXiv:astro-ph/0505266v2 p. 45.  Also = 5\log_{10}[z+\frac{1}{2}z^2] + 5\log_{10}[-\frac{c_0^2}{a_c}] + 25 from  talk at Lyon by Sanejouand
	\left\langle SB\right\rangle &\propto \left(d_A/d_L\right)^2,\\ % Ref: astro-ph/0603500
	F_\tau &= 1 + z,\\ % his talk at Lyon
	\eta_z &= 0,\\ % Ref: ?
	\Delta_\theta &= 0. % from the description of the mechanism
\end{align*}

\textbf{Discussion and comments:}

  The functional dependence of $d_A$ given above was derived from the equation $1 + z = c(t)/c_0$.  $d_L$ is taken directly from the functional relation for $m - M$ obtained by Sanejouand.

  This model introduces an ad hoc hypothesis: ``the speed of light decreases with time''.  The main argument against this hypothesis is the constancy of fine-structure and Rydberg constants.  Both of them being combinations of several physical constants their constancy imply that, if the speed of light is time-dependent, at least two other fundamental ``constants'' have to vary as well.  Although the fine-structure constant is thought to have varied in the past (a few parts in $10^6$)\cite{Webb.Barrow4.1998}, it is known that today it does not vary more than $\Delta \alpha/\alpha = -1.6\pm 2.3\times 10^{-17}$ per year\cite{Rosenband.Bergquist319.2008}.

  It is not clear why the relationship between $d_A$ and $d_L$ is not $d_L = \sqrt{1 + z} d_A$ resulting from the loss of luminosity due to the red-shift.

  This hypothesis can also be described as a non uniform flow of time (with space not expanding).  This has some similarity to the ``Spatial Change of Time Flow mechanism''~(\S\ref{sec:mec.poliakov}).
%
%                                                 ---------->>>>>>>>>>

%                                                 <<<<<<<<<<----------2005
\stepcounter{mechanism_number}
\subsection{\label{sec:mec.singh}Gravity Nullification Model}
  Singh's Gravity Nullification model\cite{Singh1.2005} integrates the missing physics of spontaneous decay of particles into the existing physics theories, specifically the general theory of relativity, without altering its original formulation.  The red-shift arises from the Heisenberg uncertainty relation applied to the energy of the photon.

  The red-shift as a function of velocity is $1 + z = \sqrt{(c+v)/(c-v)}$ and the distance is
$$d=D_H \left[\frac{z} {\sqrt{1 + z}}\right].$$

%\textbf{Conditions, limits of applicability and restrictions:}

% Conditions, limits of applicability and restrictions: Solar limb red-shift? Quasars? Discordant red-shifts? Scale: intergalactic, interstellar, interplanetary red-shift? The assumptions required to make the mechanism work, adjustable parameters, and conflicts with currently accepted theories. Adjustable parameters with density last.

\textbf{Functional relationships:}
\begin{align*}
% z &\neq f(\lambda),\\
	d_A &= \frac{D_H z} {\sqrt{1 + z}},\\ % Ref: ?
	d_L &= d_A,\\ % Ref: ?
	m - M &= 5\log_{10}[z] - 2.5\log_{10}[1 + z] + C,\\ % Ref: Mannheim340.2006, arXiv:astro-ph/0505266v2 p. 45.
	\left\langle SB\right\rangle &\propto \mbox{Cte},\\ % Ref: astro-ph/0603500
	F_\tau &= \left[ \frac{2(1 + z)}{(1 + z)^2+1} \right],\\ % Ref: ?
	\eta_z &=^? 0,\\ % Ref: ?
	\Delta_\theta &= 0. % Ref: ?
\end{align*}

\textbf{Discussion and comments:}

  This model introduces an ad hoc hypothesis: ``the missing physics of spontaneous decay of particles''.

  It is not clear why the relationship between $d_A$ and $d_L$ is not $d_L = \sqrt{1 + z} d_A$ resulting from the loss of luminosity due to the red-shift.
%
%                                                 ---------->>>>>>>>>>

%                                                 <<<<<<<<<<----------2008
\stepcounter{mechanism_number}
\subsection{\label{sec:mec.roscoe}Scalar Potential}
  In Roscoe's model, the regular electromagnetic four-vector is replaced by a 3-D invariant in 16 dimensions.  This implies the existence of a tired-light mechanism by which energy is lost without scattering\cite{Roscoe221.2009}.

%\textbf{Conditions, limits of applicability and restrictions:}

% Conditions, limits of applicability and restrictions: Solar limb red-shift? Quasars? Discordant red-shifts? Scale: intergalactic, interstellar, interplanetary red-shift? The assumptions required to make the mechanism work, adjustable parameters, and conflicts with currently accepted theories. Adjustable parameters with density last.

\noindent \textbf{Functional relationships:}
\begin{align*}
% z &\neq f(\lambda),\\
%	d_A &= \mbox{ not available},\\ % Ref: ?
	d_L &=^? \sqrt{1 + z} d_A,\\ % Ref: ?
	m - M &= 5\log_{10}[d_L/D_H] + C,\\ % Ref: Mannheim340.2006, arXiv:astro-ph/0505266v2 p. 45.
	\left\langle SB\right\rangle &\propto \left(d_A/d_L\right)^2.\\ % Ref: astro-ph/0603500
	F_\tau &= 1 + z,\\ % red-shift applies to all frequencies, even as frequency -> 0 
	\eta_z &= 0,\\ % red-shift applies to all frequencies equally
	\Delta_\theta &= 0. % nothing for light to diffuse on
\end{align*}
A functional relationship for $d_A$ is not available.

\textbf{Discussion and comments:}

  This model introduces an ad hoc hypothesis: ``the regular electromagnetic four-vector is a 3-D invariant in 16 dimensions''.
%
%                                                 ---------->>>>>>>>>>

%                                                 <<<<<<<<<<----------2010
\stepcounter{mechanism_number}
\subsection{\label{sec:mec.vlaicu}Beam Photon}
  The model presented by Vlaicu deduces the main properties of a composite photon with a linear beam structure from its specific emission mechanism.  This photon model answers some fundamental questions and explains spectral line broadening and the red-shift\cite{Vlaicu200.2010}.

%\textbf{Conditions, limits of applicability and restrictions:}

% Conditions, limits of applicability and restrictions: Solar limb red-shift? Quasars? Discordant red-shifts? Scale: intergalactic, interstellar, interplanetary red-shift? The assumptions required to make the mechanism work, adjustable parameters, and conflicts with currently accepted theories. Adjustable parameters with density last.

\noindent \textbf{Functional relationships:}
\begin{align*}
% z &\neq f(\lambda),\\
%	d_A &= \mbox{ not available},\\ % Ref: ?
	d_L &=^? \sqrt{1 + z} d_A,\\ % Ref: ?
	m - M &= 5\log_{10}[d_L/D_H] + C,\\ % Ref: Mannheim340.2006, arXiv:astro-ph/0505266v2 p. 45.
	\left\langle SB\right\rangle &\propto \left(d_A/d_L\right)^2. % Ref: astro-ph/0603500
%	F_\tau &= \mbox{ not available},\\ % Ref: ?
%	\eta_z &= \mbox{ not available},\\ % Ref: ?
%	\Delta_\theta &= \mbox{ not available}. % Ref: ?
\end{align*}
Functional relationships for $d_A, F_\tau, \eta_z$ and $\Delta_\theta$ are not available.

%\textbf{Discussion and comments:}

%
%                                                 ---------->>>>>>>>>>

%                                                 <<<<<<<<<<----------2010
\stepcounter{mechanism_number}
\subsection{\label{sec:mec.julia}Light interaction with Microwaves and Radio Waves}
  Julia's mechanism provides an explanation of two types of red-shifts: cosmological (without expansion of the universe) and intrinsic using a single tired light mechanism.  In the first case, the red-shift is produced because the light interacts with microwaves.  In the second, the interaction is with radio waves.  All this is compatible with a static universe with a space temperature of $2.7 K$~\cite{Julia97.2010}.

%\textbf{Conditions, limits of applicability and restrictions:}

% Conditions, limits of applicability and restrictions: Solar limb red-shift? Quasars? Discordant red-shifts? Scale: intergalactic, interstellar, interplanetary red-shift? The assumptions required to make the mechanism work, adjustable parameters, and conflicts with currently accepted theories. Adjustable parameters with density last.

\noindent \textbf{Functional relationships:}
\begin{align*}
% z &\neq f(\lambda),\\
%	d_A &= \mbox{ not available},\\ % Ref: ?
	d_L &=^? \sqrt{1 + z} d_A,\\ % Ref: ?
	m - M &= 5\log_{10}[d_L/D_H] + C,\\ % Ref: Mannheim340.2006, arXiv:astro-ph/0505266v2 p. 45.
	\left\langle SB\right\rangle &\propto \left(d_A/d_L\right)^2,\\ % Ref: astro-ph/0603500
	F_\tau &=^? 1. % Ref: ?
%	\eta_z &= \mbox{ not available},\\ % Ref: ?
%	\Delta_\theta &= \mbox{ not available}. % Ref: ?
\end{align*}
Functional relationships for $d_A, \eta_z$ and $\Delta_\theta$ are not available.

\textbf{Discussion and comments:}

  The cross section for this mechanism is not given.
%
%                                                 ---------->>>>>>>>>>

\clearpage

% Open Office Calc: graph 33cm x 16cm
% PSP: resize 6.9in 300pixel/in, save as .eps colour, preview
\begin{figure}
\centerline{\includegraphics[width=6.9in]{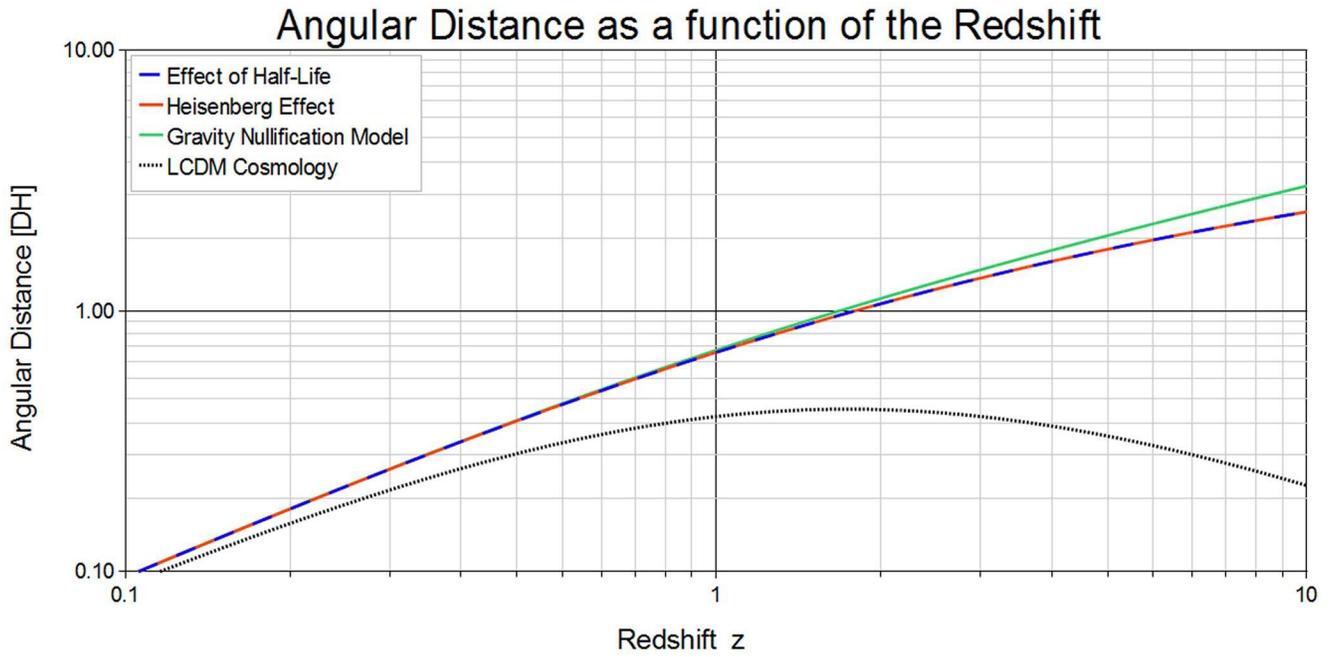}}
\caption{Angular distance in units of $D_H$ for a red-shift produced by a time-dependent property of light or an interaction of light with itself.  ``LCDM Cosmology'' is included for comparison.}
\label{fig:dAlight}
\end{figure}

\begin{figure}
\centerline{\includegraphics[width=6.9in]{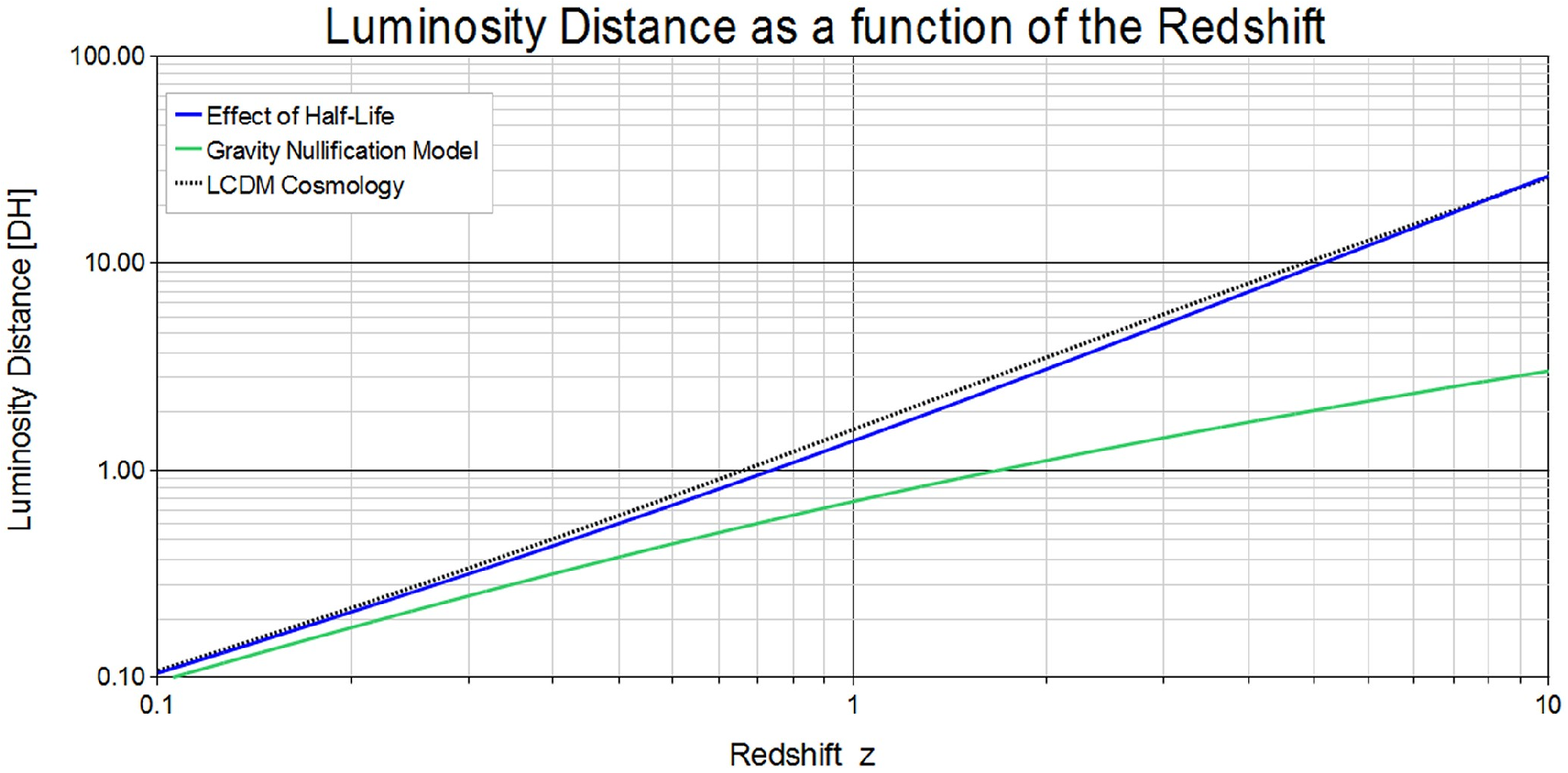}}
\caption{Luminosity distance in units of $D_H$ for a red-shift produced by a time-dependent property of light or an interaction of light with itself.  ``LCDM Cosmology'' is included for comparison.}
\label{fig:dLlight}
\end{figure}

\begin{figure}
\centerline{\includegraphics[width=6.9in]{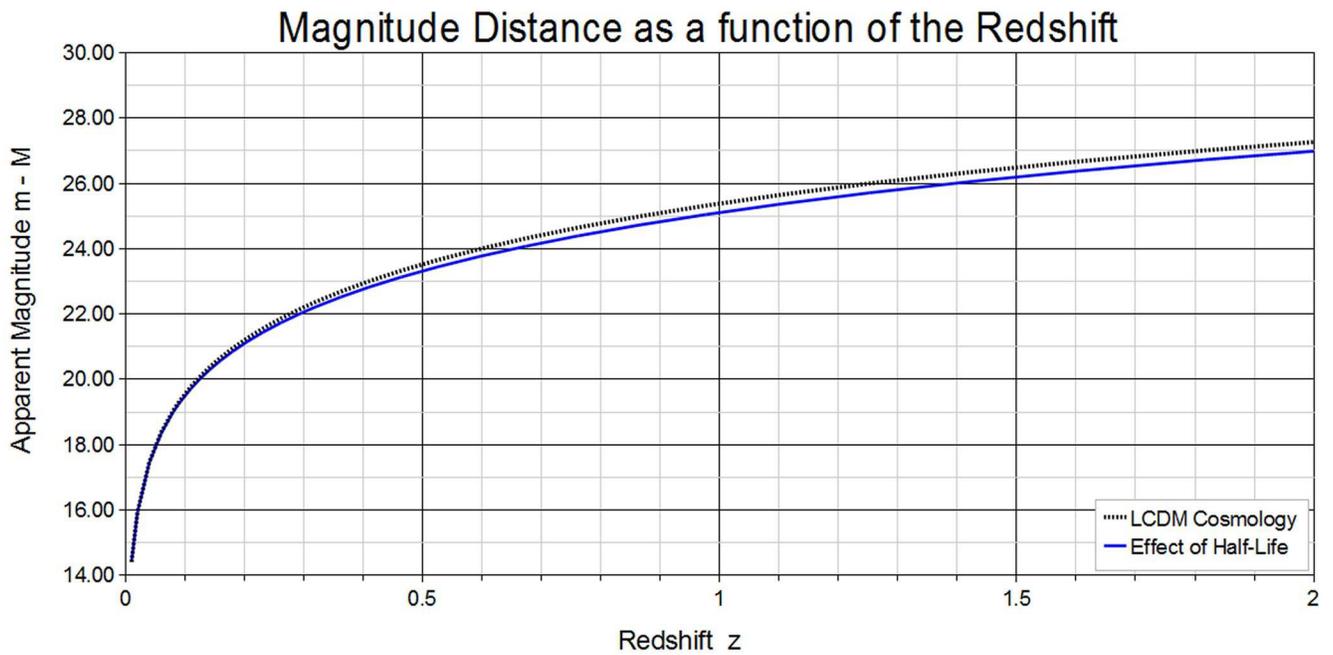}}
\caption{Magnitude distance for a red-shift produced by a time-dependent property of light or an interaction of light with itself.  ``LCDM Cosmology'' is included for comparison.}
\label{fig:dMlight}
\end{figure}

\clearpage

% ####################################################################################### %
\section{Red-Shifts produced by a time-independent geometry of space and time}
  The red-shift mechanisms listed in this section depend on a constant geometry of space and time.  For all the mechanisms in this section
\begin{align*}
	F_\tau &= 1 + z,\\ % Ref: from relativity: time dilation independent of frequency.
	\eta_z &= 0,\\ % Ref: no wavelength spread in relativity.
	\Delta_\theta &= 0. % Ref: no angular spread in relativity.  Nothing for light to diffuse on
\end{align*}

%                                                 <<<<<<<<<<----------1972
\stepcounter{mechanism_number}
\subsection{\label{sec:mec.segal}Chronometric Red-Shift Theory}
  In Segal's mechanism, the red-shift varies with distance $z=\tan^2(d_C/2R)$ where $R$ is the radius of the universe.  An alternative to the big-bang cosmology is developed in which red-shifts are explained, not by the expansion of the universe, but by an effect of conformal geometry.  The theory predicts a quadratic redshift-distance relation instead of the usual linear one\cite{Segal669.1976}
% http://math.ucr.edu/home/baez/segal.html, biography of I.E. Segal
%

%\textbf{Conditions, limits of applicability and restrictions:}

% Conditions, limits of applicability and restrictions: Solar limb red-shift? Quasars? Discordant red-shifts? Scale: intergalactic, interstellar, interplanetary red-shift? The assumptions required to make the mechanism work, adjustable parameters, and conflicts with currently accepted theories. Adjustable parameters with density last.

\noindent \textbf{Functional relationships:}
\begin{align*}
% z &\neq f(\lambda),\\
	d_A &=^? D_H \cos^{-1}\left(1/(1 + z)\right),\\ % Ref: D_A(z) = 2*R*sqrt(z)/(1 + z), Wright1996.1
	d_L &=^? (D_H/2) \left((1 + z)^2 - 1 \right),\\ % Ref: D_L(z) = sqrt[L(bol)/(4*pi*F(bol))]=sqrt(1 + z)*D_A(z) = 2*R*sqrt(z/(1 + z)), from Wright1996.1
	m - M &= 5\log_{10}[d_L/D_H] + C,\\ % Ref: Mannheim340.2006, arXiv:astro-ph/0505266v2 p. 45.
	\left\langle SB\right\rangle &\propto \left(d_A/d_L\right)^2. % Ref: astro-ph/0603500
%	F_\tau &= 1 + z,\\ % red-shift applies to all frequencies, even as frequency -> 0 
%	\eta_z &= 0,\\ % red-shift applies to all frequencies equally
%	\Delta_\theta &= 0. % nothing for light to diffuse on
\end{align*}

\textbf{Discussion and comments:}

  The chronometric theory is based on the assumption that the total energy of the photon is represented as the sum of a laboratory-measured scale-covariant component and an anti-scale-covariant component.  This is a different assumption from that used in relativity.

  For small distances, the red-shift predicted in the chronometric red-shift theory has a quadratic dependence on the distance.  This is in disagreement with observations\cite{Wright1996.1}.
%
%                                                 ---------->>>>>>>>>>

%                                                 <<<<<<<<<<----------1999
\stepcounter{mechanism_number}
\subsection{\label{sec:mec.masreliez}Scale Expanding Cosmos}
  In Masreliez's model, the length of a meter expands and the pace of time slows down, making time intervals like the second longer in proportion\cite{Masreliez399.1999}.  The wavelength of light is also made longer.  The Friedmann-Robertson-Walker line-element assumes that the pace of proper time always has been the same and is therefore different from the SEC line-element with its changing pace of time.  The SEC model cannot be modeled by General Relativity (GR) as experienced by an inhabitant because the pace of proper time is changing (the element d$s$ in GR is changing).  This changing pace of proper time models the progression of time, which cannot be modeled by GR.  The scale expanding cosmos behaves like an ordinary black body cavity without changing the cosmological temperature and the CMB results from thermal relaxation of electromagnetic radiation.

\textbf{Conditions, limits of applicability and restrictions:}

  General Relativity has to be generalized to include discrete scale adjustments that change the pace of proper time.  This means that different epochs belong to separate space-time manifolds in GR - they are not covariant.  Masreliez believes that a theory which cannot model the progression of time cannot model the universe, since the progression of time is the most important aspect of all existence, and that any model based on GR and the FRW line-element will fail.
% Conditions, limits of applicability and restrictions: Solar limb red-shift? Quasars? Discordant red-shifts? Scale: intergalactic, interstellar, interplanetary red-shift? The assumptions required to make the mechanism work, adjustable parameters, and conflicts with currently accepted theories. Adjustable parameters with density last.

\noindent \textbf{Functional relationships:}
\begin{align*}
% z &\neq f(\lambda),\\
	d_A &= D_H \ln(1 + z),\\ % Ref: de l'auteur 2009/5/21
	d_L &= (1 + z) d_A,\\ % Ref: de l'auteur 2009/5/21
	m - M &= 5\log_{10}[1 + z] + 5\log_{10}[\ln(1 + z)] + C,\\ % Ref: Mannheim340.2006, arXiv:astro-ph/0505266v2 p. 45.
	\left\langle SB\right\rangle &\propto (1 + z)^{-2}. % Ref: astro-ph/0603500
%	F_\tau &= 1 + z,\\ % red-shift applies to all frequencies, even as frequency -> 0 
%	\eta_z &= 0,\\ % red-shift applies to all frequencies equally
%	\Delta_\theta &= 0. % nothing for light to diffuse on
\end{align*}

%\textbf{Discussion and comments:}

%
%                                                 ---------->>>>>>>>>>

%                                                 <<<<<<<<<<----------2001
\stepcounter{mechanism_number}
\subsection{\label{sec:mec.ranzan}Dynamic Steady State Universe}
  Ranzan explains that while space does expand, and in doing so gives rise to the cosmic red-shift, the universe itself does not expand.  The mechanism uses a cellular structure where space expands inside the cells, but the cells themselves do not expand.  The red-shift is produced while light travels within cells but galaxies do not recede from each other.  Thus, the cellular structure interpreted by DSSU cosmology is the manifestation of dual-dynamic space involving space expansion and contraction at the edge of the cells.  A relevant cosmic-red-shift expression, applicable to this type of universe, is developed; then transformed into a cosmic redshift-distance relation\cite{Ranzan1.2001}.
% http://www.cellularuniverse.org/D1CosmicDistEq.pdf

\textbf{Conditions, limits of applicability and restrictions:}

  The red-shift produced inside one cell is $z_{UU}=0.01823$.

% Conditions, limits of applicability and restrictions: Solar limb red-shift? Quasars? Discordant red-shifts? Scale: intergalactic, interstellar, interplanetary red-shift? The assumptions required to make the mechanism work, adjustable parameters, and conflicts with currently accepted theories. Adjustable parameters with density last.

\noindent \textbf{Functional relationships:}
\begin{align*}
% z &\neq f(\lambda),\\
	d_A &= D_H \,z_{UU} \ln(1 + z)/\ln(1 + z_{UU}),\\ % Ref: http://www.cellularuniverse.org/D1CosmicDistEq.pdf
	d_L &=^? \sqrt{1 + z} d_A,\\ % Ref: ?
	m - M &= 5\log_{10}[d_L/D_H] + C,\\ % Ref: Mannheim340.2006, arXiv:astro-ph/0505266v2 p. 45.
	\left\langle SB\right\rangle &\propto \left(d_A/d_L\right)^2. % Ref: astro-ph/0603500
%	F_\tau &= 1 + z,\\ % red-shift applies to all frequencies, even as frequency -> 0 
%	\eta_z &= 0,\\ % red-shift applies to all frequencies equally
%	\Delta_\theta &= 0. % nothing for light to diffuse on
\end{align*}

\noindent \textbf{Discussion and comments:}

  This model introduces an ad hoc hypothesis: ``the cellular structure of the universe''.

  It is claimed that while aether-space within the cosmic cells expands, the boundaries between cells reverse the expansion by contracting the aether-space that constitutes the flow.  It is not clear why the wavelength of light would not be contracted when it crosses the boundaries between cells.
%
%                                                 ---------->>>>>>>>>>

%                                                 <<<<<<<<<<----------2008
\stepcounter{mechanism_number}
\subsection{\label{sec:mec.mayer}Gravitational Interaction}
% http://www.galaxyzooforum.org/index.php?topic=273364.0
  In Mayer's model, the time dimension where the light is emitted is not always parallel to the time dimension where the light will be detected.  This produces a projection reducing the observed rate, thus causing a red-shift\cite{Mayer1.2008}.

%\textbf{Conditions, limits of applicability and restrictions:}

% Conditions, limits of applicability and restrictions: Solar limb red-shift? Quasars? Discordant red-shifts? Scale: intergalactic, interstellar, interplanetary red-shift? The assumptions required to make the mechanism work, adjustable parameters, and conflicts with currently accepted theories. Adjustable parameters with density last.

\noindent \textbf{Functional relationships:}
\begin{align*}
% z &\neq f(\lambda),\\
	d_A &=^? D_H \cos^{-1}\left(1/(1 + z)\right),\\ % Ref: ?
	d_L &=^? (D_H/2) \left((1 + z)^2 - 1 \right),\\ % Ref: this can't be right?
	m - M &= 5\log_{10}[d_L/D_H] + C,\\ % Ref: Mannheim340.2006, arXiv:astro-ph/0505266v2 p. 45.
	\left\langle SB\right\rangle &\propto \left(d_A/d_L\right)^2. % Ref: astro-ph/0603500
%	F_\tau &= 1 + z,\\ % red-shift applies to all frequencies, even as frequency -> 0 
%	\eta_z &= 0,\\ % red-shift applies to all frequencies equally
%	\Delta_\theta &= 0. % nothing for light to diffuse on
\end{align*}

\textbf{Discussion and comments:}

  This model introduces an ad hoc hypothesis: ``the time dimension where the light is emitted is not always parallel to the time dimension where the light is detected''.
%
%                                                 ---------->>>>>>>>>>

%                                                 <<<<<<<<<<----------2008
\stepcounter{mechanism_number}
\subsection{\label{sec:mec.vonbrzeski}Lobachevsky Space}
  J.G. von Brzeski and V. von Brzeski use a Lobachevskian geometry of constant negative curvature to describe the universe\cite{vonBrzeski.vonBrzeski18.2008}.  The following description is limited to the red-shift in empty space as a consequence of the non-Euclidean geometry of space.  This model is based on the concepts that light follows the shortest path along geodesics, and that horospheres in Lobachevskian space are orthogonal to the geodesics and can thus be interpreted as wavefronts.  A key theorem on the rate of divergence of geodesics in Lobachevskian geometry states that the length of a segment increases exponentially with the hyperbolic distance $\lambda=\lambda_0\exp(\delta)$.  As galaxies are observed at greater distances, space is proportionally larger and the wavelength of light is longer.  The hyperbolic distance $\delta$ in Lobachevskian geometry can be mapped onto Euclidian space using $d=\rho R$ where $\rho=\tanh(\delta)$.

\textbf{Conditions, limits of applicability and restrictions:}

  Distances are measured without involving the notion of time.  The theory uses an older definition of the primary length standard as defined in the General Conference on Weights and Measures in 1983: ``The wavelength of the iodine stabilized HeNe laser is $\lambda_{HeNe} = 632.99139822$ nm''.

  The Doppler effect (which, in Lobachevskian geometry, follows from the negative curvature of velocity space) is not described here.

  The following equations were derived from published references\cite{vonBrzeski366.2007, vonBrzeski.vonBrzeski281.2003}.  The luminosity distance is derived from the ratio of intensities of electromagnetic radiation in Lobachevskian space versus Euclidean space $\sqrt{1-\rho^2}(1-\rho)$, combined with the Euclidean $(4\pi r^2)^{-1}$ law\cite{vonBrzeski.vonBrzeski169.2003}.  The reader is invited to refer directly to these papers.  The time dilation factor corresponds to the low frequency components of the electromagnetic spectrum which are also stretched by a factor $1 + z$.
% Conditions, limits of applicability and restrictions: Solar limb red-shift? Quasars? Discordant red-shifts? Scale: intergalactic, interstellar, interplanetary red-shift? The assumptions required to make the mechanism work, adjustable parameters, and conflicts with currently accepted theories. Adjustable parameters with density last.

\noindent \textbf{Functional relationships:}
\begin{align*}
% z &\neq f(\lambda),\\
	d_A &=^? D_H (1 + z)\sinh[ 2\ln(1 + z) ]/2,\\ %  or? D_H (1 + z)\tanh[ \ln(1 + z) ] % Ref: ?
	d_L &= \frac{D_H \tanh[ \ln(1 + z) ]} {\left[\sqrt{1-\rho^2}(1-\rho)\right]^{-1/2}},\\ % Ref: ?
	m - M &= 5\log_{10}[d_L/D_H] + C,\\ % Ref: Mannheim340.2006, arXiv:astro-ph/0505266v2 p. 45.
	\left\langle SB\right\rangle &\propto \left(d_A/d_L\right)^2. % Ref: astro-ph/0603500
%	F_\tau &= 1 + z,\\ % red-shift applies to all frequencies, even as frequency -> 0 
%	\eta_z &= 0,\\ % red-shift applies to all frequencies equally
%	\Delta_\theta &= 0. % nothing for light to diffuse on
\end{align*}

%\textbf{Discussion and comments:}

%
%                                                 ---------->>>>>>>>>>

%                                                 <<<<<<<<<<----------2008
\stepcounter{mechanism_number}
\subsection{\label{sec:mec.poliakov}Spatial Change of Time Flow}
  In Poliakov's model, the red-shift arises when an observer measures the frequency emitted by an object located in another part of space where time flows differently.  Since the flow of time varies continuously and monotonically, light that has traveled through space and time is seen with a red-shift.

  The mechanism is based on the premise that time is not uniform, but migrates between space-time systems to different `time flow' system\cite{Poliakov7.2008, Poliakov5.2008, Poliakov8.2008}.  Based on Noether's theorem, it is demonstrated that energy is not conserved in a non-uniform time frame.  As the flow of time does not remain constant, the pace of a process cannot remain constant either.  In the corpuscular case the velocity of a particle must grow in a zone of `slower time flow'.  However, as the speed of light obeys the Einsteinian axiom $c = \mbox{const.}$, the only possibility for a light quantum is that instead of velocity its frequency decreases in a zone of `faster time flow', i.e., the frequency observed in a zone of faster time flow is lower than the frequency of light emitted in the zone of slower time flow.  In other words, a red-shift is observed.

  This red-shift mechanism is far from artificial as it also explains the nature of Cold Dark Matter, the Cosmological Microwave Background Radiation and most importantly, the nature of gravity.  Since the inverse-square law of universal gravitation naturally appears as a consequence of the linear dependence of the flow of time on space, the red-shift as a function of space must be linear as well.

  The above does not imply that the more distant from the Earth an object is the faster it moves away.  The red-shift is observed even in the solar spectrum.  Yet, the Sun does not move away from the Earth.

\textbf{Conditions, limits of applicability and restrictions:}

  The rate of change of time flow is chosen to produce the observed $d = D_H z$ for small $z$.
% Conditions, limits of applicability and restrictions: Solar limb red-shift? Quasars? Discordant red-shifts? Scale: intergalactic, interstellar, interplanetary red-shift? The assumptions required to make the mechanism work, adjustable parameters, and conflicts with currently accepted theories. Adjustable parameters with density last.

\noindent \textbf{Functional relationships:}
\begin{align*}
% z &\neq f(\lambda),\\
	d_A &= D_H z,\\ % Ref: ?
	d_L &=^? (1 + z) d_A,\\ % Ref: ?
	m - M &= 5\log_{10}[d_L/D_H] + C,\\ % Ref: Mannheim340.2006, arXiv:astro-ph/0505266v2 p. 45.
	\left\langle SB\right\rangle &\propto \left(d_A/d_L\right)^2. % Ref: astro-ph/0603500
%	F_\tau &= 1 + z,\\ % red-shift applies to all frequencies, even as frequency -> 0 
%	\eta_z &= 0,\\ % red-shift applies to all frequencies equally
%	\Delta_\theta &= 0. % nothing for light to diffuse on
\end{align*}

\noindent \textbf{Discussion and comments:}

  This theory predicts a monotonous increase of the acceleration g with increasing depth toward the center of the Earth, due to the decrease of the time flow (which is what really causes gravity).  That prediction is not observer experimentally.
%
%                                                 ---------->>>>>>>>>>

% ####################################################################################### %
\section{Red-Shifts produced by a time-independent property of a field, gravitational or other}
  The red-shift mechanisms listed in this section depend on the presence of a constant field.  For all the mechanisms in this section
\begin{align*}
	\eta_z &= 0,\\ % red-shift applies to all frequencies equally
	\Delta_\theta &= 0. % nothing for light to diffuse on
\end{align*}

%                                                 <<<<<<<<<<----------1987
\stepcounter{mechanism_number}
\subsection{\label{sec:mec.ghosh}Velocity Dependent Inertial Induction}
  Gosh's model of inertial induction is based on a proposed extension of Mach's Principle.  According to this model the gravitational interaction between two main particles generates a force which depends not only on their separation but also on their relative velocity and acceleration\cite{Ghosh18.1987}. % Clarify: how does this produce a red-shift?

%\textbf{Conditions, limits of applicability and restrictions:}

% Conditions, limits of applicability and restrictions: Solar limb red-shift? Quasars? Discordant red-shifts? Scale: intergalactic, interstellar, interplanetary red-shift? The assumptions required to make the mechanism work, adjustable parameters, and conflicts with currently accepted theories. Adjustable parameters with density last.

\noindent \textbf{Functional relationships:}
\begin{align*}
% z &\neq f(\lambda),\\
%	d_A &= \mbox{ not available},\\ % Ref: ?
%	d_L &= \mbox{ not available},\\ % Ref: ?
	m - M &= 5\log_{10}[d_L/D_H] + C,\\ % Ref: Mannheim340.2006, arXiv:astro-ph/0505266v2 p. 45.
	\left\langle SB\right\rangle &\propto \left(d_A/d_L\right)^2. % Ref: astro-ph/0603500
%	F_\tau &= \mbox{ not available}. % Ref: ?
%	\eta_z &= 0,\\ % red-shift applies to all frequencies equally
%	\Delta_\theta &= 0. % nothing for light to diffuse on
\end{align*}
Functional relationships for $d_A, d_L$ and $F_\tau$ are not available.

\textbf{Discussion and comments:}

  According to the author, this model can be verified from other local effects predicted by this mechanism.  However, I am not aware of any experiment confirming this model.
%
%                                                 ---------->>>>>>>>>>

%                                                 <<<<<<<<<<----------1990
\stepcounter{mechanism_number}
\subsection{\label{sec:mec.stein}Photon-Graviton Interaction}
  Stein's mechanism explains the red-shift of galaxies with an interaction between photons and gravitons.  For this kind of interaction between two particles/waves, both moving with the velocity of light, Einstein's special and general theories of relativity are modified.  In special relativity, Einstein's assumption that the velocity of light is constant relative to any inertial system is extended to an absolute velocity of light.  General relativity is modified using an Euclidean space, choosing the frame of reference in which the momentum of the total system of the interacting particles/waves is zero.  As a consequence, the red-shift of galaxies is at least in part an effect of the gravitational fields the light has crossed on its way from a galaxy to the Earth\cite{Stein169.1990}.

\textbf{Conditions, limits of applicability and restrictions:}

  Einstein's special and general theories of relativity need to be modified.
% Conditions, limits of applicability and restrictions: Solar limb red-shift? Quasars? Discordant red-shifts? Scale: intergalactic, interstellar, interplanetary red-shift? The assumptions required to make the mechanism work, adjustable parameters, and conflicts with currently accepted theories. Adjustable parameters with density last.

\noindent \textbf{Functional relationships:}
\begin{align*}
% z &\neq f(\lambda),\\
	d_A &= D_H \ln(1 + z),\\ % from the properties of a tired-light theory
%	d_L &= \mbox{ not available},\\ % Ref: ?
	m - M &= 5\log_{10}[d_L/D_H] + C,\\ % Ref: Mannheim340.2006, arXiv:astro-ph/0505266v2 p. 45.
	\left\langle SB\right\rangle &\propto \left(d_A/d_L\right)^2,\\ % Ref: astro-ph/0603500
	F_\tau &= 1 + z. % Ref: ? Mechanism completely independent of light frequency
%	\eta_z &= \mbox{ not available},\\ % Ref: ?
%	\Delta_\theta &= \mbox{ not available}. % Ref: ?
\end{align*}
Functional relationships for $d_L, \eta_z$ and $\Delta_\theta$ are not available.

%\textbf{Discussion and comments:}

%
%                                                 ---------->>>>>>>>>>

%                                                 <<<<<<<<<<----------1991
\stepcounter{mechanism_number}
\subsection{\label{sec:mec.jaakkola}Electrogravitational Coupling}
  In this model proposed by Jaakkola, the universal red-shift effect (containing the cosmological red-shift, intrinsic red-shifts in QSOs and intermediate strengths of z depending on the density of a system) is interpreted as a quantized loss of energy from the photon to a vacuum composed of gravitational quanta.  The model covers consistently all the observed features of red-shift, including its quantized fine-structure, observed both in distant-dependent and distance independent red-shifts.

  Gravitation appears as a pressure effect of cosmic gravitational quanta.  The apparent two-body attraction results from mutual screening of the gravitational pressure of the background vacuum\cite{Jaakkola199.1991}.

\textbf{Conditions, limits of applicability and restrictions:}

  The K-correction $K(z)$ is observational in character.
% Conditions, limits of applicability and restrictions: Solar limb red-shift? Quasars? Discordant red-shifts? Scale: intergalactic, interstellar, interplanetary red-shift? The assumptions required to make the mechanism work, adjustable parameters, and conflicts with currently accepted theories. Adjustable parameters with density last.

\noindent \textbf{Functional relationships:}
\begin{align*}
% z &\neq f(\lambda),\\
	d_A &= D_H \ln(1 + z),\\ % from the properties of a tired-light theory
%	d_L &= \mbox{ not available},\\ % Ref: ?
	m - M &= 5\log_{10} \ln [1 + z] + 2.5\log_{10}[1 + z] + K(z) + C,\\ % Ref: Jaakkola199.1991, Eq. (8)
	\left\langle SB\right\rangle &\propto \left(d_A/d_L\right)^2,\\ % Ref: astro-ph/0603500
	F_\tau &= 1 + z. % Ref: ? Mechanism completely independent of light frequency
%	\eta_z &= \mbox{ not available},\\ % Ref: ?
%	\Delta_\theta &= \mbox{ not available}. % Ref: ?
\end{align*}
Functional relationships for $d_L, \eta_z$ and $\Delta_\theta$ are not available.

%\textbf{Discussion and comments:}

%
%                                                 ---------->>>>>>>>>>

%                                                 <<<<<<<<<<----------1992
\stepcounter{mechanism_number}
\subsection{\label{sec:mec.fischer}Gravitomagnetic Effect}
  Fischer explains that due to Lorentz invariance of General Relativity, gravitational interaction is limited to the speed of light.  Thus for particles, moving within a matter field, retardation leads to loss of energy by emission of gravitational radiation.  This 'gravitomagnetic' effect, applied to motion in homogeneous mass filled space, acts like a viscous force, slowing down every motion in the universe on the Hubble time scale.  The energy loss rate exactly equals the red shift of photons in an expanding universe, thus showing the equivalence of wavelength stretching in the wave picture and energy loss in the photon picture.  The loss mechanism is not restricted to an expanding universe, however, but would also be present in a static Einstein universe\cite{Fischer149.1992, Fischer805.2008}.

%\textbf{Conditions, limits of applicability and restrictions:}

% Conditions, limits of applicability and restrictions: Solar limb red-shift? Quasars? Discordant red-shifts? Scale: intergalactic, interstellar, interplanetary red-shift? The assumptions required to make the mechanism work, adjustable parameters, and conflicts with currently accepted theories. Adjustable parameters with density last.

\noindent \textbf{Functional relationships:}
\begin{align*}
% z &\neq f(\lambda),\\
	d_A &=^? c \ln(1 + z)/(77.5\mbox{~(km/s)/Mpc}),\\ % Ref: ?
%	d_L &= \mbox{ not available},\\ % Ref: ?
	m - M &= 5\log_{10}[d_L/D_H] + C,\\ % Ref: Mannheim340.2006, arXiv:astro-ph/0505266v2 p. 45.
	\left\langle SB\right\rangle &\propto \left(d_A/d_L\right)^2,\\ % Ref: astro-ph/0603500
	F_\tau &= 1 + z. % Ref: ? Mechanism completely independent of light frequency
%	\eta_z &= \mbox{ not available},\\ % Ref: ?
%	\Delta_\theta &= \mbox{ not available}. % Ref: ?
\end{align*}
Functional relationships for $d_L, \eta_z$ and $\Delta_\theta$ are not available.

\textbf{Discussion and comments:}

  See \ref{sec:mec.zwicky} for energy loss due to the gravitational field.  Is the magnetic effect important here?
%
%                                                 ---------->>>>>>>>>>

%                                                 <<<<<<<<<<----------1998
\stepcounter{mechanism_number}
\subsection{\label{sec:mec.gentry}New Red-Shift Interpretation}
  Gentry's model is based on a universe governed by static-space-time general relativity interprets cosmological red shifts as a combination of gravitational red shifts and ordinary Doppler shifts coming from motion in a static geometry.  While the spatial geometry is not explicitly given, a number of expressions indicate that Gentry is considering a nearly Euclidean metric.

  To be consistent with observed isotropy of matter and cosmic microwave background radiation, Gentry's model is geocentric, placing our galaxy near the center of a static spherical ball of matter with a radius $R \sim 1.4 \times 10^{10}$ light years.  This matter consists of two components: ordinary pressure-less matter (``galaxies'') and a vacuum energy (that is, a cosmological constant).  The gravitational potential varies with distance from the center, and the resulting gravitational red shift thus depends on distance from the Earth\cite{Gentry051.1998}.

\textbf{Conditions, limits of applicability and restrictions:}

  The equations are given for $u_g=0$, that is, no proper velocity.

  To explain cosmic microwave background radiation, Gentry supposes that the ball of matter is surrounded by a thin shell of hydrogen at a temperature of $5400$K.  The resulting black body radiation must be gravitationally red shifted to the observed $2.7$K at the center of the ball.
% rewrite this because it is copied from the second reference

% Conditions, limits of applicability and restrictions: Solar limb red-shift? Quasars? Discordant red-shifts? Scale: intergalactic, interstellar, interplanetary red-shift? The assumptions required to make the mechanism work, adjustable parameters, and conflicts with currently accepted theories. Adjustable parameters with density last.

\noindent \textbf{Functional relationships:}
\begin{align*}
% z &\neq f(\lambda),\\
	1 + z &= (1+d_A/D_H)/\sqrt{1-2d_A^2/D_H^2},\\ % Ref: Carlip021.1998, Eq. 2: 1 + z = (1+Hr) / (sqrt(1-(2+u_g^2)H^2r^2)) where r=d_A since he uses a nearly Euclidean, spatial metric.
	% u_g=v_\Theta / v_r, the ratio of transverse to radial velocities.  Without velocity, 1 + z = (1+H d_A) / (sqrt(1-2H^2 d_A^2))
	d_L &= (1 + z) d_A,\\ % Ref: Gentry051.1998, Eq. (2)
	m - M &= 5\log_{10}[1 + z] + 5\log_{10}[d_A/D_H] + C,\\ % Ref: Mannheim340.2006, arXiv:astro-ph/0505266v2 p. 45.
	\left\langle SB\right\rangle &\propto (1 + z)^{-2},\\ % Ref: astro-ph/0603500
	F_\tau &= 1 + z. % red-shift applies to all frequencies, even as frequency -> 0 
%	\eta_z &= 0,\\ % red-shift applies to all frequencies equally
%	\Delta_\theta &= 0. % nothing for light to diffuse on
\end{align*}

\noindent \textbf{Discussion and comments:}

  Gentry's model is inconsistent with the Einstein field equations, requires delicate fine tuning of initial conditions, is highly unstable both gravitationally and thermodynamically and its predictions disagree with observation\cite{Carlip021.1998}.

  For the calculation of the luminosity distance, Gentry explains that ``The energy of each photon is decreased by $1 + z$ because of this red-shift, and a second red-shift occurs because Doppler recession causes the rate at which photons arrive at earth to be diminished by the same factor''~\cite{Gentry051.1998}.
%
%                                                 ---------->>>>>>>>>>

%                                                 <<<<<<<<<<----------2006
\stepcounter{mechanism_number}
\subsection{\label{sec:mec.hodge}Scalar Potential Model}
John Hodge proposes a model based on a ``plenum'' where the plenum density at each point in Euclidian space is determined by the location of mass like the ``space'' of general relativity, of sources (spiral galaxies), and of sinks (elliptical galaxies)\cite{Hodge36.2005}.  SPM suggests the photon is a particle and its energy is discrete and composed of a number of basic energy/particle packets.

%  Scalar potential model of red-shift and discrete (quantized) red-shift (New Astronomy\\
%\href{http://www.elsevier.com/wps/find/journaldescription.cws_home/601274/description#description}{www.elsevier.com/wps/find/journaldescription.cws$\_$home/601274/description$\#$description}).

\textbf{Conditions, limits of applicability and restrictions:}

	$D_s = c/H_s$ and $H_s=110 \pm 20$ km s$^{-1}$ Mpc$^{-1}$.

This same model with some simplification provides the best and only fit to all the ``Pioneer Anomaly'' data (Scalar potential model of the Pioneer Anomaly\cite{Hodge13.2006}. A further simplification results in the change-of-gravitational-potential explanation of the Pound-Rebka experiment.

\noindent \textbf{Functional relationships:}
\begin{align*}
% z &\neq f(\lambda),\\
	d_A &= D_s \ln{1 + z},\\ % Ref: Eq. 12 (arXiv:astro-ph/0602344v1)
	d_L &=^? d_A,\\ % derived from Ref: Mannheim340.2006, arXiv:astro-ph/0505266v2 p. 45 and equation m-M below.
	m - M &= 5\log_{10}[\ln{1 + z}] + C,\\ % Ref: Eq. 15 (arXiv:astro-ph/0602344v1)
	\left\langle SB \right\rangle &\propto (1 + z)^{-4},\\ % Ref: astro-ph/0603500
	F_\tau &= 1 + z. % red-shift applies to all frequencies, even as frequency -> 0 
%	\eta_z &= 0,\\ % red-shift applies to all frequencies equally
%	\Delta_\theta &= 0. % nothing for light to diffuse on
\end{align*}

\noindent \textbf{Discussion and comments:}

At larger distances, the red-shift equation reduces to the Hubble Law and a static universe.  Distances are consistent with TRGB and Cepheid distance calculations.  If quasars are as Arp suggests, the correlation between Cepheid distance and red-shift distance may be improved.  This model suggests the space temperature of the CMB average is 2.718 K, is higher because of a ``hunting'' characteristic of a feedback mechanism, and is not a relic of a Big Bang (Scalar potential model of the CMB radiation temperature\cite{Hodge4.2006}.

The model has provided insight into several galaxy observations that are inconsistent with current cosmological models.% (See links to \href{http://web.comporium.net/~scjh/}{web.comporium.net/$\sim$scjh/}).
%
%                                                 ---------->>>>>>>>>>

%                                                 <<<<<<<<<<----------2007
\stepcounter{mechanism_number}
\subsection{\label{sec:mec.qcm}Quantum Celestial Mechanics Gravitational Potential}
In Potter and Preston's model, the Hamilton-Jacobi equation of the General Theory of Relativity is changed into a Schr\"odinger-like equation which predicts quantization states for any gravitationally bound system.  These Quantum Celestial Mechanics (QCM) states depend only upon the total mass and the total angular momentum of the system.  Distant redshifted SNe1a light sources from the Universe that are usually interpreted as cosmological red-shifts are shown to be universal gravitational red-shifts seen by all observers.  An increasingly negative QCM gravitational potential with distance from the observer dictates a non-linear red-shift with distance and an 'apparent' gravitational repulsion, i.e. the source clock rate is slower than at the observer.  No space expansion is necessary, nor is dark matter and dark energy needed.

The `apparent' radial velocity $v=$~d$d/\mbox{d}t$ versus coordinate distance $d$ leads to a new Hubble relation:
$$v = d \frac{c\sqrt{k}}{1-kd^2}$$
where $k = 8\pi G \rho_c/3c^2$.  Best fit when Hubble parameter is about $62$~(km/s)/Mpc.  Thus, the 'apparent' radial velocity continues to increase with increasing distance\cite{Potter.Preston1.2007}.

\textbf{Conditions, limits of applicability and restrictions:}

$a=(1 + z)^2+1$, and $b=(1 + z)^2-1$. % e-mail from drpotter@lycos.com dated 04/06/2009 7:24 PM
% Conditions, limits of applicability and restrictions: Solar limb red-shift? Quasars? Discordant red-shifts? Scale: intergalactic, interstellar, interplanetary red-shift? The assumptions required to make the mechanism work, adjustable parameters, and conflicts with currently accepted theories. Adjustable parameters with density last.

\noindent \textbf{Functional relationships:}
\begin{align*}
% z &\neq f(\lambda),\\
	d_A &= D_H \frac{a}{2b} \left[ \sqrt{1+(4 b^2/a^2)}-1 \right],\\ % Ref: e-mail from drpotter@lycos.com dated 11/06/2009 12:58 PM
	d_L &= (1 + z)^2 d_A,\\ % Ref: e-mail from drpotter@lycos.com dated 17/06/2009 3:36 PM
	m - M &=^? 10\log_{10}[1 + z] + C,\\ % Ref: ?
	\left\langle SB\right\rangle &\propto (1 + z)^{-2},\\ % Ref: e-mail from drpotter@lycos.com dated 18/06/2009 3:10 PM
	F_\tau &= 1 + z. % red-shift applies to all frequencies, even as frequency -> 0 
%	\eta_z &= 0,\\ % red-shift applies to all frequencies equally
%	\Delta_\theta &= 0. % nothing for light to diffuse on
\end{align*}

\noindent \textbf{Discussion and comments:}

One possible test of the different red-shift theories is the interpretation of the SN1a rise-fall time.  If it is truly longer for distant observers compared to the shorter rise-fall time at the source SN1a, then 'tired-light' type of cosmological red-shifts can be eliminated.  The QCM approach has a gravitational potential decreasing with distance from the observer so the clock rates are different at the source and observer, which would explain the rise-fall time difference.
%
%                                                 ---------->>>>>>>>>>

%                                                 <<<<<<<<<<----------2008
\stepcounter{mechanism_number}
\subsection{\label{sec:mec.wilson}Eternal Contracting Universe}
  In Wilson's model, the observed cosmological red-shift results from a photon losing energy through gravitational radiation.  The energy depletion is calculated using general relativity's gravitational radiation effect\cite{Wilson1.2009}.
%	This perfect cosmological principle model has a contraction rate of 0.00029 km/(sec Mpc), a density parameter of 0.10417 maintained by energy annihilation, a quantized gravitational constant (function of radius) that predicts gravitons of negative energy (-1.63 10^-33 eV), and yields a potential unification of forces.  These elements, with alternative explanations for "consensus" phenomena, form a robust global coherency.  In CCC2.

\textbf{Conditions, limits of applicability and restrictions:}

  The predicted decay rate is obtained from the binary neutron star system PSR 1913 + 16 observation.  There doesn't seem to be other contraints.  The predicted energy decay rate is equivalent to a Hubble parameter of $77.5\pm 0.2$~(km/s)/Mpc.
% Conditions, limits of applicability and restrictions: Solar limb red-shift? Quasars? Discordant red-shifts? Scale: intergalactic, interstellar, interplanetary red-shift? The assumptions required to make the mechanism work, adjustable parameters, and conflicts with currently accepted theories. Adjustable parameters with density last.

\noindent \textbf{Functional relationships:}
\begin{align*}
% z &\neq f(\lambda),\\
	d_A &=^? c \ln(1 + z)/(77.5\mbox{~(km/s)/Mpc}),\\ % Ref: ?
%	d_L &= \mbox{ not available},\\ %$$^? $$ % Ref: ?
	m - M &= 5\log_{10}[d_L/D_H] + C,\\ % Ref: Mannheim340.2006, arXiv:astro-ph/0505266v2 p. 45.
	\left\langle SB\right\rangle &\propto \left(d_A/d_L\right)^2,\\ % Ref: astro-ph/0603500
	F_\tau &= 1 + z. % Ref: ? Mechanism completely independent of light frequency
%	\eta_z &= \mbox{ not available},\\ % Ref: ?
%	\Delta_\theta &= \mbox{ not available}. % Ref: ?
\end{align*}
Functional relationships for $d_L, \eta_z$ and $\Delta_\theta$ are not available.

%\textbf{Discussion and comments:}

%
%                                                 ---------->>>>>>>>>>

% ####################################################################################### %
\section{Red-Shifts produced by an interaction between light and matter}
The red-shift mechanisms listed in this section depend on the presence of matter or particles between the emitter and the detector.  A red-shift $1 + z(N)=\lambda_1/\lambda_0$ caused by the interaction with a column density of matter $N$ is valid everywhere so that $1 + z(N) = \lambda_{i+1}/\lambda_i$ for any $i$.  Repeated red-shifts follow the relation
$$1 + z(kN) = \lambda_k/\lambda_0 = \prod_{i=1}^k \lambda_i/\lambda_{i-1} = (1 + z(N))^k$$
which is satisfied by\cite{Nernst633.1937, Chepick108.2002}
\begin{equation}
N = \rho_0 D_H \ln(1 + z),
\label{eq:tiredlightredshift}
\end{equation}
where $\rho_0$ is the density of the interacting matter and the column density is given by
$$N(d) = \int_0^d {\rho(x) \mbox{d}x}.$$

Some comments must be made about the search of a light-matter interaction.  

Absorption by any element (H, He, atoms, molecules, dust, etc...) which has an internal structure implies a wavelength dependence resulting from resonances or absorption bands.  To satisfy the Tolman surface brightness test, the absorption has to be matched in magnitude with the red-shift mechanism at all wavelengths.  However, wavelength dependent red-shift mechanisms would fail to imitate a Doppler-like red-shift\cite{Lilley172.1956}.  For this reason, a cosmological light-matter interaction producing a red-shift can only be produced in elements which do not have an internal structure, such as electrons (as Ari Brynjolfsson correctly pointed out\cite{Brynjolfsson420.2004}) because no resonance or bands exists in electrons.  Gravitational type of tired-light theories do not provide photon absorption and therefore do not solve Olbers' paradox.  Only the light-matter interactions proposed by Shelton~(\S\ref{sec:mec.shelton}), Brynjolfsson~(\S\ref{sec:mec.brynjolfsson}), Kierein~(\S\ref{sec:mec.kiereinj}), Ashmore~(\S\ref{sec:mec.ashmore}), L. Marmet~(\S\ref{sec:mec.marmetl}), Vaughan~(\S\ref{sec:mec.vaughan}), Mamas~(\S\ref{sec:mec.mamas}) and Thomson~(\S\ref{sec:mec.thomson}) use electrons as the red-shift medium.  Thomson scattering blurs images beyond recognition, causes a red-shift and the cross section is wavelength independent, thus taking care of Olbers' paradox.

  Therefore: The Tolman surface brightness test, Olbers' paradox and light-matter interaction theories producing a Doppler-like red-shift require the following:\\
1- Thomson scattering on electrons, and\\
2- A mechanism independent of wavelength, and\\
3- A tired-light mechanism involving a photon-electron interaction.

  The arguments given by Wright against tired light models\cite{Wright1996.2} are incorrect because he assumes:
\begin{itemize}
  \item ``There is no known interaction that can degrade a photon's energy without also changing its momentum''.  This is correct, but multiple coherent interactions can degrade the photon's energy while maintaining beam direction.
  \item ``The tired light model does not predict the observed time dilation of high red-shift supernova light curves''.  This is not so for the tired-light mechanisms which are completely independent of the frequency of light.  For these mechanisms, $F_\tau = 1 + z$.  Since all frequencies are 'redshifted' by the same amount, even temporal variations of the order of hours and days appear to evolve more slowly.  One also notices that supernova light curves are slower at longer wavelengths.  Coherent interactions can maintain temporal variations since the resulting irradiance is proportional to the initial irradiance.
  \item ``The tired light model can not produce a blackbody spectrum for the Cosmic Microwave Background without some incredible coincidences''.  This is not so for some of the mechanisms described below where Compton scattering in electrons provides thermal coupling between photons and electrons at the average temperature of the universe\cite{Assis.Neves79.1995}.
  \item ``The tired light model fails the Tolman surface brightness test''.  This is not so for some of the mechanisms described below where electrons are both responsible for the red-shift and Thomson scattering which reduces the intensity of observed objects.  Note that the Big Bang also fails the Tolman test unless galactic evolution is included in the model.
  \item ``The local Universe is transparent\ldots''  Water is transparent yet interacts with light in a strong way.  Some mechanisms below have a sufficiently large cross-section.
\end{itemize}
It is remarkable that the logarithmic dependence of the distance on $z$ described by Eq. (\ref{eq:tiredlightredshift}) produces the appearance of an accelerating expansion.

%                                                 <<<<<<<<<<----------1929
\stepcounter{mechanism_number}
\subsection{\label{sec:mec.zwicky}Gravitational Drag}
  In Zwicky's model, photons passing near a mass are deflected.  They transfer momentum and energy to the mass.  The photon changes it's energy and therefore it's frequency\cite{Zwicky802.1935}.

\textbf{Conditions, limits of applicability and restrictions:}

  The masses are assumed to be independent of each other, but in reality they are coupled by gravitational forces.
% Conditions, limits of applicability and restrictions: Solar limb red-shift? Quasars? Discordant red-shifts? Scale: intergalactic, interstellar, interplanetary red-shift? The assumptions required to make the mechanism work, adjustable parameters, and conflicts with currently accepted theories. Adjustable parameters with density last.

\noindent \textbf{Functional relationships:}
\begin{align*}
% z &\neq f(\lambda),\\
	d_A &= D_H \ln(1 + z),\\ % from the properties of a tired-light theory
%	d_L &= \mbox{ not available},\\ % Ref: ?
	m - M &= 5\log_{10}[d_L/D_H] + C,\\ % Ref: Mannheim340.2006, arXiv:astro-ph/0505266v2 p. 45.
	\left\langle SB\right\rangle &\propto \left(d_A/d_L\right)^2,\\ % Ref: astro-ph/0603500
	F_\tau &= 1 + z. % Ref: ? Mechanism completely independent of light frequency
%	\eta_z &= \mbox{ not available},\\ % Ref: ?
%	\Delta_\theta &= \mbox{ not available}. % Ref: ?
\end{align*}
Functional relationships for $d_L, \eta_z$ and $\Delta_\theta$ are not available.

\noindent \textbf{Discussion and comments:}

  Zwicky's original calculation was incorrect.  The effect of dynamical friction on photons or other particles moving at relativistic speeds is too small by at least $80$ orders of magnitude, if not zero\footnote{See F. Zwicky, ``On the Possibilities of a Gravitational Drag of Light,'' Phys. Rev. 34, no. 12, pp. 1623–1624, Dec. 1929 for Zwicky's answer to Eddington}.
%
%                                                 ---------->>>>>>>>>>

%                                                 <<<<<<<<<<----------1953
\stepcounter{mechanism_number}
\subsection{\label{sec:mec.shelton}Shelton Head-on Collisions on Electrons}
  Shelton proposes that photons in free space act on electrons, and other fundamental particles, in much the same way as they act on electrons in the Compton effect where the kinetic energy supplied to the electrons results in a lengthening of the wavelength of the radiation\cite{Shelton1953.84}.

\textbf{Conditions, limits of applicability and restrictions:}

  A necessary assumption is that the action takes place only for head-on collisions so that it is not accompanied by any considerable change in the direction of the photons.
% Conditions, limits of applicability and restrictions: Solar limb red-shift? Quasars? Discordant red-shifts? Scale: intergalactic, interstellar, interplanetary red-shift? The assumptions required to make the mechanism work, adjustable parameters, and conflicts with currently accepted theories. Adjustable parameters with density last.

\noindent \textbf{Functional relationships:}
\begin{align*}
% z &\neq f(\lambda),\\
	d_A &= D_H \ln(1 + z),\\ % from the properties of a tired-light theory
%	d_L &= \mbox{ not available},\\ % Ref: ?
	m - M &= 5\log_{10}[d_L/D_H] + C,\\ % Ref: Mannheim340.2006, arXiv:astro-ph/0505266v2 p. 45.
	\left\langle SB\right\rangle &\propto \left(d_A/d_L\right)^2,\\ % Ref: astro-ph/0603500
	F_\tau &= 1 + z. % Ref: ? Mechanism completely independent of light frequency
%	\eta_z &= \mbox{ not available},\\ % Ref: ?
%	\Delta_\theta &= \mbox{ not available}. % Ref: ?
\end{align*}
Functional relationships for $d_L, \eta_z$ and $\Delta_\theta$ are not available.

\noindent \textbf{Discussion and comments:}

  There is no explanation about why the action would takes place only for head-on collisions.  It is known that when a collision between a photon and an elementary particle is such that the photon loses energy, conservation of energy and momentum requires the particle to move off very nearly at right angles.  The photon must therefore always be somewhat deflected if the particle is given any velocity at all.  If one puts in the quantities, one finds that it must in fact be so much deviated that all extragalactic nebulae would become too diffuse to see\cite{Atkinson1953.159}.
%
%                                                 ---------->>>>>>>>>>

%                                                 <<<<<<<<<<----------1962
\stepcounter{mechanism_number}
\subsection{\label{sec:mec.debroglie}de Broglie's Tired-Photon}
  According to de Broglie, a quantum particle, e.g. a photon, is a real entity composed of an extended, yet finite, wave plus a singularity.  de Broglie's assumes that the behaviour of the wave devoid of singularity and the singularity itself (representing the particle) is different when interacting with the matter or the surrounding subquantum medium.

%\textbf{Conditions, limits of applicability and restrictions:}

% Conditions, limits of applicability and restrictions: Solar limb red-shift? Quasars? Discordant red-shifts? Scale: intergalactic, interstellar, interplanetary red-shift? The assumptions required to make the mechanism work, adjustable parameters, and conflicts with currently accepted theories. Adjustable parameters with density last.

\noindent \textbf{Functional relationships:}
\begin{align*}
% z &\neq f(\lambda),\\
	d_A &= D_H \ln(1 + z),\\ % from the properties of a tired-light theory
%	d_L &= \mbox{ not available},\\ % Ref: ?
	m - M &= 5\log_{10}[d_L/D_H] + C,\\ % Ref: Mannheim340.2006, arXiv:astro-ph/0505266v2 p. 45.
	\left\langle SB\right\rangle &\propto \left(d_A/d_L\right)^2,\\ % Ref: astro-ph/0603500
	F_\tau &= 1 + z. % Ref: ? Mechanism completely independent of light frequency
%	\eta_z &= \mbox{ not available},\\ % Ref: ?
%	\Delta_\theta &= \mbox{ not available}. % Ref: ?
\end{align*}
Functional relationships for $d_L, \eta_z$ and $\Delta_\theta$ are not available.

\textbf{Discussion and comments:}

  De Broglie's pilot wave model assumes that the pilot wave regenerates itself when it has reached a very small amplitude.  This has not been observed experimentally.
%
%                                                 ---------->>>>>>>>>>

%                                                 <<<<<<<<<<----------1978
\stepcounter{mechanism_number}
\subsection{\label{sec:mec.brynjolfsson}Plasma Red-Shift}
  In Brynjolfsson's model, a collective interaction between many electrons in a plasma and a photon result in a Doppler-like energy loss red-shift of light and heating of the plasma.
 
  One factor $\sqrt{1 + z}$ is due to reduction in photon energy by plasma red-shift, while the other factor $1 + z$ is due to removal of photons through Compton scattering on the free electrons in the intergalactic plasma\cite{Brynjolfsson420.2004}.
%$d_C = D_H \ln(1 + z)$. % Ref: ?

\textbf{Conditions, limits of applicability and restrictions:}

	$a=2$ due to removal of photons through Compton scattering on the free electrons.

  The theory requires a hot, low density, plasma.  It explains the red-shift on the sun's limb.  For the intergalactic red-shift, a density of $\rho_0 = 205 \mbox{electrons}/\mbox{m}^3$ is determined from dispersion measurements.
% Conditions, limits of applicability and restrictions: Solar limb red-shift? Quasars? Discordant red-shifts? Scale: intergalactic, interstellar, interplanetary red-shift? The assumptions required to make the mechanism work, adjustable parameters, and conflicts with currently accepted theories. Adjustable parameters with density last.

\noindent \textbf{Functional relationships:}
\begin{align*}
% z &\neq f(\lambda),\\
	d_A &= D_H \ln(1 + z),\\ % Ref: ?
	d_L &= \sqrt{(1 + z)^{1+a}} d_A,\\ % Ref: ?
	m - M &= 2.5(1+a)\log_{10}[1 + z] + 5\log_{10}[\ln(1 + z)] + C,\\ % Ref: ?
	\left\langle SB\right\rangle &\propto (1 + z)^{-(1+a)},\\ % Ref: ?
	F_\tau &= 1,\\ % Ref: ?
	\eta_z &=^? 0,\\  % Ref: ?
	\Delta_\theta &=^? 0.  % Ref: ?
\end{align*}

\noindent \textbf{Discussion and comments:}

  Referring to the paper\cite{Brynjolfsson420.2004}, Eq.~(1) is given as the distribution of frequencies in a photon.  In fact, Eq. (1) describes the response of an atom to an electric field.

  Eq.~(6) in the same paper lists the four poles in an equation describing the energy loss.  Pole 'd' is Compton scattering, poles 'a' and 'b' are Raman scattering at the plasma frequency.  Pole 'c' is a pure imaginary number, that is, the energy loss at a null frequency (i.e. dc polarisation).  Why this is interpreted as a red-shift of the photons is not clear.  Every term in Eq. (6) corresponds to some exchange of energy: in case 'a', the plasma frequency is added (anti-Stokes Raman), in case 'b' the plasma frequency is subtracted (Stokes Raman), in case 'd' the frequency of the photon appears as the photon is scattered (Compton scattering).  For case 'c', there is only an imaginary number, the real part is zero.  The equation says that the plasma takes energy from the electromagnetic wave when it is dc polarized.  (Note that photons do not have a dc component.)  No quantitative value is given for the amount of red-shift.  The equation is interpreted as the energy of the photon which is reduced by some small amount but in fact, it means that few photons are simply absorbed.
%
%                                                 ---------->>>>>>>>>>

%                                                 <<<<<<<<<<----------1981
\stepcounter{mechanism_number}
\subsection{\label{sec:mec.marmetp}Atomic Secondary Emission}
  In P. Marmet's mechanism, a photon loses a small amount of energy each time it interacts with an atom\cite{MarmetP64.1981, MarmetP24.1988}.  The momentum of the photon polarizes the atom and causes it to emit a small amount of energy. (This was described in terms of a Bremstrahlung mechanism by the author.)  The incident photon is re-emitted with a small energy loss.  The coherence time of the radiation needs to be short for the effect to work.  The spectral width is described by the spectrum of a blackbody radiator at temperature $T$.

  Each interaction with an atom produces a red-shift $\delta z = 2.73\times 10^{-13} T_{10k}^2$, where $T_{10k}$ is the normalized temperature $T_{10k} = T / 10000\mbox{~K}$.

\textbf{Conditions, limits of applicability and restrictions:}

  Dependent on the polarizability of the atoms (or ions) in the intergalactic medium.  Not very sensitive on the specific atomic species.  Does not work at high density due to the increased collective mass of the medium.  Explains the red-shift on the sun's limb.  Frequency shift $\Delta\lambda/\lambda$ independent of $\lambda$ in the wavelength region where the index of refraction of hydrogen is constant.

  An atomic hydrogen density $\rho_0 = 1.25 \times 10^6\mbox{~m}^{-3} h_{100}/T_{10k}^2$ produces $d_C = D_H \ln(1 + z)$ which approximates $d_C = D_H z$ for small $z$.  In this model, the comoving and the angular distances are equal.
% Conditions, limits of applicability and restrictions: Solar limb red-shift? Quasars? Discordant red-shifts? Scale: intergalactic, interstellar, interplanetary red-shift? The assumptions required to make the mechanism work, adjustable parameters, and conflicts with currently accepted theories. Adjustable parameters with density last.

\noindent \textbf{Functional relationships:}
\begin{align*}
% z &\neq f(\lambda),\\
	d_A &= D_H \ln(1 + z),\\ % from the properties of a tired-light theory
	d_L &= \sqrt{(1 + z)} d_A,\\ % Ref: ?
	m - M &= 2.5\log_{10}[1 + z] + 5\log_{10}[\ln(1 + z)] + C,\\ % Ref: ?
	\left\langle SB\right\rangle &\propto (1 + z)^{-1},\\ % Ref: ?
	F_\tau &= 1,\\ % Ref: ?
	\eta_z &= 5.22\times 10^{-7} T_{10k} \sqrt{z},\\ % Ref: ?
	\Delta_\theta &= 5.22\times 10^{-7} T_{10k} \sqrt{z}. % Ref: ?
\end{align*}

\noindent \textbf{Discussion and comments:}

  This mechanism is very similar to ``Electronic Secondary Emission'' described in \S\ref{sec:mec.ashmore} but in atoms instead of electrons.  Because atoms have electronic resonances, this mechanism is dependent on the wavelength.  The required atomic density is much higher than the required density of electron in \S\ref{sec:mec.ashmore}.  This mechanism also predicts a small angular broadening.

  The difficulty with this mechanism is that the interaction in atoms is necessarily dependent on atomic resonances and would have a strong dependence on wavelength near these atomic resonances.  Also, this mechanism needs about $2.5\times 10^4 \mbox{atoms}/m^3$ while the estimated mass density in the universe is four orders of magnitude smaller than that value: the mechanism does not have a sufficiently large cross section.
%
%                                                 ---------->>>>>>>>>>

%                                                 <<<<<<<<<<----------1988
\stepcounter{mechanism_number}
\subsection{\label{sec:mec.kiereinj}Compton Effect Interpretation}
  In Kierein's model, the Compton effect interpretation of astrophysical red shifts provides the mechanism for intrinsic red shifts of quasi-stellar sources, the solar limb, and other objects.  The observed red shift is explained classically by considering the electrons to act as centers of Huygens' secondary wavelets that reconstruct the wave front.  The Compton effect does not depend on the electric charge of the electron, but rather is a consequence of conservation of momentum and energy.  Thus, the $E \times H$ vector of the photon need not be altered by the Compton process. It is the $E \times H$ vector of the wave front that is seen and that contains the information that defines the wave front's initial velocity vector\cite{Kierein1.1988}.
%Check Kierein http://www.angelfire.com/az/BIGBANGisWRONG/index.html

%\textbf{Conditions, limits of applicability and restrictions:}

% Conditions, limits of applicability and restrictions: Solar limb red-shift? Quasars? Discordant red-shifts? Scale: intergalactic, interstellar, interplanetary red-shift? The assumptions required to make the mechanism work, adjustable parameters, and conflicts with currently accepted theories. Adjustable parameters with density last.

\noindent \textbf{Functional relationships:}
\begin{align*}
% z &\neq f(\lambda),\\
	z &=^? d_A/(D_H - d_A/2),\\ % Ref: ?
%	d_L &= \mbox{ not available},\\ % Ref: ?
	m - M &= 5\log_{10}[d_L/D_H] + C,\\ % Ref: Mannheim340.2006, arXiv:astro-ph/0505266v2 p. 45.
	\left\langle SB\right\rangle &\propto \left(d_A/d_L\right)^2,\\ % Ref: astro-ph/0603500
	F_\tau &= 1 + z. % Ref: ? Mechanism completely independent of light frequency
%	\eta_z &= \mbox{ not available},\\ % Ref: ?
%	\Delta_\theta &= \mbox{ not available}. % Ref: ?
\end{align*}
Functional relationships for $d_L, \eta_z$ and $\Delta_\theta$ are not available.

\textbf{Discussion and comments:}

  This mechanism is very similar to ``Forward Scattering by Relativistic Electrons'' described in \S\ref{sec:mec.vaughan}.

  However, for photons in the visible the Compton scattering cross section is $\sigma \approx 6.65\times 10^{-29}\mbox{m}^2$, much too small by a factor one million for any significant red-shift produced by the low density of electrons in intergalactic space. 
%
%                                                 ---------->>>>>>>>>>

%                                                 <<<<<<<<<<----------1992
\stepcounter{mechanism_number}
\subsection{\label{sec:mec.corriveau}Interaction with the Intergalactic Gas}
  The mechanism, presented by Corriveau, assumes an infinite universe in space and time filled throughout with a diffuse gas (molecular hydrogen), the intergalactic gas.  In such universe the red-shift of spectral lines from distant galaxies results from the interaction of light with the intergalactic gas.  Photons lose energy with distance in the gas.  Olbers' paradox is solved by the red-shift of stellar radiation.  Finally, the 3 K background emission is believed to be the thermal emission of the intergalactic gas\cite{Corriveau39.1992}.

%\textbf{Conditions, limits of applicability and restrictions:}

% Conditions, limits of applicability and restrictions: Solar limb red-shift? Quasars? Discordant red-shifts? Scale: intergalactic, interstellar, interplanetary red-shift? The assumptions required to make the mechanism work, adjustable parameters, and conflicts with currently accepted theories. Adjustable parameters with density last.

\noindent \textbf{Functional relationships:}
\begin{align*}
% z &\neq f(\lambda),\\
	d_A &= D_H \ln(1 + z),\\ % from the properties of a tired-light theory
%	d_L &= \mbox{ not available},\\ % Ref: ?
	m - M &= 5\log_{10}[d_L/D_H] + C,\\ % Ref: Mannheim340.2006, arXiv:astro-ph/0505266v2 p. 45.
	\left\langle SB\right\rangle &\propto \left(d_A/d_L\right)^2,\\ % Ref: astro-ph/0603500
	F_\tau &=^? 1. % Ref: ?
%	\eta_z &= \mbox{ not available},\\ % Ref: ?
%	\Delta_\theta &= \mbox{ not available}. % Ref: ?
\end{align*}
Functional relationships for $d_L, \eta_z$ and $\Delta_\theta$ are not available.

%\textbf{Discussion and comments:}

%
%                                                 ---------->>>>>>>>>>

%                                                 <<<<<<<<<<----------1994
\stepcounter{mechanism_number}
\subsection{\label{sec:mec.ashmore}Electronic Secondary Emission}
  In Ashmore's model, photons from distant galaxies are absorbed and reemitted by electrons in the intergalactic (IG) space.  On absorption and reemission, the electron recoils and the photon loses energy.  The emitted photon is therefore redshifted from it's initial wavelength.  The Hubble relation arises from the proportionality of the number of interactions to the distance traveled.  The recoiling electron emits two low-energy photons each time an interaction occurs, producing radiation contributing to the Cosmological Microwave Background (CMB).

  The collision cross section for an electron interacting with a photon is $\sigma = 2r_e\lambda$, where $r_e$ is the classical electron radius.  Using $\rho_0$ as the electron density in IG space, an expression for the number of collisions experienced by the photon in traveling a distance $d$ is derived, and hence the red-shift $z$.  Comparing this with $z = H_0 d/c$ gives an expression for $H_0$ by this theory of $H_0=2\rho_0 hr_e/m$.
  
  Each scattering event on an electron produces a red-shift $\delta z = 2.43\times 10^{-12}/\lambda$.  Because the cross section increases with $\lambda$, the total red-shift resulting from this mechanism is independent of $\lambda$~\cite{Ashmore3.2006, Ashmore456.2015}.

\textbf{Conditions, limits of applicability and restrictions:}

  The relevant column density is calculated from the density of electrons.

  An electron density $\rho_0 = 0.80h_{100} \mbox{m}^{-3}$ produces the observed $d = D_H z$ for small $z$.  No angular spread of the light occurs as this is similar to transmission in a transparent medium.
% Conditions, limits of applicability and restrictions: Solar limb red-shift? Quasars? Discordant red-shifts? Scale: intergalactic, interstellar, interplanetary red-shift? The assumptions required to make the mechanism work, adjustable parameters, and conflicts with currently accepted theories. Adjustable parameters with density last.

\noindent \textbf{Functional relationships:}
\begin{align*}
% z &\neq f(\lambda),\\
	d_A &= D_H \ln(1 + z),\\ % Ref: ?
	d_L &=^? \sqrt{1 + z} d_A,\\ % Ref: ?
	m - M &= 5\log_{10}[d_L/D_H] + C,\\ % Ref: Mannheim340.2006, arXiv:astro-ph/0505266v2 p. 45.
	\left\langle SB\right\rangle &\propto \left(d_A/d_L\right)^2,\\ % Ref: ?
	F_\tau &= 1,\\ % Ref: ?
	\eta_z &= \sqrt{2.43\times 10^{-12}/\lambda} \ \frac{\sqrt{z}} {1 + z},\\ % Ref: ?
	\Delta_\theta &\approx 0. % Ref: ?
\end{align*}

\noindent \textbf{Discussion and comments:}

  A test of this mechanism is to evaluate Hubble's constant from measurements of the density of electrons.  For $\rho_0$, a value is taken from: ``Outside the galaxy, in intergalactic space, the number density of particles is thought to fall off to about one atom or molecule per cubic meter ($10^{-6}/\mbox{cm}^3$)''~\cite{JPL1.2011}.  Since the hydrogen is fully ionized the same density is used for $\rho_0$.  This gives $H_0 = 4.1\times 10^{-18}/\mbox{s}$ or $h_{100} = 1.24$.

  In the same way one can use this theory to predict the CMB by using conservation of momentum to calculate the energy transferred to the electron on each recoil and the energy and wavelength of the CMB photons emitted.  It is found that light/UV photons produce CMB photons in the microwave.

  This mechanism is very similar to ``Atomic Secondary Emission'' described in \S\ref{sec:mec.marmetp} but in electrons instead of atoms.  Because electrons do not have any internal degrees of freedom, this mechanism is truly independent of the wavelength.  The required electron density is much lower than the required density of hydrogen in \S\ref{sec:mec.marmetp}.  The mechanism produces functional relationships that are almost equal to the relationships for the ``Spectral Transfer Red-Shift'' described in \S\ref{sec:mec.marmetl}.  Accurate measurements of the line broadening are necessary to distinguish between the two mechanisms.

  The statement ``The collision cross section for an electron interacting with a photon is $\sigma = 2r_e\lambda$'' is not justified and largely overestimates the cross section for such an interaction.
%
%                                                 ---------->>>>>>>>>>

%                                                 <<<<<<<<<<----------1995
\stepcounter{mechanism_number}
\subsection{\label{sec:mec.standardqed}Red-Shift Theorem}
Accardi, Laio, Lu and Rizzi use quantum field theory to derive a general red-shift theorem.  They model the interstellar medium by a low density Fermi gas and its interaction with light by standard QED.  Applying the techniques of the stochastic limit of quantum field theory, they prove a general red-shift theorem.  This result fits satisfactorily both Pioneer 6 and solar limb red-shift data.

  The relevant column density is the path integral of the product of the plasma density %check: ions or/and electrons
and the temperature
$$N_T(d) = B_\omega \int_0^d {\rho(x) T(x) \mbox{d}x},$$
with $B_\omega=2 \times 10^{-7}$~\cite{Accardi1995.Rizzi278}.

\textbf{Conditions, limits of applicability and restrictions:}

  The relation is valid for electromagnetic radiation at the frequency $10^9\mbox{~Hz} < \omega < 10^{16}\mbox{~Hz}$.  The value of the constant $B_\omega$ depends on a divergent integral; the value given above produces the observed $d = D_H z$ for small $z$.
%  A density $\rho_0 = ?.?? \times 10^?\mbox{m}^{-3}$ of plasma produces the observed $d = D_H z$ for small $z$.
% Conditions, limits of applicability and restrictions: Solar limb red-shift? Quasars? Discordant red-shifts? Scale: intergalactic, interstellar, interplanetary red-shift? The assumptions required to make the mechanism work, adjustable parameters, and conflicts with currently accepted theories. Adjustable parameters with density last.

\noindent \textbf{Functional relationships:}
\begin{align*}
% z &\neq f(\lambda),\\
	d_A &= D_H \ln(1 + z),\\ % Ref: ?
	d_L &=^? \sqrt{1 + z} d_A,\\ % Ref: ?
	m - M &= 5\log_{10}[d_L/D_H] + C,\\ % Ref: Mannheim340.2006, arXiv:astro-ph/0505266v2 p. 45.
	\left\langle SB\right\rangle &\propto \left(d_A/d_L\right)^2,\\ % Ref: astro-ph/0603500
	F_\tau &=^? 1. % Ref: ?
%	\eta_z &= \mbox{ not available},\\ % Ref: ?
%	\Delta_\theta &= \mbox{ not available}. % Ref: ?
\end{align*}
Functional relationships for $\eta_z$ and $\Delta_\theta$ are not available.

%\textbf{Discussion and comments:}

%
%                                                 ---------->>>>>>>>>>

%                                                 <<<<<<<<<<----------2000
\stepcounter{mechanism_number}
\subsection{\label{sec:mec.smid}Thomas Smid's plasma red-shift theory}
Smid's previous research has already revealed several hitherto unknown effects in the area of ionospheric physics which clearly show the importance of temporal and spatial characteristics of the random plasma fluctuation field for the emission and propagation of electromagnetic waves, and he is suggesting that it is the plasma 'micro'-field which is also responsible for the red-shift of galaxies\cite{Smid1.2000}.

Due to the low plasma density in intergalactic space, the associated electric field can be considered to be quasi-static and quasi-homogeneous for most electromagnetic waves, but these conditions are not valid any more for sufficiently low frequencies/large wavelengths.  For a plasma density $n_p = 10^{-6} cm^{-3}$ for instance, quasi- homogeneity breaks down for wavelengths of the order of $1m$ and more. Even observations in the cm-region could already reveal deviations from the known values which on the other hand should obviously hold throughout the spectrum if the observed red-shift is velocity related.

%\textbf{Conditions, limits of applicability and restrictions:}

% Conditions, limits of applicability and restrictions: Solar limb red-shift? Quasars? Discordant red-shifts? Scale: intergalactic, interstellar, interplanetary red-shift? The assumptions required to make the mechanism work, adjustable parameters, and conflicts with currently accepted theories. Adjustable parameters with density last.

\noindent \textbf{Functional relationships:}
\begin{align*}
% z &\neq f(\lambda),\\
	d_A &= D_H \ln(1 + z),\\ % from the properties of a tired-light theory
%	d_L &= \mbox{ not available},\\ % Ref: ?
	m - M &= 5\log_{10}[d_L/D_H] + C,\\ % Ref: Mannheim340.2006, arXiv:astro-ph/0505266v2 p. 45.
	\left\langle SB\right\rangle &\propto \left(d_A/d_L\right)^2,\\ % Ref: astro-ph/0603500
	F_\tau &=^? 1. % Ref: ?
%	\eta_z &= \mbox{ not available},\\ % Ref: ?
%	\Delta_\theta &= \mbox{ not available}. % Ref: ?
\end{align*}
Functional relationships for $d_L, \eta_z$ and $\Delta_\theta$ are not available.

\textbf{Discussion and comments:}

  If spectral features of galaxies in the radio region are too weak to be measured, the effect could possibly still be demonstrated in the laboratory, by examining the propagation of light through a (sufficiently strong) static electric field.
%
%                                                 ---------->>>>>>>>>>

%                                                 <<<<<<<<<<----------2003
\stepcounter{mechanism_number}
\subsection{\label{sec:mec.gallo}Thermalization}
  In Gallo's model, ``hot'' radiation thermalizes with the ``cold'' intergalactic medium.  The effect would depend on the density of the intergalactic medium and its temperature\cite{Gallo7.2006}.

%\textbf{Conditions, limits of applicability and restrictions:}
% Conditions, limits of applicability and restrictions: Solar limb red-shift? Quasars? Discordant red-shifts? Scale: intergalactic, interstellar, interplanetary red-shift? The assumptions required to make the mechanism work, adjustable parameters, and conflicts with currently accepted theories. Adjustable parameters with density last.

\textbf{Functional relationships:}
\begin{align*}
% z &\neq f(\lambda),\\
	d_A &= D_H \ln(1 + z),\\   % from the properties of a tired-light theory
%	d_L &= \mbox{ not available},\\ % Ref: ?
	m - M &= 5\log_{10}[d_L/D_H] + C,\\ % Ref: Mannheim340.2006, arXiv:astro-ph/0505266v2 p. 45.
	\left\langle SB\right\rangle &\propto \left(d_A/d_L\right)^2,\\ % Ref: astro-ph/0603500
	F_\tau &= 1. % Ref: ?
%	\eta_z &= \mbox{ not available},\\ % Ref: ?
%	\Delta_\theta &= \mbox{ not available}. % Ref: ?
\end{align*}
Functional relationships for $d_L, \eta_z$ and $\Delta_\theta$ are not available.

\textbf{Discussion and comments:}

  No precise model is given for the interaction.  Measurements of neutrino red-shift are proposed\cite{Gallo1230.2003} as an experimental test of Doppler versus non-Doppler red-shifts.
%
%                                                 ---------->>>>>>>>>>

%                                                 <<<<<<<<<<----------2006
\stepcounter{mechanism_number}
\subsection{\label{sec:mec.marmetl}Spectral Transfer Red-Shift}
  In L. Marmet's model, optical forces transfer energy to the electrons and atoms in the intergalactic medium while removing energy from one of the two photons interacting with the medium.  Electrons cause a red-shift through the ponderomotive force with $z$ independent of the wavelength.  Atoms produce a larger red-shift near resonances via the dipole force.  Stimulated emission ensures that all interacting photon-pairs are emitted in the same direction of an existing light ray - thus this mechanism does not blur images.  Spectral transfer occurs directly if the source is extended.  At cosmological distances, the red-shift occurs because light from the observed object interacts with the background light intensity $I_{bkg}$ emitted by all other galaxies.

  Each interaction of a photon with a second photon, incident at a relative angle $\theta$ on an electron, produces a red-shift of the photon which is then emitted in the direction of the second photon.  Conservation of momentum and energy implies that the red-shift produced by each interaction decreases with increasing $\lambda$, but the cross section $\sqrt{I_{bkg}/I}~[(\alpha_F~h)/(2\pi m_e\nu)]~sin(\theta/2)$ increases with $\lambda$.  Hence the total red-shift resulting from this mechanism is independent of $\lambda$~\cite{MarmetL268.2009}.  This mechanism is not to be confused with Stimulated-Compton-Scattering(SCS) which depends on the square of the intensity of the light.  The ponderomotive force is linear in total intensity which gives it a large interaction cross section even at the low intensities encountered in intergalactic space.  Of importance is the dependence on $sin(\theta/2)$, the angle between the observed light and the light from other sources.  The presence of bright sources at an angle with respect to the observed light direction will produce a large red-shift, as is observed in the case of discordant red-shifts of one galaxy in a compact group of galaxies.
  
\textbf{Conditions, limits of applicability and restrictions:}

  $a\approx 2$ is a constant to be determined theoretically.  The constant $a$ appears because Thomson scattering decreases the intensity of the light.

  An electron density $\rho_0 = 3.7 h_{100} /\mbox{m}^3$ produces the observed $d = D_H z$ for small $z$.
% Conditions, limits of applicability and restrictions: Solar limb red-shift? Quasars? Discordant red-shifts? Scale: intergalactic, interstellar, interplanetary red-shift? The assumptions required to make the mechanism work, adjustable parameters, and conflicts with currently accepted theories. Adjustable parameters with density last.

\noindent \textbf{Functional relationships:}
\begin{align*}
% z &\neq f(\lambda),\\
	d_A &= D_H \ln(1 + z),\\ % Ref: derived from fundamental relations.  This is correct.
	d_L &= \sqrt{(1 + z)^{1+a}} d_A,\\ % Ref: from fundamental relations, and Thomson scattering included
	m - M &= 2.5(1+a)\log_{10}[1 + z] + 5\log_{10}[\ln(1 + z)] + C,\\
                                               % Ref: from above and m - M &= 5\log_{10}[d_L/D_H] + C, Ref: Mannheim340.2006, arXiv:astro-ph/0505266v2 p. 45.
	\left\langle SB\right\rangle &\propto (1 + z)^{-(1+a)},\\ % Ref: astro-ph/0603500
	F_\tau &= 1,\\ % Ref: ?
	\eta_z &= \sqrt{1.54\times 10^{-12}/\lambda} \ \frac{\sqrt{z}} {1 + z},\\ % Ref: ?
	\Delta_\theta &= 0. % Ref: ?.
\end{align*}

\noindent \textbf{Discussion and comments:}

  The mechanism produces functional relationships that are almost equal to the relationships for the ``Electronic Secondary Emission'' mechanism described in \S\ref{sec:mec.ashmore}.  Accurate measurements of the line broadening are necessary to distinguish between the two mechanisms.

  This mechanism is based on the dipole force and the ponderomotive force, both demonstrated in laboratory experiments with atomic trapping.  Objections against tired light theories\cite{Wright1996.2} do not hold against this mechanism because of the unique two-photon nature of the mechanism.
%
%                                                 ---------->>>>>>>>>>

%                                                 <<<<<<<<<<----------2007
\stepcounter{mechanism_number}
\subsection{\label{sec:mec.santilli}Santilli Isoredshift}
%From Santerelli:Institute for Basic Research preprint IBR-Ex-09-35, June 27, 2009
%Revised July 7, 2-009, July 12, 2009 and August 19, 2009
%'Santilli interpretation of the cosmological red-shift is that of an isoredshift caused by the propagation of light through light.'
  Santilli's red-shift mechanism is explained by an interaction between light and the energy density of the medium through which it propagates.  The energy density of the medium is determined by the density of cosmic rays, hydrogen and other matter as well as the density of radiation.  Thus, this model involves an interaction of light with matter but also with light itself\cite{Santilli203.2007}.
%Institute for Basic Research
%P. O. Box 1577, Palm Harbor, FL 34682, U.S.A.
%ibr@ibr.net, http://www.i-b-r.org, www.santilli-foundation.org

%\textbf{Conditions, limits of applicability and restrictions:}

% Conditions, limits of applicability and restrictions: Solar limb red-shift? Quasars? Discordant red-shifts? Scale: intergalactic, interstellar, interplanetary red-shift? The assumptions required to make the mechanism work, adjustable parameters, and conflicts with currently accepted theories. Adjustable parameters with density last.

\noindent \textbf{Functional relationships:}
\begin{align*}
% z &\neq f(\lambda),\\
	d_A &= D_H \ln(1 + z),\\   % from the properties of a tired-light theory
%	d_L &= \mbox{ not available},\\ % Ref: ?
	m - M &= 5\log_{10}[d_L/D_H] + C,\\ % Ref: Mannheim340.2006, arXiv:astro-ph/0505266v2 p. 45.
	\left\langle SB\right\rangle &\propto \left(d_A/d_L\right)^2,\\ % Ref: astro-ph/0603500
	F_\tau &=^? 1. % Ref: ?
%	\eta_z &= \mbox{ not available},\\ % Ref: ?
%	\Delta_\theta &= \mbox{ not available}. % Ref: ?
\end{align*}
Functional relationships for $d_L, \eta_z$ and $\Delta_\theta$ are not available.

\textbf{Discussion and comments:}

  Although Santilli claims that his mechanism has been demonstrated experimentally, the results clearly show a strong attenuation of the light even for a small red-shift.  Such an attenuation would preclude any observation of galaxies at any large value of z.

  It is also claimed that the redness of ``direct'' sunlight at sunset is due to the IsoRedShift based on the loss of energy by light to air at low temperature without any relative motion between the source, the air and the observer\cite{Santilli.Amato141.2012}.  This ignores the difference between ``red-shift'' (wavelength shift) and ``reddening'' (absorption of the red part of the spectrum), raising doubts on the authors' understanding of the physics involved in the cosmological red-shift.
%
%                                                 ---------->>>>>>>>>>

%                                                 <<<<<<<<<<----------2008
\stepcounter{mechanism_number}
\subsection{\label{sec:mec.vaughan}Forward Scattering by Relativistic Electrons}
  Vaughan's mechanism describes light propagation through a high temperature plasma.  The red-shift results from the  relativistic transverse Doppler effect which reduces the frequency of light on forward scattering off high speed plasma electrons.  A larger electron temperature produces a larger transverse Doppler effect.  The energy and momentum is transferred from the radiation to the medium itself.  The strength of the mechanism is related to the `extinction distance', which is wavelength dependent.  When all these factors are taken into account, a red-shift is obtained which is independent of the wavelength.  Huygen's principle is invoked to keep the light from dispersing.

  The relevant column density is given by the dynamic pressure of the plasma
$$N_T(d) = \int_0^d {\rho(x) T(x) \mbox{d}x},$$
where $T(x)$ is the electron temperature\cite{Bonn365.2008}.

\textbf{Conditions, limits of applicability and restrictions:}

  $a=1$.

  A density $\rho_0 = 5\times 10^9~\mbox{Km}^{-3}/T_0$ of free electrons produces the observed $d = D_H z$ for small $z$.
% Conditions, limits of applicability and restrictions: Solar limb red-shift? Quasars? Discordant red-shifts? Scale: intergalactic, interstellar, interplanetary red-shift? The assumptions required to make the mechanism work, adjustable parameters, and conflicts with currently accepted theories. Adjustable parameters with density last.

\noindent \textbf{Functional relationships:}
\begin{align*}
% z &\neq f(\lambda),\\
	d_A &= D_H \ln(1 + z),\\ % from book
	d_L &= \sqrt{(1 + z)^{1+a}} d_A,\\ % from book
	m - M &= 2.5(1+a)\log_{10}[1 + z] + 5\log_{10}[\ln(1 + z)] + C,\\ % Ref: ?
	\left\langle SB\right\rangle &\propto (1 + z)^{-(1+a)},\\ % Ref: astro-ph/0603500
	F_\tau &= 1,\\ % Ref: ?
%	\eta_z &= \mbox{ not available},\\ % Ref: ?
	\Delta_\theta &= 0. % Ref: ?
\end{align*}
A functional relationship for $\eta_z$ is not available.

\noindent \textbf{Discussion and comments:}

  This mechanism is very similar to ``Compton Effect Interpretation'' described in \S\ref{sec:mec.kiereinj} but includes an additional transverse relativistic Doppler effect.

  For photons in the visible the Compton scattering cross section is $6.65\times 10^{-29} m^2$, much too small by a factor one million for any significant red-shift produced by the density of electrons in intergalactic space.  The physics of the emission of redshifted radiation is incorrect: a correction (transverse Doppler) from the frame of the electron to the frame of the observer is considered, but mistakenly the correction (aberration) from the frame of the emitter to the frame of the electron is ignored.  Electrons re-emit light at the same frequency as they are receiving it.  If they have a high velocity, then in their reference frame, the radiation is shifted to the higher frequencies due to the aberration angle.  This shift compensates exactly the red-shift due to the transverse Doppler effect.
%
%                                                 ---------->>>>>>>>>>

%                                                 <<<<<<<<<<----------2011
\refstepcounter{mechanism_number} % for the last mechanism, \ref is defined as mechanism_number and used by \label below
\label{count:mechanism_number}
\subsection{\label{sec:mec.mamas}Deep Space Electrons}
  In Mamas' model the photon is viewed as an electromagnetic wave whose electric field component causes oscillations in deep space free electrons which then re-radiate energy from the photon, causing a red-shift. The predicted red-shift coincides with the data of the Hubble diagram\cite{Mamas326.2010}.
  % Copied from abstract

%\textbf{Conditions, limits of applicability and restrictions:}

% Conditions, limits of applicability and restrictions: Solar limb red-shift? Quasars? Discordant red-shifts? Scale: intergalactic, interstellar, interplanetary red-shift? The assumptions required to make the mechanism work, adjustable parameters, and conflicts with currently accepted theories. Adjustable parameters with density last.

\noindent \textbf{Functional relationships:}
\begin{align*}
% z &\neq f(\lambda),\\
	d_A &= D_H \ln(1 + z),\\   % from the properties of a tired-light theory
%	d_L &= \mbox{ not available},\\ % Ref: ?
	m - M &= 5\log_{10}[d_L/D_H] + C,\\ % Ref: Mannheim340.2006, arXiv:astro-ph/0505266v2 p. 45.
	\left\langle SB\right\rangle &\propto \left(d_A/d_L\right)^2,\\ % Ref: astro-ph/0603500
%	F_\tau &= \mbox{ not available},\\ % Ref: ?
%	\eta_z &= \mbox{ not available},\\ % Ref: ?
	\Delta_\theta &=^? 0. % Ref: ?
\end{align*}
Functional relationships for $d_L, F_\tau$ and $\eta_z$ are not available.

\textbf{Discussion and comments:}

  The predicted red-shift expression allows for the first time distance measurements to the furthest observable objects, without having to rely on their apparent magnitudes which may be subject to cosmic dust.  This new theoretical model is not the same as, and is fundamentally different from, Compton scattering, and therefore avoids any problems associated with Compton scattering such as the blurring of images.
%
%                                                 ---------->>>>>>>>>>

% Open Office Calc: graph 33cm x 16cm
% PSP: resize 6.9in 300pixel/in, save as .eps colour, preview
\begin{figure}
\centerline{\includegraphics[width=6.9in]{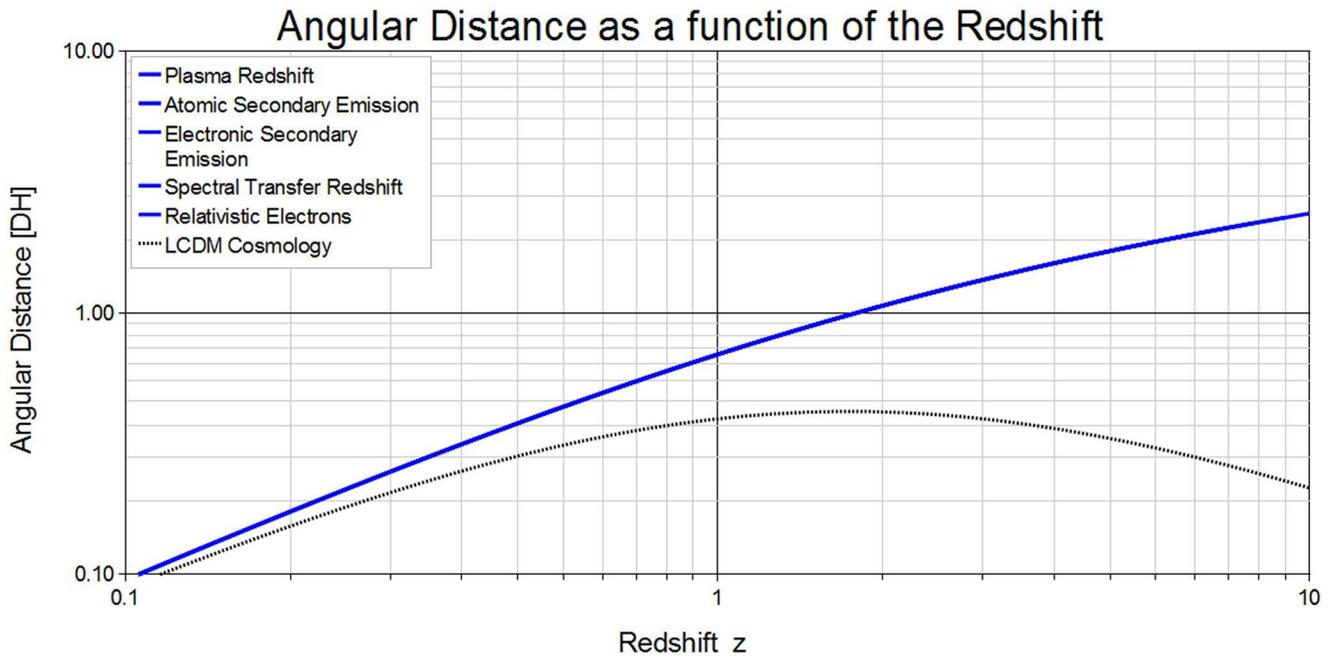}}
\caption{Angular distance in units of $D_H$ for a red-shift produced by an interaction between light and matter.  ``LCDM Cosmology'' is included for comparison.}
\label{fig:dAlhtmat}
\end{figure}

\begin{figure}
\centerline{\includegraphics[width=6.9in]{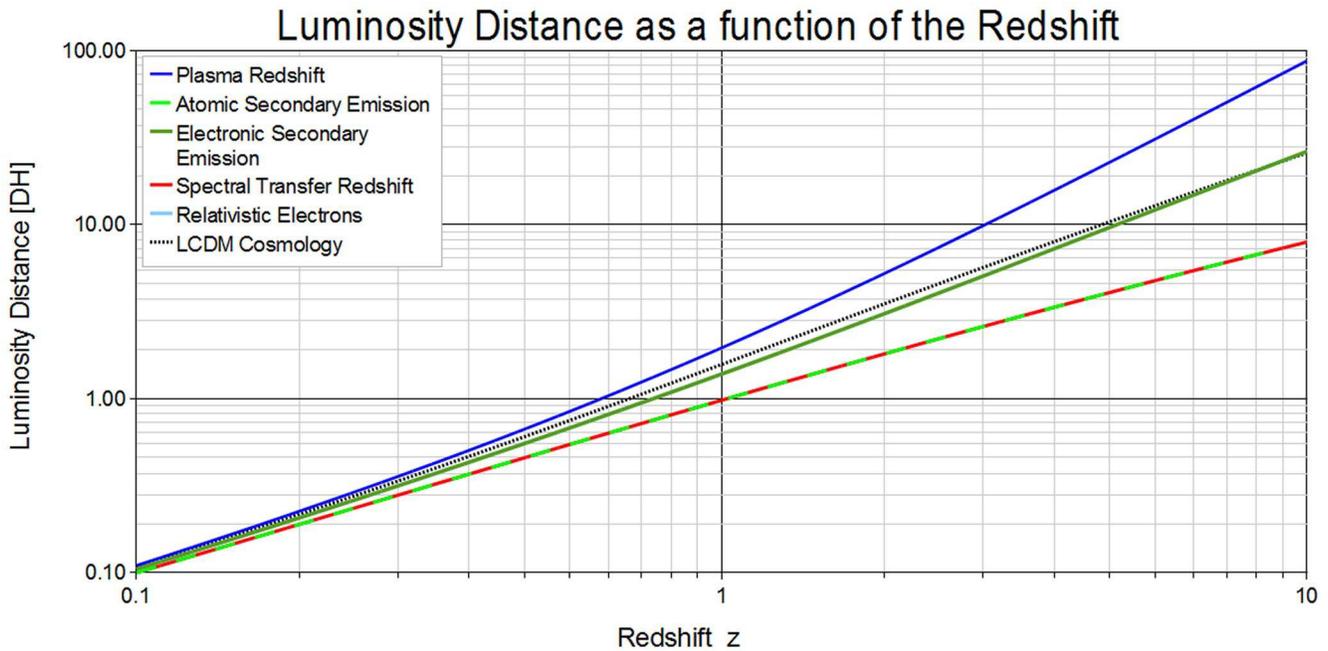}}
\caption{Luminosity distance in units of $D_H$ for a red-shift produced by an interaction between light and matter.  ``LCDM Cosmology'' is included for comparison.}
\label{fig:dLlhtmat}
\end{figure}

\begin{figure}
\centerline{\includegraphics[width=6.9in]{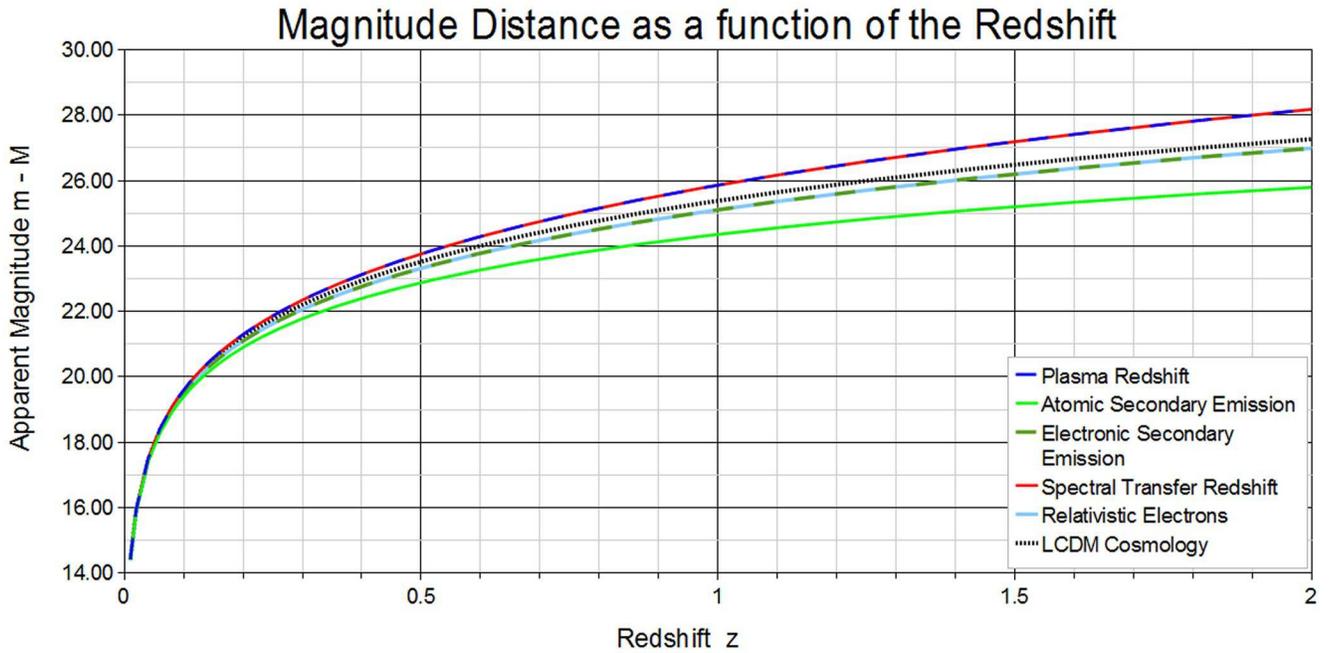}}
\caption{Magnitude distance for a red-shift produced by an interaction between light and matter.  ``LCDM Cosmology'' is included for comparison.}
\label{fig:dMlhtmat}
\end{figure}

\newpage

% ####################################################################################### %

\section{Conclusions}
This paper was written to collect as many red-shift mechanisms as possible in a single, coherent presentation.  Many questions arise: ``Which one of these best describes the observations?'' ``Which one, if any, is right?''  There are so many proposed mechanisms with even more adjustable parameters, it is possible that a few might fit experimental results within measurement errors.  However, this doesn't mean that the model is right.  Another method might be required in the future to decide which, if any, of these provides a good explanation for the red-shift.

During the great depression of the 1930's, it was observed that men wearing gold watches were in better health than other people.  The correlation between gold and the weight of these men was explained with some assumed property of gold.  Although the theory explained the observations very well, it was certainly incorrect - richer men who could afford to buy gold had enough money to feed themselves.  This anecdote is to be remembered when considering the above models and how well they fit experimental data.

\section{Acknowledgements}
The author is grateful to Paul~Marmet, Halton~Arp, Chuck~Gallo, Philip~Mannheim, Lyndon~Ashmore, Ari~Brynjolfsson and William~Lama for working outside the boundaries of accepted ideas.  Their ideas and many discussions improved my understanding of cosmology and helped me discover and develop the ``Spectral Transfer Red-Shift''.  Many thanks also to the authors of the theories presented in this paper, J.~Georg von~Brzeski, R.~Caswell, J.~Masreliez,  A.~Mayer, J.~Moret-Bailly, E.~Poliakov, F.~Potter, Y.-H. Sanejouand, H.A.~Schmitz, A.~Singh, R.F.~Vaughan, and R.G.~Vishwakarma for their patience in answering my questions, as well as D.~Shaw for his classical approach to the subject which resulted in fruitful discussions.  David Dilworth convinced me to add my own critique for each mechanism instead of giving an otherwise neutral observation.  After all, if my colleagues can't stand the heat of critique, particularly from a sympathetic ear, how can their idea ever stand the test of open review?  Many of these mechanisms were presented at the \emph{$2^{nd}$ Crisis in Cosmology Conference} in September 2008\cite{CCC2.2009}.  Many thanks to H.~Workman who provided audio recordings of the talks at the CCC2.  The recordings helped my memory for many enjoyable hours and helped me gain a better understanding of some of the mechanisms.

\vspace{0.75in}

\begin{center}
  \textbf{APPENDICES}
\end{center}

\appendix

\section{Detection of particles}

While light is used in most observations, particles such as protons, alpha particles, electrons, etc. are also detected.

\begin{itemize}
	\item \textbf{The frequency}:  In the case of particles, the detector measures the energy of the particles $E_0$ which is often proportional to the size of the output pulses.
	\item \textbf{The spectral irradiance}:  When particles are detected, the spectral irradiance is defined in this paper as the energy received by the detector per unit area in the energy range $E_0$ to $E_0+\mbox{d}E_0$ as $I_0(E_0)$~[W~m$^{-2}$~J$^{-1}$].  The total irradiance is the spectral irradiance integrated over all energies $I_0 = \int_0^\infty{I_0(E_0) \mbox{d}E_0}$.
\end{itemize}

The red-shift can also be defined from the energy of a particle.  However, the interpretation requires that the object at a cosmological distance has the same properties as those of a known object:
\begin{itemize}
	\item \textbf{The red-shift} $z$:  It is difficult to determine the red-shift from the energy of particles because several uncontrolled interactions can modify their energy.  Expressed as a function of the energy, the red-shift is $z = (E - E_0)/E_0$, with $E$ the energy of the emitted particle and $E_0$ the observed energy of the detected particle.
\end{itemize}

\section{Red-Shifts that are not necessarily an indication of cosmological distance}
Other mechanisms have been proposed to explain the red-shift but they do not predict a relationship between distance and red-shift, or they remain so far without a precise physical model.  Therefore they do not offer any quantitative predictions for the dependence of the red-shift as a function of distance.  However, these mechanisms might affect the light we are receiving or explain intrinsic red-shifts of quasars.  These are listed here for completeness and comparisons.

%                                                 <<<<<<<<<<----------1871
\subsection{\label{sec:mec.rayleigh}Rayleigh Scattering}
  Light scattering on bound electrons loses energy.
% discuss density required for red-shift, and absorption

\noindent \textbf{Conditions, limits of applicability and restrictions:}

Scattering process changes the direction of the light.
% Conditions, limits of applicability and restrictions: Solar limb red-shift? Quasars? Discordant red-shifts? Scale: intergalactic, interstellar, interplanetary red-shift? The assumptions required to make the mechanism work, adjustable parameters, and conflicts with currently accepted theories. Adjustable parameters with density last.

\noindent \textbf{Discussion and comments:}

Rayleigh scattering would diffuse the light from remote galaxies, thus preventing any possibility of imaging.  Not applicable to cosmological red-shift. %  CMB (?)
%
%                                                 ---------->>>>>>>>>>

%                                                 <<<<<<<<<<----------
\subsection{\label{sec:mec.thomson}Thomson/Compton Scattering}
Light scattered on free electrons loses energy.  The Thomson scattering cross section is evaluated for comparison with the mechanisms described above.  The frequency shift caused by Thomson scattering off an electron is 
$$\delta z = \frac{h \nu} {m c^2} \sin{(\theta /2)}.$$
For light in the visible, this is $\delta z \approx 4.4\times 10^{-6}$.  At high energies, the cross section is given by the Klein-Nischina formula (Compton scattering).  For photons in the visible the effect reduces to the classical Thomson scattering with a cross section $\sigma \approx 6.65\times 10^{-29}\mbox{m}^2$.
% discuss density required for red-shift...
% discuss density required for absorption

\noindent \textbf{Conditions, limits of applicability and restrictions:}

Scattering process changes the direction of the light, therefore remote galaxies could not be imaged.  Not applicable to cosmological red-shift.
% Conditions, limits of applicability and restrictions: Solar limb red-shift? Quasars? Discordant red-shifts? Scale: intergalactic, interstellar, interplanetary red-shift? The assumptions required to make the mechanism work, adjustable parameters, and conflicts with currently accepted theories. Adjustable parameters with density last.

\noindent \textbf{Discussion and comments:}

This process cannot contribute to the cosmological microwave background: the cross section is too small.
%
%                                                 ---------->>>>>>>>>>

%                                                 <<<<<<<<<<----------
\subsection{\label{sec:mec.raman}Raman effect}
Light interacting with atoms lose energy by a coherent process which transfers it to the atom's higher energy electronic states.  Molecules can also gain some rotational or vibrational energy.

%\textbf{Conditions, limits of applicability and restrictions:}

% Cosmological microwave background?  No cross section is too small
% Conditions, limits of applicability and restrictions: Solar limb red-shift? Quasars? Discordant red-shifts? Scale: intergalactic, interstellar, interplanetary red-shift? The assumptions required to make the mechanism work, adjustable parameters, and conflicts with currently accepted theories. Adjustable parameters with density last.

\noindent \textbf{Discussion and comments:}

This process cannot contribute to the cosmological microwave background: the cross section is too small for the weak intensities of light in intergalactic space.
%
%                                                 ---------->>>>>>>>>>

%                                                 <<<<<<<<<<----------
\subsection{\label{sec:mec.wolf}Wolf Effect}
Coherent effect resulting from the coupling of two partially coherent sources.  May explain some features encountered in quasars and the solar limb red-shift, but cannot account for the majority of observed shifts of extra-galactic objects.  Although the strength of the mechanism is proportional to the density of gas, a relationship between distance and red-shift cannot be obtained.  The red-shift is produced at the source and appears as an intrinsic red-shift in the case of quasars\cite{Wolf1370.1986}.

\noindent \textbf{Conditions, limits of applicability and restrictions:}

Dependent on the density for a shift larger than the linewidth of the spectrum.
% Conditions, limits of applicability and restrictions: Solar limb red-shift? Quasars? Discordant red-shifts? Scale: intergalactic, interstellar, interplanetary red-shift? The assumptions required to make the mechanism work, adjustable parameters, and conflicts with currently accepted theories. Adjustable parameters with density last.

\noindent \textbf{Discussion and comments:}

The Wolf effect was demonstrated for acoustic waves.  However, a special distribution of particles acting as scatterers is required for the effect to be produced.  In Wolf's examples of red-shift in quasars, a very complex geometry is required to block the non-shifted light while allow the redshifted light to escape from the system.  The Wolf effect also produces blue-shifts which are never observed.  Because of all the requirements needed to produce the Wolf effect, it is unlikely that it is the mechanism behind quasar red-shift.
%
%                                                 ---------->>>>>>>>>>

%                                                 <<<<<<<<<<----------1998
\subsection{\label{sec:mec.moretbailly}Coherent Raman Effect on Incoherent Light}
  Moret-Bailly explains that when the propagation of light through a medium is described by Rayleigh scattering and a Huygens construction, the description provides an explanation for the index of refraction and predicts that the frequency of light is not affected.  The Coherent Raman Effect on Incoherent Light (CREIL) considers the additional coherent process of the Raman effect to the index of refraction.  The Raman effect produces frequency shifts $\omega$ on light ($\omega <0$ for a Stokes scattering).  The propagated light is then described with an incident and a transmitted beam which interfere to produce a beat on the outgoing wave described approximately by $\sin(\Omega t)+ k_\omega \sin[(\Omega +\omega)t]$.  For short pulses $\omega t << 1$ and weak scattering $k_\omega << 1$, the wave simplifies to nearly a simple sine wave at a shifted frequency $\sin[(\Omega + k_\omega \omega)t]$.

  The CREIL frequency shifting occurs as the result of several coherent Raman scatterings to produce light at a single shifted frequency.  The CREIL is space-coherent (the geometry of the light beams is not changed in homogeneous matter) and a parametric interaction (matter acts as a catalyst) between several light beams.  As a result, the images are not blurred.

  In astrophysics, it seems that CREIL frequency shifts are only produced in hydrogen excited to $n=2$ which has convenient spin coupling resonances.  These excited atoms are generated thermally (e.g. in a hot plasma such as the solar wind), or by a Ly-$\alpha$ excitation in a ``cold'' plasma ($10 000$K - $30 000$K).  Under laboratory conditions ($10$fs pulses), the CREIL is inversely proportional to the cube of the pulse duration.  With $1$ns pulses the path is increased by a factor $10^{15}$, so that the CREIL is only observed at astronomical distances\cite{MoretBaillyL35.1998, MoretBailly1215.2003}.

\noindent \textbf{Conditions, limits of applicability and restrictions:}

  The relative shift $z$ is nearly independent of $\lambda$ since $k_\omega \propto \omega$ (the calculation makes an approximation which neglects dispersion).  To fulfill Lamb's conditions, %G. L. Lamb jr., Rev. Mod. Phys., 43, 99 (1971)
the CREIL requires short pulses (such as the pulses of time-incoherent light which are $\sim 1\mbox{~ns}$ long), a low pressure gas, and a Raman active resonance in the radio frequency range ($<1\mbox{~GHz}$).  The effect is dependent on the specific atomic species and the state of the gas.

  The relevant column density is given by the density of atomic hydrogen $\mbox{H}^*$ excited in the $n=2$ state by heat or radiation.

  A density $\rho_0 = 14300 h_{100}/\mbox{~m}^3$ of $\mbox{H}^*$ and a temperature of $3\mbox{~K}$ produce the observed $d = D_H z$ for small $z$.  However, since excited atomic hydrogen is not generally found in intergalactic space, the effect will be seen locally around some objects such as quasars.
% Conditions, limits of applicability and restrictions: Solar limb red-shift? Quasars? Discordant red-shifts? Scale: intergalactic, interstellar, interplanetary red-shift? The assumptions required to make the mechanism work, adjustable parameters, and conflicts with currently accepted theories. Adjustable parameters with density last.

%\noindent \textbf{Functional relationships:}

\noindent \textbf{Discussion and comments:}

  The following effects result from the CREIL:
\\- a correlation between high red-shift and high temperature of a bright star,
\\- a ``proximity effect'' caused by the excitation of ``cold'' atomic hydrogen by the far UV radiated from a hot star,
\\- a Karlsson periodicity in the Ly-$\alpha$ forest of the quasars (also, possibly a Tifft-Napier periodicity of the red-shifts of the galaxies and ``voids'' in the Universe),
\\- an ``anomalous acceleration'' such as the one seen for the Pioneer probes, where the solar wind is cold enough to generate excited atoms,
\\- in supernovae, a red-shift along the path of emission inside a Str\"omgren sphere (such as the Ly-$\alpha$ emission seen inside the equatorial ring of SNR1987A), but a blueshift where superradiant light is emitted tangentially to the sphere where the intensity of Ly-$\alpha$ is large.  There is no red-shift where atoms are de-excited by superradiant emission,
\\- a CMB bound to the ecliptic because of the presence of thermal radiation at that location,
\\- dispersion observed in the multiplets of the quasars.

%- CREIL is not a sufficient acronym, because a simple coherent Raman effect would excite the atoms. CREIL is a parametric effect in which the atoms are catalysts, several simultaneous coherent Raman effects being involved, so that the principles of thermodynamics apply.
%
%                                                 ---------->>>>>>>>>>

%                                                 <<<<<<<<<<----------2003
\subsection{\label{sec:mec.junwang}Dispersive Extinction}
  In Wang's model, the dispersive scattering and absorption of starlight by the space medium causes a shift of Gaussian lines\cite{Wang1.2005}.

  The relevant column density is given by
$$N(d) = \int_0^d {\lambda \beta^2 \rho(x) \mbox{d}x}.$$

\textbf{Conditions, limits of applicability and restrictions:}

  The frequency shift $\Delta\lambda/\lambda$ of a spectral line is proportional to $\lambda$ and the square of its linewidth $\beta$.  Derivation assumes $z << 1$.
% Conditions, limits of applicability and restrictions: Solar limb red-shift? Quasars? Discordant red-shifts? Scale: intergalactic, interstellar, interplanetary red-shift? The assumptions required to make the mechanism work, adjustable parameters, and conflicts with currently accepted theories. Adjustable parameters with density last.

\noindent \textbf{Discussion and comments:}

  Because the red-shift results from absorption of the blue components of a spectral line, the luminosity distance increases at an exponential rate.  Extinction is very significant for large distances.  The dependence of the effect on the wavelength is problematic.  The solar limb red-shift may show a dependence on wavelength, but we know that the extinction is not as large as required by this theory.

%                                                 ---------->>>>>>>>>>

\vspace{5mm}

%                                                 <<<<<<<<<<----------2008
\subsection{\label{sec:mec.prevenslik}Quantum Electrodynamics Induced Electromagnetic Radiation in Cosmic Dust}
  Prevenslik proposes a red-shift caused by absorption of energy in submicron cosmic dust particles by the mechanism of QED induced EM radiation.  Only a single photon absorption is necessary to produce the red-shift, thus $z$ is not an indicator of distance.  QED induced red-shift may be understood by treating the absorbed photon as EM energy confined within the dust particle by total internal reflection.  Also explains the production of the CMB radiation\cite{Prevenslik11.2008}.

%\noindent \textbf{Conditions, limits of applicability and restrictions:}
% Conditions, limits of applicability and restrictions: Solar limb red-shift? Quasars? Discordant red-shifts? Scale: intergalactic, interstellar, interplanetary red-shift? The assumptions required to make the mechanism work, adjustable parameters, and conflicts with currently accepted theories. Adjustable parameters with density last.

%\noindent \textbf{Functional relationships:}

\noindent \textbf{Discussion and comments:}

  Scattering is not considered.

%
%                                                 ---------->>>>>>>>>>

%                                                 <<<<<<<<<<----------2011
\subsection{\label{sec:mec.lindsay}Quaternion Formulation of Dark Energy}
According to Lindsay, the universe is not expanding and the red-shift is not necessarily related to distance\cite{Lindsay17.2011}.  The red-shift is explained when the energy is represented as a quaternion $W = [-vh/r, c\vec{p}~]$, where $-vh/r=-GMm/r$ is the scalar potential energy and $c\vec{p}$ is the vector momentum energy, the dark energy.

  The force F is calculated from the derivative of the energy $W$:
$$F = \left[ \frac{\mbox{d}}{\mbox{d}r},~\vec{\nabla}~\right]~W = \left[ \frac{vh}{r^2} - c\vec{\nabla}\cdot\vec{p},~c\frac{\mbox{d}\vec{p}}{\mbox{d}r} - \vec{\nabla}\frac{vh}{r} + c\vec{\nabla}\times\vec{p}~\right].$$
  The presence of momentum energy produces a centrifugal force.

  Under the \textbf{Continuity Condition}, the centrifugal force $c\vec{\nabla}\cdot\vec{p}$ balances the centripetal force of gravity $vh/r^2 = vp/r$ and the force can be expressed:
$$F = \left[ \frac{vp}{r} - \frac{cp}{r}\cos(\angle\vec{p}\,\vec{r}),~c\frac{\mbox{d}\vec{p}}{\mbox{d}r} + \frac{vp}{r}~\vec{r} + \frac{cp}{r}\sin(\angle\vec{p}\,\vec{r})~\vec{r}\times \vec{p}~\right] = 0,$$
  where $\cos(\angle\vec{p}\,\vec{r})$ and $\sin(\angle\vec{p}\,\vec{r})$ are the cosine and sine, respectively, of the angle between $\vec{p}$ and $\vec{r}$, and $\vec{H}=\vec{r}\times \vec{p}$ is the angular momentum vector.

  The red-shift is $z = v/c = \cos(\angle\vec{p}\,\vec{r}).$  For equilibrium in a static universe $\cos(\angle\vec{p}\,\vec{r}) = 1$ and $v = c$ is the speed of light. 

Newton's Gravitational Constant $G$ holds the key to the properties of the Universe.  Of interest here are the radius of the Universe $R_U \approx 1.58\times 10^{26}$~m $= 16.7$~Gly, its mass $M_U = 2.133\times 10^{53}$~kg and Hubble's Constant $H_0 = c/R_U = 1.8987\times 10^{-18}$/s $= 58.48$~km/s/Mpc.

\noindent \textbf{Conditions, limits of applicability and restrictions:}

  The scalar energy is $vh/r = ch/r$ for photons, $-vh/r = -GMm/r$ for matter and $vh/r = Z\alpha ch/r$ for charged particles.

%\noindent \textbf{Functional relationships:}
%	F_\tau &= 1 + z,\\ % Ref: confirmed by author 2013/1/6
%	\eta_z &= 0,\\ % Ref: confirmed by author 2013/1/6
%	\Delta_\theta &= 0. % Ref: confirmed by author 2013/1/6

\vspace{1cm}

\noindent \textbf{Discussion and comments:}
 
  This theory gives an average density of matter of
$$\frac{M_U}{\frac{4}{3}\pi R_U^3}\frac{1}{m_P} = 7.7\mbox{~protons/m}^3.$$
%
%                                                 ---------->>>>>>>>>>

% http://arxiv.org/abs/0706.2885 is a joke.  It uses the index of refraction of the IGM as $1+\epsilon$ to calculate an increase of wavelengths, then adds them up over colmological distances!!

---

\bibliography{mechanisms}

\begin{thebibliography}{100}
\bibitem{note1}
The units of the variables are given in brackets to clarify the meaning of the
  variable.
\end{thebibliography}
\bibliographystyle{unsrt}

\end{document}